\documentclass[10pt,twocolumn,unsortedaddress,showpacs,prd,aps,amsmath,amssymb,nofootinbib]{revtex4-1}
\usepackage{graphicx}
\usepackage{amssymb}
\usepackage{bm}
\usepackage{color}
\usepackage{cancel}
\usepackage{soul}
\newcommand{\beq}{\begin{equation}} 
\newcommand{\eeq}{\end{equation}} 
\newcommand{\beqn}{\begin{eqnarray}} 
\newcommand{\eeqn}{\end{eqnarray}} 

\newcommand{\na}{\nabla}

\newcommand{\zD}{{\raise1.0ex\hbox{${}^{\ \circ}$}}\!\!\!\!\!D}
\newcommand{\alone}{{\raise0.5ex\hbox{${}^{\ 1}$}}\!\!\!\!\alpha}

\newcommand{\Dl}{\Delta}

\newcommand{\nalam}{\mathrel{\raise0.9ex\hbox{$^\lambda$}\mkern-14mu
\lower0.0ex\hbox{$\nabla$}}}

\newcommand{\Nrf}{{N_r^{\rm f}}}
\newcommand{\Nrm}{{N_r^{\rm m}}}

\newcommand{\zeroD}{{\raise1.0ex\hbox{${}^{\ \circ}$}}\!\!\!\!\!D}

\newcommand{\zLap}{{\raise1.0ex\hbox{${}^{\ \circ}$}}\!\!\!\!\Delta}
\newcommand{\zna}{{\raise1.0ex\hbox{${}^{\ \circ}$}}\!\!\!\!\!\nabla}
\newcommand{\zS}{{\raise1.0ex\hbox{${}^{\ \circ}$}}\!\!\!\!\!S}

\newcommand{\cocal}{{\sc cocal }}
\newcommand{\lorene}{{\sc lorene }}

\newcommand{\GA}{\alpha}
\newcommand{\GB}{\beta}
\newcommand{\GG}{\gamma}
\newcommand{\GD}{\delta}
\newcommand{\GE}{\epsilon}

\newcommand{\GL}{\lambda}
\newcommand{\GR}{\rho}
\newcommand{\GS}{\sigma}

\newcommand{\GC}{\psi}
\newcommand{\GX}{\chi}
\newcommand{\GO}{\omega}
\newcommand{\GZ}{\zeta}
\newcommand{\GP}{\phi}

\newcommand{\GU}{\theta}
\newcommand{\BS}{\vec{\boldsymbol{S}}}

\newcommand{\pd}{\partial}

\newcommand{\be}{\begin{equation}}
\newcommand{\ee}{\end{equation}}
\newcommand{\AJ}[3]{\emph{Astrophys. J.} \textbf{#1}, {#2} ({#3})}

\newcommand{\AJS}[3]{\emph{Astrophys. J. Suppl.} \textbf{#1}, {#2} ({#3})}

\newcommand{\ASAS}[3]{\emph{Astron. Astrophys.} \textbf{#1}, {#2} ({#3})}

\newcommand{\CQG}[3]{\emph{Class. Quantum Grav.} \textbf{#1}, {#2} ({#3})}
\newcommand{\GRG}[3]{\emph{Gen. Rel. Grav.} \textbf{#1}, {#2} ({#3})}

\newcommand{\IJMPD}[3]{\emph{Int. J. Mod. Phys. D} \textbf{#1}, {#2} ({#3})}
\newcommand{\JEM}[3]{\emph{J. Eng. Math.} \textbf{#1}, {#2} ({#3})}

\newcommand{\JMP}[3]{\emph{J. Math. Phys.} \textbf{#1}, {#2} ({#3})}

\newcommand{\LRR}[3]{\emph{Living Rev. Rel.} \textbf{#1}, {#2} ({#3})}
\newcommand{\MNRAS}[3]{\emph{Mon. Not. Roy. Astron. Soc.} \textbf{#1}, {#2} ({#3})}

\newcommand{\NATU}[3]{\emph{Nature} \textbf{#1}, {#2} ({#3})}

\newcommand{\PLA}[3]{\emph{Phys. Lett. A} \textbf{#1}, {#2} ({#3})}

\newcommand{\PRD}[3]{\emph{Phys. Rev. D} \textbf{#1}, {#2} ({#3})}
\newcommand{\PREV}[3]{\emph{Phys. Rev.} \textbf{#1}, {#2} ({#3})}

\newcommand{\PRL}[3]{\emph{Phys. Rev. Lett.} \textbf{#1}, {#2} ({#3})}
\newcommand{\PCPS}[3]{\emph{Proc. Cambridge Philos. Soc.} \textbf{#1}, {#2} ({#3})}

\newcommand{\SCIE}[3]{\emph{Science} \textbf{#1}, {#2} ({#3})}
\usepackage[]{amsmath}
\usepackage[]{dsfont}
\usepackage[]{amsfonts}
\usepackage[]{amssymb}
\usepackage[]{mathrsfs}
\usepackage{times}
\usepackage{hyperref}
\hypersetup{
  colorlinks=true,        % false: boxed links; true: colored links
  linkcolor=blue,         % color of internal links
  citecolor=cyan,         % color of links to bibliography
}
\newcommand{\cf}{cf.,~}
\newcommand{\ie}{i.e.,~}
\newcommand{\eg}{e.g.,~}

\begin{document}

\title{New code for quasiequilibrium initial data of binary neutron
  stars: Corotating, irrotational and slowly spinning systems}

\author{Antonios Tsokaros}
\affiliation{Institut f\"ur Theoretische Physik, Johann Wolfgang Goethe-Universit\"at, Max-von-Laue-Strasse 1, 
60438 Frankfurt am Main, Germany}
%\email{tsokaros@th.physik.uni-frankfurt.de}

\author{K\=oji Ury\=u}
\affiliation{Department of Physics, University of the Ryukyus, Senbaru, Nishihara, 
Okinawa 903-0213, Japan}
%\email{uryu@sci.u-ryukyu.ac.jp}

\author{Luciano Rezzolla}
\affiliation{Institut f\"ur Theoretische Physik, Johann Wolfgang Goethe-Universit\"at, Max-von-Laue-Strasse 1, 
60438 Frankfurt am Main, Germany}
\affiliation{Frankfurt Institute for Advanced Studies, Goethe University, D-60438 Frankfurt am Main, Germany}
%\email{rezzolla@th.physik.uni-frankfurt.de}

\date{\today}

\begin{abstract}
We present the extension of our \cocal~- Compact Object CALculator - code
to compute general-relativistic initial data for binary compact-star
systems. In particular, we construct quasiequilibrium initial data for
equal-mass binaries with spins that are either aligned or antialigned
with the orbital angular momentum. The Isenberg-Wilson-Mathews formalism
is adopted and the constraint equations are solved using the
representation formula with a suitable choice of a Green's function.  We
validate the new code with solutions for equal-mass binaries and explore
its capabilities for a wide range of compactnesses, from a white dwarf
binary with compactness $\sim 10^{-4}$, up to a highly relativistic
neutron-star binary with compactness $\sim 0.22$. We also present a
comparison with corotating and irrotational quasiequilibrium sequences
from the spectral code \lorene 
[Taniguchi and Gourgoulhon, Phys. Rev. D {\bf 66}, 104019 (2002)] and
%\cite{TG2002b} 
with different compactness,
showing that the results from the two codes agree to a precision of the
order of $0.05\%$. Finally, we present equilibria for spinning
configurations with a nuclear-physics equation of state in a piecewise
polytropic representation.
\end{abstract}

\maketitle

\section{Introduction}
\label{sec:intro}

With a compactness slightly smaller than that of a black hole, neutron
stars are most probably nature's ultimate compact matter configuration
before gravitational collapse and black-hole formation. As such, they
present an invaluable tool to astrophysicists in order to study a
plethora of problems and test the limits of existing knowledge, from
general relativity, via the emission of gravitational waves, to nuclear
physics, via the input on the equation of state of nuclear matter
\cite{SU2000, BGR2008, AHLLMNPT2008, LSRT2008, BTB2012}. For example, the
leading (but not unique) candidate to explain one of the most luminous
explosions in the universe, the so-called short gamma-ray bursts
\cite{NPP1992,ELPS89} (see \cite{EB2014} for a recent review) is the
merger of two neutron stars (or of one neutron star and one black hole)
with the subsequent formation of a black hole, an accretion torus, and a
jet structure of ultrastrong magnetic field
\cite{RGBGKA2011,PRS2014}. Yet another example has to do with the
production site of the heaviest elements in the universe through the
so-called, rapid neutron capture (r-process)
\cite{LS1974,LP1998,T2013,BFC2013,TH2013,RKATP2014}.

Central to the processes described above \cite{RS1999,Aea2010} is a
binary neutron star system which in addition constitutes a prime source
of gravitational waves for ground-based laser interferometric
gravitational-wave detectors such as LIGO, Virgo, KAGRA, and ET
\cite{A1992,Accadia2011,K2010,AMSAMSTY2013,P2010}. The advanced
generation of these detectors will become operational in a few years, and
they will be able to observe a volume of the universe a thousand times
more than their predecessors. According to present estimates
\cite{Abadie2010} it may be possible to detect $\sim 1-100$ events per
year, making the study of such systems an important step toward a
practical verification of general relativity in the strong field regime,
as well as an exploration of its limits. At the same time,
gravitational-wave observations are expected to constrain the
neutron-star equation of state \cite{FH2008, H2008, HLLR2010, DNV2012,
  BJ12, RBCFGKMRST2013, TRB2014a, BNBDU2014, TRB2014b}.

The broad-brush picture for the two-body problem in general relativity can
be divided into three phases: the inspiral, the merger, and the
ring-down, with each one having its own methods and tools of
investigation. The purpose of this work lies in the interface between the
first and the second phases, the so-called quasiequilibrium stage, and
the solutions presented are meant as ``snapshots'' at particular instants
of the binary system. The purpose is twofold: on the one hand to provide
initial data for the simulation of the merging phase and, on the other
hand, to provide evolutionary information about the system studied by
constructing quasiequilibrium sequences. In this way, we can learn how
much the star shape is deformed as the orbit shrinks, where an
instability sets in, where the mass-shedding limit is and what the
angular velocity of the system is there. All of this information can be
computed with a modest computational infrastructure, thus allowing for
the exploration of a wide parameter space.

Because the orbital decay time scale due to gravitational wave emission is
much shorter than the synchronization time scale due to the neutron star
viscosity, it is unlikely that the two stars will be tidally locked
before merger \cite{K1992,BC1992}. For such slowly rotating
configurations, the assumption of an irrotational flow is physically
reasonable and mathematically simple to impose. An irrotational flow is
also called a potential flow, since the fluid velocity is the gradient of
a potential \cite{RZ2013}. A formalism to compute initial data within a
conformally flat geometry, the so-called Isenberg-Wilson-Mathews (IWM)
formalism \cite{I1980,WM1989}, was presented in Refs. \cite{BGM1997, S1998,
  T1998, A1998}, and numerical implementations for a variety of physical
assumptions have been discussed by several groups \cite{BGM1999, MMW1999,
  UE2000, USE2000, GGTMB2001, TG2002b, TG2003, LRG2005, BRGHTZ2005,
  2009T,TS2010}. Nonconformally flat formulations, have also been implemented
\cite{SUF2004, ULFGS2006, ULFGS2009, SG2004, BGGN2004, YBRUF2006}, where
the full system of Einstein equations is being solved. These more
computationally expensive solutions are expected to respect the
circularity of an orbit better than the ones coming from conformally flat
initial data, which in addition seem to suppress tidal effects as the
compactness of the stars increases. A conformally flat geometry can still
be used to produce low eccentricity initial data if one uses ideas
similar to those applied to the binary black hole problem
\cite{PBKLLS2007}, as they were implemented in \cite{KST2014,MMMTB2014}.

As it was pointed in Ref. \cite{T2011}, the double pulsar PSR J0737-3039,
has one of its stars reaching the merging epoch with a spin of $\sim 27$
ms, hence in a state that cannot be considered
irrotational. Since we only know less than a dozen binary systems
\cite{L2008}, and hundreds of millisecond pulsars, it is reasonable to
simulate arbitrary spinning binary neutron stars and assess the
impact that the stellar spin has on the gravitational-wave
signal. Initial data for binary neutron stars with intermediate (and
arbitrary) rotation states are more difficult to calculate, since there
is no self-consistent scheme to incorporate the fluid equations with the
rest of the elliptic gravitational equations. Various schemes have been
proposed recently in Refs. \cite{T2011, MS2003, BS2009, T2012, TM2013},
which introduce some additional approximations, while evolution of
spinning binaries have been performed in \cite{TM2013, KGARF2013,
  BDTB2014}.

In this work we continue the \cocal~program for computing equilibrium
configurations of single \cite{HMSU2008,UGMFTE2014} and binary systems
building on the infrastructure introduced in
Refs. \cite{UT2012,UTG2012,TU2013} for binary black holes. Here, we
describe how to calculate initial data for binary stars, concentrating on
neutron stars. The ability to compute configurations with a wide range of
compactness was one of the goals of this work. At present, \cocal~
makes use of a piecewise polytropic description to represent the equation
of state (EOS), but fully tabulated EOSs can also be implemented. As in
the vacuum case, we employ the Komatsu-Eriguchi-Hachisu (KEH) method
\cite{OM1968,H1986a,H1986b,HEN1986a,HEN1986b,KEH1989} on multiple patches
\cite{TU2007} in order to be able to treat binaries consisting of
different compact objects. 
The multiple coordinates systems used in this work are not necessary
  for the computation of initial data coming from equal-mass binaries
  with spins either aligned or antialigned. In this case, in fact, it is
  possible to use a coordinate system positioned at the center of one
  compact object and employ a $\pi$ symmetry to acquire the complete
  solution. Nevertheless, we here use three different coordinate systems
  (\ie we solve all equations separately in three patches) as a first
  step toward solving for asymmetric binary systems, which will be
  presented in a future work.
The gravitational equations are solved using the \cocal~Poisson solvers
(with appropriate Green's functions), while for the conservation of rest
mass, we follow \cite{UE1998b} and employ the least-squared algorithm
demonstrating the versatility of the methods used by our code. In this
way we can calculate sequences of corotating, irrotational, as well as
spinning binaries, where for the last case we use the formulation of
\cite{T2011} after small modifications to adapt it to our numerical
  methods.

The paper is organized as follows: In Sec. \ref{sec:equations} we discuss
the equations to be solved and the assumptions made, both for the
gravitational field in Sec. \ref{ssec:grav}, as well as for the fluid
part in Sec. \ref{ssec:fluideq}. For the latter we present the forms used
in \cocal~code for corotating, irrotational, and spinning cases. Section
\ref{sec:nm} represents the core of this work. In Sec. \ref{ssec:grid} we
briefly review the gravitational multipatch coordinate systems used in
\cite{UT2012,UTG2012} and discuss additional changes that are related to
the neutron-star surface. In Sec. \ref{ssec:dimnorm} we describe the
removal of dimensions from the equations, in conjunction with the scaling
introduced in Sec. \ref{ssec:grid}. Section \ref{ssec:esgp} describes the
Green's function used for the star patch, while in Sec. \ref{ssec:esfp}
the least-squared method for solving a spinning configuration is
introduced. Tests for our new code are presented in
Sec. \ref{ssec:corotsol} for corotating binaries and in
Sec. \ref{ssec:irrotsol} for irrotational ones, while spinning solutions
with piecewise polytropes are presented in Sec. \ref{sec:spinsol}.  A
number of appendixes provides more technical details on several
topics. More specifically, Appendix \ref{sec:mam} reports the expressions
used for the calculation of the mass and angular momentum of the binary,
Appendix \ref{sec:iTOV} illustrates a different approach to obtain a solution
of the Tolmann-Oppenheimer-Volkoff (TOV) equations, while
Appendix \ref{ssec:is} describes in detail the full iteration scheme and
Appendix \ref{sec:WDs} shows tests of \cocal~in a very different regime of
compactness by considering binaries of white dwarfs. Finally,
Appendix \ref{sec:pn} reports the post-Newtonian expressions for the binding
energy and orbital angular momentum of a binary system in quasicircular
orbit, which are used as a reference.

Hereafter, spacetime indices will be indicated with Greek letters, and
spatial indices with Latin lowercase letters. The metric has signature
$-+++$, and we use a set of geometric units in which $G=c=M_\odot=1$,
unless stated otherwise.

\section{Quasiequilibrium equations}
\label{sec:equations}

In this section we review the basic equations that need to be solved to
obtain binary equilibrium configurations. Details of the initial-data
formalism can be found in \cite{MTW1973, Y1979,C2000,G2012}. Here, we
only mention the points that are relevant to the \cocal's new
developments.

\subsection{The gravitational equations}
\label{ssec:grav}

One of the most fruitful ideas in simulating the circular motion of two
bodies in general relativity was the introduction of helical-symmetry
approximation \cite{BD1992,D1994}. Solutions with such symmetry are
stationary in the corotating frame and have a long history, starting from
the electromagnetic two-body problem \cite{S1963}. Analogous solutions in
the post-Minkowski approximation have been derived in
\cite{FU2006,GU2007}. Helical symmetry was also used to obtain the first
law of binary star thermodynamics \cite{FUS2002}, as well as to produce
equilibrium configurations of binary black holes
\cite{GGB2002a,GGB2002b}.

Neglecting the loss of energy due to gravitational radiation and assuming
closed orbits for the binary system, results in the existence of a
helical Killing vector
\begin{equation} 
k^\mu := t^\mu + \Omega\GP^\mu\,,
\label{eq:hkv}
\end{equation} 
such that
\begin{equation} 
\mathscr{L}_{\boldsymbol k}g_{\GA\GB}= 0\,,
\end{equation} 
where $\mathscr{L}_{\boldsymbol k}$ is the Lie derivative along
${\boldsymbol k}$ and $\GP^i$ is the generator of rotational symmetry. In
a Cartesian coordinate system, but without loss of generality, we can
assume the generator of rotational symmetry to have components
\begin{equation}
\label{eq:phii}
\GP^i=(-y,x,0)\,. 
\end{equation}
Writing the spacetime metric in 3+1 form as
\cite{A2008,BS2010,G2012,RZ2013}
\begin{equation}
ds^2 = -\GA^2 dt^2 + \GG_{ij}(dx^i+\GB^i dt)(dx^j+\GB^j dt)\,,
\label{eq:4metric}
\end{equation}
where $\GA,\ \GB^i,\ \GG_{ij}$ are, respectively, the lapse, the shift
vector, and the three-metric on $\Sigma_t$, the generator of time
translations in the rotating frame can be expressed as
\begin{equation} 
k^\mu := \GA n^\mu + \GO^\mu\, .
\label{eq:hkv31}
\end{equation}
Here, $\GO^\mu:=\GB^\mu+\Omega\GP^\mu$ is the corotating shift, and
$n^\mu$ the unit normal to $\Sigma_t$, $n_\mu := -\GA\nabla_\mu t$. Since
$\mathscr{L}_{\boldsymbol k} \GG_{ij} = 0 = \mathscr{L}_{\boldsymbol k}
K_{ij}$, the evolution equation of $\GG_{ij}$ specifies the extrinsic
curvature in terms of the shift and the lapse
\begin{equation}
K_{ij} = \frac{1}{2\GA} (D_i\GO_j + D_j\GO_i)\,,
\label{eq:SFF}
\end{equation}
where $D$ the derivative operator associated with $\GG_{ij}$. The
traceless extrinsic curvature 
\begin{equation}
A_{ij} := K_{ij} - \frac{1}{3} K^m_{\ \,m} \gamma_{ij} = K_{ij} - \frac{1}{3} K \gamma_{ij}\,,
\label{eq:Aij}
\end{equation}
can be written in terms of the longitudinal operator $\mathbb{L}$
\begin{eqnarray}
A_{ij} & = & \frac{1}{2\GA} \left(D_i\GO_j + D_j\GO_i -\frac{2}{3}\GG_{ij}D_k \GO^k\right)   \nonumber \\
       & := & \frac{1}{2\GA} (\mathbb{L}\GO)_{ij}\,.
\label{eq:Lij}
\end{eqnarray}

Assuming a maximal and conformally flat slice \cite{I1980,WM1989}
\begin{equation}
\GG_{ij}=\GC^4 \GD_{ij},
\label{eq:3metric}
\end{equation}
the traceless extrinsic curvature Eq. (\ref{eq:Lij}) is written in terms of the shift
\begin{eqnarray}
A^{ij} & = & \frac{\GC^{-4}}{2\GA} \left(\pd^i\GB^j + \pd^j\GB^i -\frac{2}{3}\GD^{ij}\pd_k \GB^k\right)   \nonumber \\
       & = & \frac{\GC^{-4}}{2\GA} (\widetilde{\mathbb{L}}\GB)^{ij} \,.
\label{eq:Lijcf}
\end{eqnarray}
The tilde symbol on the longitudinal operator $\mathbb{L}$ denotes the fact that it
is related to the conformally flat geometry. In deriving Eq. (\ref{eq:Lijcf})
use has been made of the fact that
$\pd_i\GP^i=0=\pd_{(i}\GP_{j)}$ [\cf Eq. \eqref{eq:phii}]. 
In the conformally flat geometry, the contravariant
components of the shift remain the same as in the original
spatial geometry (\ie $\tilde{\GB}^i=\GB^i$), while this is 
not true for the covariant components. 

With the help of Eq. (\ref{eq:Lijcf}),
the constraint equations and the spatial trace of the time derivative of
the extrinsic curvature result in five elliptic equations for the
conformal factor $\GC$, the shift $\GB^i$, and the lapse
function $\GA$
\begin{eqnarray}
\nabla^2 \GC & = & -\frac{\GC^5}{32\GA^2}(\widetilde{\mathbb{L}}\GB)^{ab}(\widetilde{\mathbb{L}}\GB)^{ij}\GD_{ia}\GD_{jb} 
                    - 2\pi E\GC^5 \nonumber \\
					   & := & S^g_\GC + S^f_\GC\,,   \label{eq:hamcon} \\
\nabla^2 (\GA\GC) & = & \frac{7\GC^5}{32\GA}(\widetilde{\mathbb{L}}\GB)^{ab}(\widetilde{\mathbb{L}}\GB)^{ij}\GD_{ia}\GD_{jb} 
                        + 2\pi\GA\GC^5(E+2S)  \nonumber \\
								  & := & S^g_\GA + S^f_\GA\,,  \label{eq:trdotkij} \\
\nabla^2 \GB^i & = & -\frac{1}{3}\pd^i\pd_j \GB^j + \pd_j\ln\left(\frac{\GA}{\GC^6}\right)(\widetilde{\mathbb{L}}\GB)^{ij} 
                    + 16\pi\GA\GC^4 j^i  \nonumber \\
						 & := & S^g_\GB + S^f_\GB\,. \label{eq:momcon}
\end{eqnarray} 
We have denoted with $S^f_{\alpha}$ the sources of the Poisson-type
equations that come from the energy-momentum tensor, while with
$S^g_{\alpha}$ are the sources that come from the nonlinear part of the
Einstein tensor $G_{\GA\GB}$. The matter sources in
Eqs. \eqref{eq:hamcon}--\eqref{eq:momcon}, $S^f_\GC,\ S^f_\GA,\ S^f_\GB$,
are related to the corresponding projections of the energy-momentum
tensor
\begin{eqnarray}
 E   & := &     n_\GA n_\GB  T^{\GA\GB}\,,     \label{eq:rhoham} \\
 S   & := &     \GG_{\GA\GB} T^{\GA\GB}\,,     \label{eq:S} \\
 j^i & := & -\GG^i_\GA n_\GB T^{\GA\GB}\,, \label{eq:ji}
\end{eqnarray}
where $E$ is the energy density as measured by a ``normal'' observer,
that is, an observer with four-velocity $\boldsymbol{n}$. 
Note also that since we use $G=c=M_\odot=1$, all quantities in equations
(\ref{eq:hamcon})--(\ref{eq:momcon}) are dimensionless.  This is contrary
to some previous works (\eg \cite{UE2000}), where only $G=c=1$ was
assumed, and a procedure to remove units was applied through the use of
the adiabatic constant $K$. Here the normalization scheme used is
explained in detail in Sec. \ref{ssec:dimnorm}.

The above set of equations must be supplied with conditions on the
boundary of our computational region. Since we will consider only binary
stars, our boundary for the gravitational equations will be only at
spatial infinity, where we impose asymptotic flatness, \ie
\begin{equation}
\lim_{r\rightarrow\infty} \GC = 1\,,\qquad 
\lim_{r\rightarrow\infty} \GA = 1\,,\qquad 
\lim_{r\rightarrow\infty} \GB^i = 0\,.
\label{eq:bcpsal}
\end{equation}

We recall that a helically symmetric spacetime cannot be asymptotically
flat, because a helically symmetric binary produces an infinite amount of
radiation. Therefore conditions (\ref{eq:bcpsal})
seem to contradict assumption (\ref{eq:hkv}). In reality, the helical
symmetry is only an approximation that is valid either for long times
when the binary is widely separated, or for only a short time when the
binary is tight. In practice, the emission of gravitational radiation
reaction will lead to an inspiral, thus breaking the symmetry.

\subsection{The fluid equations}
\label{ssec:fluideq}

Let $u^\GA$ be the four-velocity of the fluid. We consider a perfect
fluid with energy-momentum tensor \cite{RZ2013,FS2013}
\begin{equation}
T_{\GA\GB}=(\GE+p)u_\GA u_\GB + pg_{\GA\GB}=\GR h u_\GA u_\GB + pg_{\GA\GB}\,, \label{eq:set}
\end{equation}
where $\GR, \GE, h$, and $p$ are, respectively, the rest-mass density, the
total energy density, the specific enthalpy, and the pressure as measured
in the rest frame of the fluid. The specific internal energy $e$ is
related to the enthalpy through\footnote{Note that many authors use $e$
  and $\epsilon$ to indicate the energy density and the specific internal
  energy, respectively.}

\begin{equation}
h:=\frac{\GE+p}{\GR}=1+e+\frac{p}{\GR}\,. 
\label{eq:enthalpy}
\end{equation}

The first law of thermodynamics, $d\GE=\GR Tds+hd\GR$, where $s$ is the
specific entropy, written in terms of the specific enthalpy $h$, reads
$dh=Tds+{dp}/{\GR}$. For isentropic configurations, like the
quasiequilibrium solutions we are seeking, given an EOS which
relates, for example, the pressure $p$ with the rest-mass energy density
$\GR$, we can see that the extra variables that enter our problem are
$\GR$ (or $p$) and the four-velocity $u^\GA$. For these new variables,
extra equations need to be used exploiting conservation laws. In
particular, from the conservation of the energy-momentum tensor
\begin{eqnarray}
0= \nabla_\GA T^{\GA\GB}& = & \GR[u^\GA\nabla_\GA(hu^\GB)+\nabla^\GB h] \nonumber \\
 & \phantom{=} & + hu^\GB\nabla_\GA(\GR u^\GA) -\GR T\nabla^\GB s\,,
\end{eqnarray}
assuming isentropic configurations and local conservation of rest
mass\footnote{More precisely, the Euler equation is the projection
  orthogonal to the fluid flow of the conservation of the energy-momentum
  tensor and leads to three distinct spatial equations. On the other
  hand, the projection along the flow of $\nabla_\GA T^{\GA\GB}=0$ yields
  a single equation expressing energy conservation \cite{RZ2013}.}
\begin{equation}
\nabla_\GA(\GR u^\GA)=0\,,  \label{eq:cbm}
\end{equation}
we arrive at the relativistic Euler equation
\begin{equation}
u^\GA\nabla_\GA(hu_\GB)+\nabla_\GB h =0\,. \label{eq:ree}
\end{equation}
Although we use four-dimensional indices, this is a fully spatial
equation, since the projection along the fluid flow is trivially
satisfied. Equations (\ref{eq:cbm}) and (\ref{eq:ree}) provide us with
four more equations for the fluid variables. If one of them is used for
the determination of $\GR$, we are left with three equations that must
determine the four-velocity $u^\GA$, which has only three independent
components.

Expressing the four-velocity as $u^\GA=u^t(1,v^i)$ and in analogy with a
Newtonian decomposition, we can split the spatial component $v^i$ into
two parts: one that follows the orbital path $\GP^i$, and one that
represents the velocity in the corotating frame $V^i$. Using the helical
Killing vector, Eq. (\ref{eq:hkv}), we can write
\begin{equation}
u^\GA=u^t (k^\GA + V^\GA)\,,  \label{eq:f4v}
\end{equation}
where the spatial part, $V^\GA=(0,V^i)$ can be considered to be the
``nonrotating'' part of the fluid flow (\ie the velocity in the
corotating frame).  The conservation of rest mass (\ref{eq:cbm}), and the
spatial projection of the Euler equation (\ref{eq:ree}), written in 3+1
form translate to
\begin{eqnarray}
\mathscr{L}_{\boldsymbol k}(\GR u^t) + \frac{1}{\GA}D_i(\GA\GR u^t V^i) = 0,   \label{eq:cbm3}\\
\GG_i^\GA \mathscr{L}_{\boldsymbol k}(hu_\GA) + D_i\left(\frac{h}{u^t} + hu_j V^j\right) \qquad\qquad \nonumber  \\
           + V^j(D_j (hu_i) - D_i (hu_j)) = 0\,. \label{eq:ree3}
\end{eqnarray}
The last term of Eq. (\ref{eq:ree3}) involves
the relativistic vorticity tensor \cite{RZ2013}
\begin{equation}
\GO_{\GA\GB}:=\nabla_{\GA}(hu_\GB) - \nabla_{\GB}(h u_\GA)\,   \label{eq:rvt}
\end{equation}
and is zero for an irrotational flow. It is not difficult to show that in
the presence of a generic Killing vector field (\eg the helical Killing
field) $\boldsymbol{k}$, the following identity holds \cite{RZ2013}
\begin{equation}
\label{eq:symmetry}
\mathscr{L}_{\boldsymbol u}(h \boldsymbol{u} \cdot \boldsymbol{k}) = 0\,.
\end{equation}
In the case of a rigid corotation of the binary system, $\boldsymbol{u} =
\boldsymbol{k}$, so that the Lie derivative along the fluid four-velocity
$\boldsymbol{u}$ in Eq. \eqref{eq:symmetry} can be replaced by the Lie
derivative along the helical Killing vector $\boldsymbol{k}$. 
This yields $\mathscr{L}_{\boldsymbol k}h = 0$ and expresses that in the
corotating frame the fluid properties do not change. When
the fluid four-velocity does not coincide with the helical Killing field,
but the two vector fields are not too different, \ie when $\boldsymbol{u}
\simeq \boldsymbol{k}$, expression \eqref{eq:symmetry} can still be true
and indeed for the flows we will consider hereafter we will assume 
the following assumption
\begin{equation}
\GG_i^\GA\mathscr{L}_{\boldsymbol k}(hu_\GA) = 0 =
\mathscr{L}_{\boldsymbol k}(\GR u^t)\,.
\label{eq:statass}
\end{equation}
While Eq. \eqref{eq:statass} is an assumption, its correctness can only
be assessed a posteriori and could indeed not represent a valid
approximation if the stars are spinning very rapidly, a case we will not
investigate here.

In the following we will specialize Eqs. (\ref{eq:cbm3}) and
(\ref{eq:ree3}) under the assumptions (\ref{eq:statass}) for corotating,
irrotational and slowly rotating flows. Before closing we introduce a
quantity that will be used often in subsequent sections, namely, the
spatial projection of the specific enthalpy current
\begin{equation}
\hat{u}_i := \GG_i^\GA hu_\GA\,.
\end{equation}

\subsubsection{Corotating binaries}
\label{sssec:corot}

The corotating case, also called of rigid rotation \cite{B1965,BW1971},
is the simplest case, since the spatial fluid velocity $V^i$ vanishes,
$u^\alpha = u^t k^\alpha$, and thus the fluid is at rest in a corotating
frame. This means that apart from the gravitational variables
$\GC,\GA,\GB^i$, we have only two extra fluid variables, for example
$\GR$ and $u^t$ once an EOS is fixed. The conservation of rest mass
(\ref{eq:cbm3}) is trivially satisfied, while the Euler equation
(\ref{eq:ree3}) becomes a single integral equation that, together with
the normalization condition $u^\GA u_\GA=-1$, will determine all our
fluid variables.

In particular, the specific enthalpy current becomes
\begin{equation}
\hat{u}^i = hu^t \GO^i\,,   \label{eq:corotuih}
\end{equation}
and from the four-velocity normalization condition we have
\begin{equation}
u^t = \frac{1}{\sqrt{\GA^2 - \GO_i \GO^i}}\,.
\label{eq:corotut}
\end{equation}
Equation (\ref{eq:ree3}), on the other hand, has the first integral
\begin{equation}
\frac{h}{u^t} = C\,,  \label{eq:corotei}
\end{equation}
where $C$ is a constant to be determined. Equations (\ref{eq:corotut})
and (\ref{eq:corotei}), together with the gravitational potentials,
completely determine the solution for this case. We note that in all
equations to be solved (the gravitational ones included) two constants are
involved. One is $C$, the constant that comes from the Euler integral,
and one is $\Omega$, the orbital angular velocity. Thus, in order to be
able to achieve a solution for our system, a self-consistent scheme that
involves the determination of both $C$and $\Omega$ must be employed. As
we will elaborate later on, this will be achieved in conjunction with the
determination of the length scale $R_0$ of our problem.

For the corotating case the matter sources in
Eqs. (\ref{eq:hamcon})--(\ref{eq:momcon}) are
\begin{eqnarray}
E      & = & \GR [h(\GA u^t)^2-q]\,,   \label{eq:corot_sou_psi}  \\
E + 2S & = & \GR[h[3(\GA u^t)^2-2]+5q]\,,  \label{eq:corot_sou_alpha} \\
j^i              & = & \GR\GA u^t \hat{u}^i\,,   \label{eq:corot_sou_shift}
\end{eqnarray}
where $q:=p/\GR$. As we have already
mentioned, all quantities appearing above are dimensionless, while in
previous studies, where geometric units were used,
Eqs. (\ref{eq:corot_sou_psi})--(\ref{eq:corot_sou_shift}) had units of
length$^{-2}$.

\subsubsection{Irrotational and spinning binaries}
\label{sssec:irsp}

Irrotational configurations have $\GO_{\GA\GB}=0$, so that the specific
enthalpy current $hu_\GA$ can be derived from a potential \cite{RZ2013}
\ie
\begin{equation}
\label{eq:hua_irr}
hu_\GA=\nabla_\GA \Phi=D_\GA\Phi +
n_\GA\mathscr{L}_{\boldsymbol n}\Phi\,, 
\end{equation}
so that $\hat{u}_i = D_i\Phi$ since $\boldsymbol{\gamma}\cdot
\boldsymbol{n}=0$. To allow for spinning configurations we need to extend
expression \eqref{eq:hua_irr} and we do this following Ref. \cite{T2012}
and introducing a four-vector $\boldsymbol{s}$ (not to be confused with
the specific entropy $s$)
\begin{equation}
\label{eq:hua_irr_spin}
hu_\GA=\nabla_\GA \Phi + s_{\alpha} \,, 
\end{equation}
so that $\hat{u}^i$ is decomposed as
\begin{equation}
\hat{u}^i = D^i\Phi + s^i\,,  \label{eq:uih_dec}
\end{equation}
where the $D^i\Phi$ part corresponds to the ``irrotational part'' of the
flow and the $s^i$ part to the ``spinning part'' of the flow. In what
follows we will present expressions for spinning binaries ($s^i\neq 0$)
and one can recover the irrotational ones by setting the spinning
component $s^i$ equal to zero.

Using the decomposition (\ref{eq:uih_dec}) and the assumption
(\ref{eq:statass}), the Euler equation (\ref{eq:ree3}) can be rewritten
as
\begin{equation}
\mathscr{L}_{_{\boldsymbol V}}s_i + D_i\left(\frac{h}{u^t} + 
V^jD_j\Phi\right)= 0\,, 
\label{eq:ree3a}
\end{equation} 
and hereafter we will assume
\begin{equation}
\mathscr{L}_{_{\boldsymbol V}}s_i = 0\,,
\label{eq:rotass}
\end{equation}
which is likely to be a very good approximation in the case of slowly and
uniformly rotating stars, for which $s_i$ is intrinsically small.\footnote{
In practice we will consider stars with spin period down to $0.6\ {\rm ms}$,
but this is still "slowly" spinning when compared to the minimum period.} 
Hence, the Euler equation for generic binaries \eqref{eq:ree3}
\begin{equation}
\GG_i^\GA \left[\mathscr{L}_{\boldsymbol k}(hu_\GA) + 
\mathscr{L}_{\boldsymbol V}(s_\GA) \right] + 
D_i\left(\frac{h}{u^t} + V^j D_j\Phi\right) = 0\,, 
\end{equation}
under the assumptions \eqref{eq:statass}$_1$
and \eqref{eq:rotass}, yields the reduced Euler integral
\begin{equation}
\frac{h}{u^t} + V^j D_j\Phi = C\,, 
\label{eq:irspei}
\end{equation}
where again $C$ is a constant to be determined. A few remarks should be
made at this point. First, it is not difficult to obtain the following
identity
\begin{eqnarray}
\label{eq:equiv}
\GG_i^\GA [ 
\mathscr{L}_{\boldsymbol k}(hu_\GA) &+& 
\mathscr{L}_{\boldsymbol V}(s_\GA) ] =  \nonumber \\
\GG_i^\GA  [ \mathscr{L}_{\boldsymbol k }(\nabla_\GA\Phi) &+&
\mathscr{L}_{\boldsymbol{\nabla}\Phi/(hu^t)}(s_\GA)  +
\mathscr{L}_{\boldsymbol{s}/(hu^t)}(s_\GA)  ]\,,
\end{eqnarray}
so that our assumptions \eqref{eq:statass}$_1$ and \eqref{eq:rotass}
\begin{equation}
\label{eq:conds_tur}
\gamma^\alpha_i \mathscr{L}_{_{\boldsymbol k}} (h u_{\alpha}) = 0 =
\gamma^\alpha_i \mathscr{L}_{_{\boldsymbol V}}(s_\alpha) \,,
\end{equation}
are equivalent to setting the left-hand side of Eq. \eqref{eq:equiv} to
zero. In turn, this implies that also the right-hand side of
\eqref{eq:equiv} is zero, which is true if, for instance, each of the
three terms is zero, \ie if
\begin{equation}
\label{eq:conds_tichy}
\GG_i^\GA \mathscr{L}_{\boldsymbol{k}}(\nabla_{\alpha}\Phi) = 0 = 
\GG_i^\GA \mathscr{L}_{\boldsymbol{\nabla}\Phi/(hu^t)}(s_\alpha) = 
\GG_i^\GA \mathscr{L}_{\boldsymbol{s}/(hu^t)}(s_\alpha) \,.
\end{equation}
The three conditions in \eqref{eq:conds_tichy} coincide with the
assumptions made in \cite{T2011}. Stated differently, because the
conditions \eqref{eq:conds_tur} are compatible with the conditions
\eqref{eq:conds_tichy}, it does not come as a surprise that we obtain the
same Euler integral \eqref{eq:irspei} as in  \cite{T2011} despite making
apparently different assumptions [\cf \eqref{eq:conds_tur} vs
  \eqref{eq:conds_tichy}]. Second, using the decomposition
\eqref{eq:hua_irr_spin}, it follows that
\begin{equation}
\GG_i^\GA  \mathscr{L}_{\boldsymbol k }(hu_\GA) = 
\GG_i^\GA [ \mathscr{L}_{\boldsymbol k }(\nabla_\GA\Phi) +
\mathscr{L}_{\boldsymbol k }(s_\GA)  ]\,,
\end{equation}
and hence the question about $\mathscr{L}_{\boldsymbol k }(hu_\GA)=0$
  depends on both $\mathscr{L}_{\boldsymbol k }(\nabla_\GA\Phi)$ and
  $\mathscr{L}_{\boldsymbol k}(s_\GA)$ being zero\footnote{Note that even
    when the spins are aligned with the orbital angular momentum
    $\mathscr{L}_{\boldsymbol k}(s_\GA) \neq 0$.}. The second term is
  essentially an input to our problem, while the first one comes from the
  conservation of rest mass (\ref{eq:cbm4}), which depends on the spin
  input $s_\GA$.
%
%% \re{and hence the question about $\mathscr{L}_{\boldsymbol k }(hu_\GA)=0$
%%   being a valid assumption or not, depends on both terms:
%%   $\mathscr{L}_{\boldsymbol k }(\nabla_\GA\Phi)$, and
%%   $\mathscr{L}_{\boldsymbol k}(s_\GA)$. The second one is essentially an
%%   input to our problem, while the first comes from the conservation of
%%   rest mass (\ref{eq:cbm4}), which depends on the spin input $s_\GA$. In
%%   that sense we consider \eqref{eq:conds_tichy} "equivalent" to
%%   \eqref{eq:conds_tur} and although further research at this point is
%%   needed, this will leave our solutions unaffected.}  \st{and hence that
%%   considering a flow with $\mathscr{L}_{\boldsymbol k
%%   }(s_\GA)=0$,\footnote{Note that even when the spins are aligned with
%%     the orbital angular momentum $\mathscr{L}_{\boldsymbol k}(s_\GA) \neq
%%     0$.}  is not sufficient for having $\GG_i^\GA
%%   \mathscr{L}_{\boldsymbol k }(hu_\GA) = 0$, but the additional
%%   assumption $\GG_i^\GA \mathscr{L}_{\boldsymbol k }(\nabla_\GA\Phi) =0$
%%   is necessary.}  
%
Finally, although the Euler integral has the same form
for both irrotational and spinning binaries, it produces a different
equation since the three-velocity $V^i$ is different in these two
cases. More specifically, it is
\begin{equation}
\hat{u}^i = hu^t (\GO^i + V^i)\,,
\end{equation}
so that
\begin{equation}
V^i=\frac{D^i\Phi+s^i}{hu^t}-\GO^i\,. \label{eq:irspuih}
\end{equation}

In this case, the fluid variables are $\GR$ (or equivalently $p$ or $h$),
$u^t$, and the fluid potential $\Phi$. The equations that will determine
them are the normalization condition $u_\GA u^\GA=-1$, the Euler integral
(\ref{eq:irspei}) [with the use of Eq. (\ref{eq:irspuih})], and the
conservation of rest mass (\ref{eq:cbm3}).

In particular, from the norm of $\hat{u}^i$ we get  
\begin{equation}
h = \sqrt{\GA^2(hu^t)^2 - (D_i\Phi+s_i)(D^i\Phi+s^i)}\,, 
\label{eq:irsph}
\end{equation}
therefore, the Euler integral (\ref{eq:irspei}) takes
the following form quadratic in $hu^t$
\begin{equation}
\GA^2(hu^t)^2 - \GL (hu^t) - s_i (D^i\Phi + s^i)=0\,,  \label{eq:irsphut}
\end{equation}
where $\GL=C+\GO^i D_i\Phi$. Thus
\begin{equation}
hu^t=\frac{\GL+\sqrt{\GL^2+4\GA^2 s_i(D^i\Phi + s^i)}}{2\GA^2}\,,   
\label{eq:irsphutsol}
\end{equation}
where we take the positive root since the negative one is incorrect at
least in the limit of $s^i = 0$, when it yields $h u^t=0$.

Having computed $\GL$, we first calculate $hu^t$ from
Eq. (\ref{eq:irsphutsol}), and then $h$ from Eq. (\ref{eq:irsph}). For
purely irrotational binaries $hu^t=\GL/\GA^2$ and $h=\sqrt{\GL^2/\GA^2 -
  D_i\Phi D^i\Phi}$.

Although we will not make immediate use of $u^t$ and $h$ separately, we
report below their form for completeness
\begin{eqnarray}
h   & = & \sqrt{L^2-(D_i\Phi+s_i)(D^i\Phi+s^i)}\,, \\
u^t & = & \frac{\sqrt{h^2+(D_i\Phi+s_i)(D^i\Phi+s^i)}}{h\GA}\,,
\end{eqnarray}
where
\begin{eqnarray}
L^2 & := & \frac{\GL^2+2\GA^2 W+\GL\sqrt{\GL^2+4\GA^2 W}}{2\GA^2}\,, \\
W & := & s_i(D^i\Phi+s^i)\,.
\end{eqnarray}

The potential $\Phi$ will be computed from the conservation of rest mass
(\ref{eq:cbm3}), which under Eq. (\ref{eq:irspuih}), and after expressing
the spin velocity as a power law \cite{T2011}
\begin{equation}
s^i = \GC^{A} \tilde{s}^i\,,\qquad \qquad A\in\mathbb{R}   
\label{eq:wi}
\end{equation}
will produce an extra elliptic equation
\begin{eqnarray}
\nabla^2\Phi & = & -\frac{2}{\GC}\pd_i\GC\pd^i\Phi + \GC^4 \GO^i\pd_i(hu^t)  \nonumber \\
             & \phantom{=} &+   [\GC^4hu^t \GO^i-\pd^i\Phi]\pd_i\ln\left(\frac{\GA\GR}{h}\right)  \nonumber \\
						 & \phantom{=} & -\GC^{4+A}\left[\pd_i \tilde{s}^i + \tilde{s}^i 
						       \pd_i\ln\left(\frac{\GA\GR\GC^{6+A}}{h}\right)\right]
= S_\Phi\,. \nonumber \\
\label{eq:cbm4}
\end{eqnarray}
The boundary for the fluid is represented by the surface of the star;
hence the boundary condition for Eq. (\ref{eq:cbm4}) will be of
von Neumann type, that is, in terms of derivatives of the rest-mass
density and of $\Phi$
\begin{equation}
\left[\left(\GC^4hu^t \GO^i-\pd^i\Phi-\GC^{4+A}\tilde{s}^i\right)
       \pd_i\GR \right]_{{\rm surf.}} =0\,. 
\label{eq:bcphi}
\end{equation}

A possible and convenient choice for the parameter $A$ that will be used
in Sec.~\ref{sec:spinsol} is $A=-6$, as it removes the last term in
Eq. \eqref{eq:cbm4}\footnote{More precisely, $A=-6$ makes the spin have
  zero divergence in the three-geometry ($D_i s^i =0$) if we choose it to
  have zero divergence in the conformal three-geometry ($\tilde{D}_i
  \tilde{s}^i =0$).}. Any other value will not change the character of
the equation or the boundary condition, although it will change the
detailed properties of the flow velocity and therefore of the binary. We
will comment on this point in Sec.~\ref{sec:spinsol}, where we will also
illustrate the results for $A=0$.

For the spinning case, the matter sources in
Eqs. (\ref{eq:hamcon})--(\ref{eq:momcon}) are
\begin{eqnarray}
E      & = & \GR \left[\frac{\GA^2}{h}(hu^t)^2-q\right]\,,   \label{eq:irsp_sou_psi}  \\
E + 2S & = & \GR \left[3\frac{\GA^2}{h}(hu^t)^2-2h+5q\right]\,,  \label{eq:irsp_sou_alpha} \\
j^i              & = & \GR\GA u^t \hat{u}^i = 
          \GR \frac{\GA}{h} (hu^t) \left[\GC^{-4}\pd^i\Phi+\GC^{A}\tilde{s}^i\right]\,,   \label{eq:irsp:sou_shift}
\end{eqnarray}
where we have used Eq. (\ref{eq:irsphutsol}) to simplify the
calculations.

\subsection{Equation of state}
\label{ssec:eos}

The EOS used in this work is represented by a sequence of polytropes
called a piecewise polytrope. This is proven to be a good approximation
for a great variety of models
\cite{RLOF2009,RMSUCF2009,DH2001,APR1998}. If $N$ is the number of such
polytropes, in each piece the pressure and the rest-mass density are
\begin{equation}
p = K_i \GR^{\Gamma_i}\,,\qquad i=1,2,\ldots,N\,. 
\label{eq:pp}
\end{equation}
The order of the polytropes is $i=1$ for the crust, and $i=N$ for the
core, and Eq. (\ref{eq:pp}) holds for
\begin{equation}
\GR_{i-1}\ \leq\ \GR\ <\ \GR_i\,.
\label{eq:ppint}
\end{equation}
As we have discussed in Sec. \ref{ssec:fluideq}, the first law of
thermodynamics for isentropic configurations gives $dh=dp/\GR$, which can
be expressed in terms of $q$ to yield
\begin{equation}
  dh = \frac{\Gamma_i}{\Gamma_i-1}dq\,,
\end{equation}
or equivalently 
\begin{equation}
h-h_i=\frac{\Gamma_i}{\Gamma_i-1}(q-q_i)\,,  
\label{eq:h}
\end{equation}
where $h_i,q_i$ correspond to values at the
right end point (the one closest to the core) of the $i$-th interval. In
terms of $q$, we can express the rest of thermodynamic variables as
\begin{eqnarray}
\GR &\ =\ &  K_i^{1/(1-\Gamma_i)} q^{{1}/{(\Gamma_i-1)}} \,, \label{eq:rho0}\\
p &\ =\ &  K_i^{1/(1-\Gamma_i)} q^{{\Gamma_i}/(\Gamma_i-1)} \,, \label{eq:P}\\
	 \epsilon &\ =\ & \GR h -p\,. \label{eq:ened}
\end{eqnarray}
Enforcing the continuity of the pressure at the $N-1$ interfaces of each
interval constraints all adiabatic constants $K_i$ but one
\begin{equation}
      K_i \GR_i^{\Gamma_i}\ =\ K_{i+1}\GR_i^{\Gamma_{i+1}} \,. 
\label{eq:contP}
\end{equation} 
As a result, the free parameters are: one adiabatic constant, $N-1$
rest-mass densities, and $N$ adiabatic indices, a total of $2N$
parameters.

\section{Numerical method}
\label{sec:nm}

The \cocal~code for binary black holes has been described in detail in
Refs. \cite{UT2012,UTG2012,TU2013}. Here, we will review the most salient
features of the grids used for the solution of the field equations and
discuss the differences that arise from the treatment of binary stars.

When treating binary systems, \cocal~employs two kinds of coordinate
systems. The first kind is compact object coordinate
patch (COCP) and has exactly two members, one centered on each star. The
second kind is asymptotic region coordinate patch (ARCP)
and can have in principle any number of members in an onion type of
structure. In our computations, the ARCP patch has only one member, which
is centered on the center of mass of the system. All coordinate systems
use spherical coordinates $(r,\GU,\GP)\in [r_a,r_b] \times [0,\pi] \times
[0,2\pi]$, but the components of field variables (like the shift) are
written with their Cartesian components $(\GB^x,\GB^y,\GB^z)$. The values
of $r_a,\ r_b$ depend on the compact object (black hole or neutron star)
and the coordinate patch (COCP or ARCP). For the binary systems treated
here, $r_a$ of the COCP patch will always be zero, while for the ARCP
patch $r_a\approx \mathcal{O}(10 M)$, $M$ being the mass of the star. As
for $r_b$, the values are kept the same as in the binary black hole
computation, \ie $\mathcal{O}(10^2 M)$ for the COCP patch, and
$\mathcal{O}(10^6 M)$ for the ARCP patch.

The orientation of the coordinate patches is as follows: the ARCP patch
has the familiar $(x,y,z)$ orientation, the first COCP patch, which is
centered on the negative $x$ axis of the ARCP patch, has the same
orientation as the ARCP patch, while the second COCP patch, which is
centered on the positive $x$ axis of the ARCP patch, has negative $(x,y)$
orientation, and positive $z$ orientation with respect to the ARCP
patch and to the first COCP patch. In other words the coordinate system
of the second COCP patch is obtained from the first COCP patch by
a rotation through an angle of $\pi$.

\begin{table}
\begin{tabular}{rll}
\hline
\hline
$r_a$: & Radial coordinate where the radial grids start. For      \\
\phantom{:}& the COCP patch it is $r_a=0$. \\
$r_b$: & Radial coordinate where the radial grids end. \\
$r_c$: & Center of mass point. Excised sphere is located   \\
\phantom{:}& at $2r_c$ in the COCP patch. \\
$r_e$: & Radius of the excised sphere. Only in the COCP patch. \\
$r_s$: & Radius of the sphere bounding the star's surface. \\
\phantom{:}& It is $r_s\leq 1$. Only in COCP. \\
$N_{r}$: & Number of intervals $\Dl r_i$ in $r \in[r_a,r_{b}]$. \\
$N_{r}^{1}$: & Number of intervals $\Dl r_i$ in $r \in[0,1]$. Only \\
\phantom{:}& in the COCP patch. \\
$\Nrf$: & Number of intervals $\Dl r_i$ in $r \in[0,r_s]$ in the COCP patch \\
\phantom{:}& or $r \in[r_a,r_a+1]$ in the ARCP patch. \\
$\Nrm$: & Number of intervals $\Dl r_i$ in $r \in[r_a,r_{c}]$. \\
$N_{\theta}$: & Number of intervals $\Dl \theta_j$ in $\theta\in[0,\pi]$. \\
$N_{\phi}$: & Number of intervals $\Dl \phi_k$ in $\phi\in[0,2\pi]$. \\
$d$: & Coordinate distance between the center of $S_a$ ($r=0$) \\
\phantom{:}& and the center of mass. \\
$d_s$: & Coordinate distance between the center of $S_a$ ($r=0$) \\
\phantom{:}& and the center of $S_e$. \\
$L$: & Order of included multipoles. \\
\hline
\hline
\end{tabular}  
\caption{Summary of grid the parameters used for the binary systems
  computed here.}
\label{tab:grid_param}
\end{table}

The geometry of the ARCP patch (or any number of them) is that of a solid
spherical shell with inner radius $r_a$ and outer radius $r_b$. On the
other hand, the COCP patch geometry is that of a sphere of radius $r_b$,
with another sphere of radius $r_e$ at distance $d_s$ from the center,
being removed from its interior. This second sphere whose boundary we
call the excised sphere $S_e$, is centered on the $x$ axis around the
\textit{other} compact object. For the first COCP patch, its excised
sphere $S_e$ is centered on the position of the second star, while for
the second COCP patch, its excised sphere $S_e$ is centered on the
position of the first star. The size of every sphere $S_e$ is slightly
larger than the star resolved with an opening half-angle of $\sim \pi/3$
as seen from the origin of the COCP patch. This is done to resolve
accurately the region around the \textit{other} star and 
reduce the number of multipoles used in the Legendre expansion. In this work we typically use
12 multipoles in our computations.
Table \ref{tab:grid_param}
summarizes the properties of the various coordinate patches used and
which are illustrated schematically in Fig. \ref{fig:cocp_radial}.
%% \re{Note that all quantities appearing in Fig. \ref{fig:cocp_radial} are
%%   rescaled as discussed in detail in Sec. \ref{ssec:dimnorm}. Hence, for
%%   consistency they should be denoted with a hat, but we do not do it for .}

On the coordinate grids described above we solve the Poisson-type
  nonlinear equations \eqref{eq:hamcon}, \eqref{eq:trdotkij}, and
  \eqref{eq:momcon}. These equations are solved using the representation
  formula with a suitable choice of a Green's function (this is also
  known as the KEH method \cite{KEH1989}). The Green's function is
  expanded in terms of spherical harmonics and the integrals (both volume
  and surface) are performed on the spherical coordinate grids described
  above and in more detail in Sec. \ref{ssec:grid} for the case of binary
  stars.

The numerical approximations introduced in \cocal are of different
  types: First, from the truncation of the series of the Legendre
  expansion; second, from the solution of the equations on discretized
  grids via finite-difference methods \cite{RZ2013}. Typically, we use
  second-order centered stencils for the numerical differentiations and
  integrations \cite{UT2012}. An exception is the use of a third-order
  finite-difference stencil for the radial derivatives and the use of a
  fourth-order integration in the polar coordinate \cite{UTG2012}.  We
  also typically use second-order interpolations of scalar functions from
  grid points to midpoints. Furthermore, when a function needs to be
  interpolated from one coordinate system to another, we use a
  fourth-order Lagrange interpolation. This usually happens when we
  compute the surface integral at the excised sphere $S_e$ for in that
  case we need the potential and its derivative on $S_e$ as seen from the
  center of $S_e$.%% These values are computed by interpolating the
  %%nearby $64$%% points of the other COCP.

\subsection{The numerical grids}
\label{ssec:grid}

In Refs. \cite{UT2012,UTG2012} we described in detail optimal numerical
grids that we constructed in order to lower the error of the potentials
for both close and large separations, for any kind of mass ratio. In
particular, when we computed sequences, instead of keeping the radii of
the black holes the same and increasing their separation, we kept the
separation fixed and decreased their radii. By choosing the interval
separation near the black holes according to their excised radius, we
were able to obtain sequences comparable to the ones produced by spectral
methods.

\begin{figure*}
  \includegraphics[height=100mm]{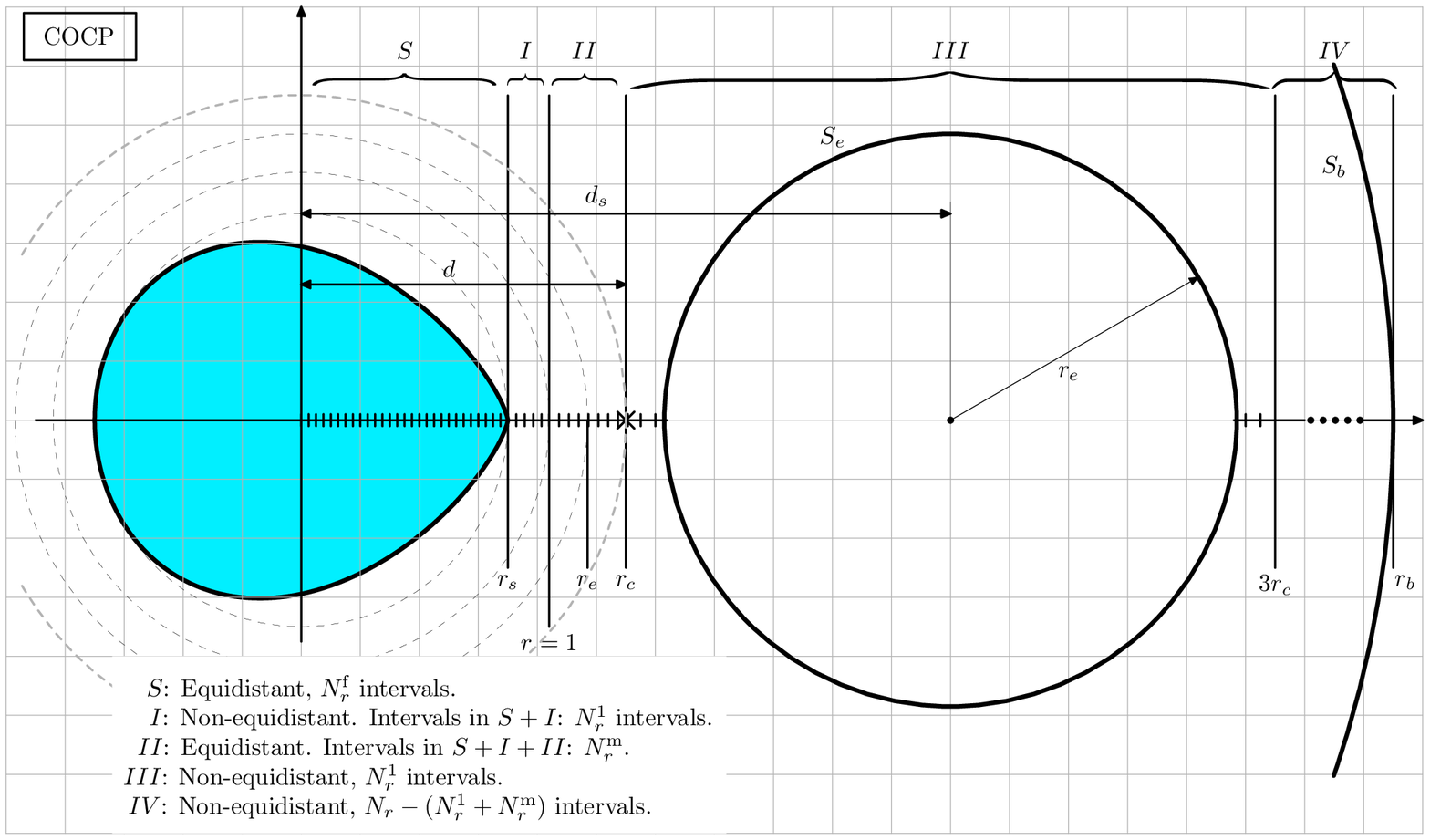}
	\caption{Structure of the radial grid for the COCP coordinate
          system for binary stars. $r_s$ can take any value in $(0,1]$,
        but typical values used are in the range $[0.5,1]$. Decreasing
        $r_s$ amounts to a larger effective distance between the two
        stars.}
	\label{fig:cocp_radial}
\end{figure*}

We adopt here the same philosophy for the computation of binary
stars. Contrary to previous studies \cite{UE2000}, in order to compute
sequences of binary stars we let the maximum radius of the star be
variable and we denote by $r_s$ the infimum of the radii of all spheres
bounding the star that are centered on the origin of the COCP patch. By
continuously diminishing $r_s$ while keeping the distance between the
stars constant, we can compute sequences of stars with a continuously
increasing separation. In this way we can control the region around the
excised sphere as described in \cite{UT2012,UTG2012}, while maintaining
the accuracy in the area covered by the neutron star.
   
The \cocal~radial grid for binary stars can be seen in
Fig. \ref{fig:cocp_radial}, where all the radial distances
are the normalized quantities discussed in Sec. \ref{ssec:dimnorm}.
In that sense, they should be denoted with a hat, for example $\hat{r}$,
which is omitted here for simplicity.
When comparing with Fig. 2 of
\cite{UT2012}, we can observe that there is also an important difference
in the notation. This regards the quantity $\Nrf$, which previously was
used to denote the number of intervals in $[r_a,1]$, while here is used
to denote the number of intervals in $[r_a,r_s]=[0,r_s]$, with $r_s\leq
1$. The number of intervals in $[0,1]$ is denoted by a new variable
called $N_r^1$ (this plays the role of the old $\Nrf$ for grid
comparisons). This change was necessary since in all previous studies
\cite{UE2000}, the surface of the star was bounded by the fixed $r=1$
sphere, and therefore the fluid extended until that point. To compute
stars at larger separation while satisfying this constraint, we would have
to increase the position of the excised sphere $S_e$ and therefore expand
all grid quantities analogously. To avoid such a complication, we
introduce a variable $r_s$ that effectively mimics the change in
separation. By varying the end point of the fluid (point $r_s$) we achieve
the same result as varying the distance between the stars, but we
maintain the good convergence properties that were established in
\cite{UTG2012} while maintaining our fluid code essentially unchanged.

As we can see from Fig. \ref{fig:cocp_radial}, there are five regions in
the COCP patch of a star that are denoted by $S,I,II,III,$ and $IV$. The
star is resolved by a constant grid spacing $\Dl r=r_s/\Nrf$, as region
$II$, which has spacing $\Dl r_2=1/N_r^1$. Setting $\Dl r_i := r_i -
r_{i-1}$, the grid intervals in each of them are
\begin{eqnarray}
\Dl r_{i\phantom{+1}} &=& \phantom{h_s} \Dl r\,,\ \ \ \mbox{for}\ \  i = 1, \cdots, \Nrf-1\,,    \\
\Dl r_{i+1} &=& h_1 \Dl r_i\,,\ \ \mbox{for}\ \  i = \Nrf, \cdots, N_r^1 -1\,,  \\
\Dl r_{i\phantom{+1}} &=& \phantom{h_2} \Dl r_2,\ \ \mbox{for}\ \  i = N_r^1, \cdots, \Nrm\,,  \\
\Dl r_{i+1} &=& h_3 \Dl r_i\,,\ \ \mbox{for}\ \  i = \Nrm, \cdots, \Nrm+N_r^1-1\,, \ \   \\
\Dl r_{i+1} &=& h_4 \Dl r_i\,,\ \ \mbox{for}\ \  i = \Nrm+N_r^1, \cdots, N_r-1\,, \ \  
\label{eq:dri} 
\end{eqnarray} 
which correspond to regions $S, I, II, III$, and $IV$, respectively. The
ratios $h_i(> 1)$ $(i=1,3,4)$ are, respectively, determined from the relations
\begin{eqnarray} 
1-r_s &=& \Dl r \frac{h_1(h_1^{N_r^1-\Nrf}-1)}{h_1-1}\,, \\ 2 r_c &=&
\Dl r \frac{h_3(h_3^{\Nrf}-1)}{h_3-1}\,, \\ 
r_b - 3r_c &=& \Dl r \frac{h_4(h_4^{N_r-\Nrm-\Nrf}-1)}{h_4-1}\,.
\label{eq:coord_r_ratio_k}
\end{eqnarray}

For the ARCP coordinate system, there are in general two regions, one
with constant grid spacing and one with increasing spacing. The grid
intervals in these regions are defined by
\begin{align}
\Dl r_{i} & = \phantom{h} \Dl r_1\,,& &\mbox{for}\ \  i = 1, \cdots, \Nrm\,,  \\
\Dl r_{i+1} & =  k \Dl r_i\,,&  &\mbox{for}\ \  i = \Nrm, \cdots, N_r-1\,, 
\end{align}
where $\Dl r_1=1/\Nrf$, and  the ratio $k$ is determined from 
\begin{equation}
r_b - r_c =: \Dl r \frac{k(k^{N_r-\Nrm}-1)}{k-1}\,. 
\label{eq:arcp_ratio_k}
\end{equation}

As regards the angular resolution, we keep the same grid interval in the
$\GU$ and $\GP$ directions and therefore
\begin{eqnarray}
\Delta\GU_j &=& \theta_j - \theta_{j-1} \,=\, \Dl \theta \,=\, 
\frac{\pi}{N_\theta}\,, \\
\Delta\GP_k &=& \phi_k - \phi_{k-1} \,=\, \Dl \phi \,=\, \frac{2\pi}{N_\phi}\,. 
\label{eq:dang}
\end{eqnarray}

One of the additional complications of having to deal with the fluid of a
star, instead of a vacuum spacetime, is the need to accurately find its
surface. This surface may contract or expand during the calculation,
creating significant problems in close binary configurations. One very
effective solution to these issues \cite{UE1998} is the use of
\textit{surface-fitted coordinates} (SFC) that exist only inside each
fluid and are normalized by the radius of the star. We denote this extra
spherical coordinate system as $\left(r_f,\GU_f,\GP_f\right)$, where
\begin{equation}
r_f := \frac{r}{R(\GU,\GP)},\qquad \GU_f:=\GU,\qquad \GP_f:=\GP\,,
\label{eq:sfc}
\end{equation} 
and where the surface of the star is denoted by $R(\GU,\GP)$.

By construction, the domain of these fluid coordinates is $[0,1]\times
[0,\pi]\times [0,2\pi]$, and $R(\GU_f,\GP_f)$ is a function that will be
determined at the end of the self-consistent iterative method (see
Appendix \ref{ssec:is}). The advantage of SFC in the computation of
derivatives on the star's surface, as well as the implementation of the
boundary condition Eq. (\ref{eq:bcphi}), will be discussed in
Sec. \ref{ssec:esfp}.

\subsection{Dimensionless and normalized variables}
\label{ssec:dimnorm}

Having removed the dimensions from our equations by using units in which
$G=c=M_\odot=1$, we perform a normalization of all variables in order to
introduce a scale in our problem that is intimately related to the
variable $r_s$, introduced in Sec. \ref{ssec:grid}. 

We normalize variables by demanding that the intersection of the star's
surface with the positive $x$ axis be at $r=r_s$. If $R_0$ is the scaling
parameter, we impose\footnote{Note that $r_s$, but also $r_f$ in
  Eq. \eqref{eq:sfc}, are ratios of two radial coordinates and thus
  dimensionless for any choice of units; in this respect, they do not
  need to be indicated with a hat.}
\begin{equation}
\frac{R(\pi/2,0)}{R_0}\ =\ r_s\ =\ \frac{R(\pi/2,\pi)}{R_0}\,,
\label{eq:R0}
\end{equation}
so that $r_s R_0$ is the real semimajor radius of the star. Hereafter,
we will denote normalized variables with a hat and thus define
\begin{equation}
\hat{x}^i := \frac{x^i}{R_0}\,,
\label{eq:normal}
\end{equation}
from which it follows that the normalized version of
Eq. (\ref{eq:hamcon}) is\footnote{Similar normalized equations hold for
  Eqs. (\ref{eq:trdotkij}) and (\ref{eq:momcon}).}
\begin{equation}
\hat{\nabla}^2\GC = \hat{S}_\GC^g+ R_0^2 S_\GC^f\,,
\label{eq:psi_normal}
\end{equation}
where $\hat{\nabla}$ is the Laplacian operator associated with the
variables $\hat{x}^i$, and similarly $\hat{S}_\GC^g$ has all derivatives
with respect to the normalized variables. Note also that
\begin{equation}
\GO^i=\GB^i+\Omega\GP^i=\GB^i+\hat{\Omega}\hat{\GP}^i \,,
\label{eq:ome_normal}
\end{equation}
where $\hat{\Omega}:=\Omega R_0$. 
If we now define
\begin{equation}
\hat{\Phi} := \frac{\Phi}{R_0},
\label{eq:phi_normal}
\end{equation}
we observe that all scaling factors in Eq. (\ref{eq:cbm4}) drop
out. Because this is also true for the boundary condition, \ie
Eq. (\ref{eq:bcphi}), these equations are each coded in the same form but
with normalized quantities replacing the original ones.

Before proceeding further with our normalization scheme, let us comment
that the surface-fitted coordinates of Eq. (\ref{eq:sfc}) are already
normalized coordinates and their radial range is $[0,1]$ irrespective of
the fluid scaling profile $r_s$. Since $R(\GU,\GP)\leq R(\pi/2,0)=R_0
r_s$, we have $0\leq r \leq R(\GU,\GP) \leq R_0 r_s$, so that the range
for $\hat{r}$ is
\begin{equation}     
\hat{r}\leq \hat{R}(\GU,\GP) \leq r_s\,,   \qquad\mbox{or}\qquad 
r_f=\frac{\hat{r}}{\hat{R}(\GU,\GP)}\leq 1\,. 
\end{equation}
Changing the scale by modifying $R_0$ will affect the conformal factor
and the lapse function since they scale as \cite{UE2000,BCSST1998a}
\begin{equation}
\GC=\hat{\GC}^{R_0^2}\,,\qquad  \GA=\hat{\GA}^{R_0^2}\,.
\label{eq:alpsi_scale}
\end{equation}
 
As mentioned earlier, the system of partial differential equations that
we have to solve \ie the normalized versions of
Eqs. (\ref{eq:hamcon})--(\ref{eq:momcon}) and Eq. (\ref{eq:cbm4}),
involve three constants: $R_0,\ \hat{\Omega}$, and $C$. To find them we
will use the Euler integral evaluated at three arbitrary points to
construct a nonlinear $3\times 3$ system that will be solved with a
typical Newton-Raphson method. This procedure will be repeated every time
we solve for any of the unknown variables $\GC,\GB^i,\GA,\Phi$, and $q$,
since any change of them will affect the Euler integral and thus the
three constants.

The arbitrary points we choose to evaluate the Euler integral are along
the $x$ axis and in spherical coordinates are defined as
\begin{eqnarray}
r_1=r_s\,,\quad &\quad  \GU_1=\pi/2\,,\quad  &\quad  \GP_1=0\,,     
\label{eq:p1}\\
r_2=0\,,\quad\  &\quad  \GU_2=0\,,\quad\ \ \ &\quad  \GP_2=0\,,           
\label{eq:p2}  \\
r_3=r_s\,,\quad &\quad  \GU_3=\pi/2\,,\quad  & \quad \GP_3=\pi\,.   
\label{eq:p3}  
\end{eqnarray}

In the corotating case, Eqs. (\ref{eq:corotei}) and (\ref{eq:corotut})
will become
\begin{eqnarray}
&& F_c(\hat{\Omega},R_0,C) = -\ln C + R_0^2\ln\hat{\GA} + \ln h
\qquad\qquad\qquad \nonumber\\
&&\qquad + \frac{1}{2}\ln\left[1-\left(\frac{\hat{\GC}^2}{\hat{\GA}}\right)^{2R_0^2}
		(\GB^y+\hat{\Omega}\hat{\GP}^y)^2\right]=0\,,
\label{eq:fcorot}
\end{eqnarray}
while for the spinning case, Eqs. (\ref{eq:irspei}), (\ref{eq:irsph}),
and (\ref{eq:irsphutsol}) yield
\begin{eqnarray}
&& F_{is}(\hat{\Omega},R_0,C) = -\ln\GL + R_0^2\ln\hat{\GA} + \ln h
\qquad\qquad\qquad \nonumber\\
&& - \ln\left[\frac{1}{2}+\sqrt{\frac{1}{4}+\frac{A(R_0)}{\GL^2}}\right]
         + \frac{1}{2}\ln\left[1+\frac{B(R_0)}{h^2}\right]=0\,,
\label{eq:firsp}
\end{eqnarray}
with
\begin{eqnarray*}
A(R_0) &:=& (\hat{\GA}^2\hat{\GC}^{A})^{R_0^2}\ \tilde{s}^i \hat{\pd}_i \hat{\Phi}
               + (\hat{\GA}^2\hat{\GC}^{4+2{A}})^{R_0^2}\ \GD_{ij}\tilde{s}^i \tilde{s}^j ,\\
B(R_0) &:=& (\hat{\GC}^{-4})^{R_0^2} \hat{\pd}_i \hat{\Phi} \hat{\pd}^i \hat{\Phi}
              + 2(\hat{\GC}^{A})^{R_0^2}\ \tilde{s}^i \hat{\pd}_i \hat{\Phi}\,, \\
        		 &\phantom{=}& + (\hat{\GC}^{4+2{A}})^{R_0^2}\ \GD_{ij}\tilde{s}^i \tilde{s}^j   \\
\GL    &:=& C+(\GB^y + \hat{\Omega}\hat{\GP}^y)\hat{\pd}_{\hat{y}} \hat{\Phi} \,.
\end{eqnarray*}

Evaluating either Eq. (\ref{eq:fcorot}) or Eq. (\ref{eq:firsp}) at the
three points given by (\ref{eq:p1})--(\ref{eq:p3}) will produce a system
of three equations in the three unknowns $\hat{\Omega},\ R_0,$ and $C$;
the solution of the system will determine these constants.
These unknowns are computed separately for the two stars, although they yield
  the same solution for the case of equal-mass binaries considered
  here. We also note that the star's surface remains fixed and the values
  of the specific enthalpy $h$ at the three points
  (\ref{eq:p1})--(\ref{eq:p3}) is unchanged, with $h=1$ at (\ref{eq:p1})
  and (\ref{eq:p3}).

\subsection{Elliptic solver for the gravitational part}
\label{ssec:esgp}

As discussed in detail in Refs. \cite{UT2012,UTG2012,TU2013},
Eqs. (\ref{eq:hamcon})--(\ref{eq:momcon}) are solved using the
representation theorem of partial differential equations in a
self-consistent way.  Starting from 
\begin{equation}
\nabla^2 f = S\, 
\end{equation}
where $S$ is a nonlinear function of $f$, and using the Green's function
without boundary $G(x,x')=1/|x-x'|$ that satisfies
\begin{equation}
\nabla^2 G(x,x') = -4\pi \GD(x-x')\,,   
\end{equation}
a solution for $f$ is obtained as 
\begin{align}
f(x) &= -\frac1{4\pi}\int_{V} G(x,x')S(x') d^{3}x' 
\nonumber  \\
&\!\!+ \frac{1}{4\pi} \int_{\pd V} \left[G(x,x')\na'^{a} f(x')
- f(x')\na'^a G(x,x') \right]dS'_a\,. 
\end{align}
where $V$ is the domain of integration, $x,x' \in V\subseteq\Sigma_0$,
the initial spacelike hypersurface. The volume $V$ and its boundary $\pd
V$ depend on the coordinate system we are solving for, \ie COCP or ARCP
as described in Sec.  \ref{ssec:grid}.  
This is the principle of the KEH
method \cite{KEH1989} and will be suitably modified in order to account
for the specific boundary conditions that exist in the new
\cocal~coordinate systems. For example, the conformal factor will be
expressed as
\begin{equation} 
\GC(x)=\GX(x)+\GC_{\rm INT}(x)\,,
\label{eq:psisol}
\end{equation}
where
\begin{eqnarray}
\GC_{\rm INT}(x)= -\frac1{4\pi}\int_{V} \frac{S^g_\GC(x') + S^f_\GC(x')}{|x-x'|} d^{3}x' 
\qquad\qquad\qquad\nonumber \\
+ \frac{1}{4\pi} \int_{\pd V} \left[\frac{\na'^{a} \GC(x')}{|x-x'|}
- \GC(x')\na'^a \frac{1}{|x-x'|} \right]dS'_a\,,  \qquad
\label{eq:psi_int}
\end{eqnarray}
and 
\begin{eqnarray}
\chi(x) &=&  \frac{1}{4\pi} \int_{S_a \cup S_b} 
\left[G^{\rm BC}(x,x')\na'^{a} (\GC_{\rm BC} - \GC_{\rm INT}(x')  \right. \nonumber\\
&& \left. - (\GC_{\rm BC} - \GC_{\rm INT})(x')\na'^a G^{\rm BC}(x,x') \right]dS'_a\,. 
\label{eq:psi_chi}
\end{eqnarray}
Note that $G^{\rm BC}$ is the Green's function associated with the
boundary conditions applied on the corresponding field $\GC_{\rm BC}$ at
the boundaries $S_a$ and $S_b$. Formulas
(\ref{eq:psisol})--(\ref{eq:psi_chi}) will be applied separately on every
coordinate patch. If, for example, we have one ARCP patch (as it happens
in our computations) it means that the equations above will be applied
three times: two for the COCP patches and one for the ARCP patch. Of
course, the domains of integration vary according to the different
patches considered. More specifically, if we denote by $B(R)$ a sphere of
radius $R$ in each of the COCP patch, then the integration domain of
Eq. (\ref{eq:psi_int}) will be $V=B(r_b)-B(r_e)$ and $\pd V=S_e\cup
S_b=\pd B(r_e)\cup\pd B(r_b)$, while that of Eq. (\ref{eq:psi_chi}) will
be $S_a\cup S_b=S_b$, since $r_a=0$ for star configurations in the COCP
patch. Similarly, in the ARCP patch the integration domain of
Eq. (\ref{eq:psi_int}) will be $V=B(r_b)-B(r_a)$, and that of
Eq. (\ref{eq:psi_chi}) $\pd V=S_a\cup S_b$. 

We recall that in Ref. \cite{UT2012} we have introduced a number of
Green's functions $G^{\rm BC}(x,x')$ suitable for various boundary
conditions. Here we add one more Green's function used in the COCP patch
\begin{eqnarray} 
G^{\rm SD}(x,x') &:=& \sum_{\ell=0}^\infty g^{\rm SD}_\ell(r,r') \sum_{m=0}^\ell \epsilon_m \, \frac{(\ell-m)!}{(\ell+m)!}
\nonumber  \\
&&\!\!\!\!\!\!\!\!\!\!\!\!\!\!\!\!\!
\times P_\ell^{~m}(\cos\theta)\,P_\ell^{~m}(\cos\theta') \cos [m(\phi-\phi')]\,, 
\label{eq:green_sd}
\end{eqnarray}
where 
\begin{eqnarray}
&&g^{\rm SD}_\ell(r,r') := \frac{r_{<}^\ell}{r_b^{\ell+1}} 
\left[\left(\frac{r_b}{r_{>}}\right)^{\ell+1}-\left(\frac{r_{>}}{r_b}\right)^{\ell}\right]\,, 
\label{eq:radial_sd}
\end{eqnarray} 
and $\epsilon_0 = 1$, $\epsilon_m = 2$ for $m\ge 1$, while $P_\ell^{~m}$
are the associated Legendre polynomials, and $r_> := \sup\{r,r'\}$, $r_<
:= \inf\{r,r'\}$.

%Note that although a star does not have an inner boundary ($S_a$ is
%absent), the use of a flat Green's function in the COCP coordinate system
%is inconsistent, since the COCP patch is no longer a global chart. 

In the ARCP patch we use a Green's function $G^{\rm DD}(x,x')$, whose radial
part satisfies the Dirichlet-Dirichlet boundary conditions on $S_a$ and $S_b$
\begin{eqnarray}
&&g^{\rm DD}_\ell(r,r') \,=\,\left[1-\left(\frac{r_a}{r_b}\right)^{2\ell+1}\right]^{-1}\frac{r_a^\ell}{r_b^{\ell+1}}\times
\nonumber\\
&&\left[\left(\frac{r_{<}}{r_a}\right)^{\ell} -\left(\frac{r_a}{r_{<}}\right)^{\ell+1}\right]
\left[\left(\frac{r_b}{r_{>}}\right)^{\ell+1}-\left(\frac{r_{>}}{r_b}\right)^{\ell}\right]\,.\ 
\label{eq:radial_dd}
\end{eqnarray}

\subsection{Elliptic solver for the fluid part}
\label{ssec:esfp}

Next, we describe the method used to solve Eq. (\ref{eq:cbm4}) and which
is therefore valid only for the spinning binaries. The boundary condition
for $\Phi$, Eq. (\ref{eq:bcphi}), is of von Neumann type and therefore we
could apply the Poisson solver of Sec. \ref{ssec:esgp} to obtain a
solution. Instead, and as a demonstration of the versatility of the
methods employed by \cocal, we will adapt the procedure discussed in
Ref. \cite{UE2000} and solve this boundary-value problem as an
application of the least-squares algorithm.

First, we assume that the solution of Eq. (\ref{eq:cbm4}) can be written
in the form
\begin{equation}
\Phi(x)=-\frac{1}{4\pi} \int_V \frac{S_\Phi (x')}{|x-x'|} dV + \GZ(x)
= \Phi_V(x) + \GZ(x)\,,
\label{eq:phisol}
\end{equation}
with $\GZ(x)$ obeying the Laplace equation
\begin{equation}
\nabla^2 \GZ (x) =0\,. \label{eq:zetalap}
\end{equation}
Using the decomposition of Eq. (\ref{eq:phisol}), the boundary condition
(\ref{eq:bcphi}) is written as
\begin{equation}
\GC^4 h u^t \GO^i m_i - m^i \pd_i\Phi_V - \GC^{4+A}\tilde{s}^i m_i = m^i\pd_i\GZ\,,
\label{eq:bcphi1}
\end{equation}
where we used the normal to the surface unit vector $m^i=(\hat{n}_s)^i$
instead of the gradient of the rest-mass density. The equation above is
evaluated on the surface of the star, $R(\GU,\GP)$, and the velocity
potential satisfies the following symmetries
\begin{eqnarray}
\Phi(r,\pi-\GU,\GP) &\ =\ & \phantom{-} \Phi(r,\GU,\GP) \,, \label{eq:sym1} \\
\Phi(r,\GU,2\pi-\GP) &\ =\ & -\Phi(r,\GU,\GP) \,,  \label{eq:sym2}
\end{eqnarray}
which in turn imply that the homogeneous solution, which is regular at
the stellar center, can be expanded as
\begin{equation}
\GZ(r,\GU,\GP) = \sum_{\ell=1}^{L} \sum_{m=1}^{\ell} 
        a_{\ell m} r^\ell [1+(-1)^{\ell+m}] Y_\ell^m (\cos\GU) \sin(m\GP)\,,
\label{eq:zeta}
\end{equation}
where $Y_\ell^m$ are the spherical harmonics and $a_{\ell m}$
coefficients to be determined.  
When the spins of the stars are in
  arbitrary directions, the symmetries \eqref{eq:sym1} and
  \eqref{eq:sym2} no longer apply and expression \eqref{eq:zeta} will
  contain also cosine terms.

On the other hand, if $R(\GU,\GP)$ is the surface of the star, the
spatial vector connecting any point on it with the center of coordinates
is given by
\begin{equation}
\vec{\boldsymbol{x}}(\GU,\GP)= R(\GU,\GP)(\sin\GU\cos\GP,\sin\GU\sin\GP,\cos\GU)\,,
\end{equation}
thus the unit normal vector will be 
\begin{equation}
\hat{\boldsymbol{n}}_s(\GU,\GP) := 
\left({\frac{\partial\vec{\boldsymbol{x}}}{\partial\GU}\times\frac{\partial\vec{\boldsymbol{x}}}{\partial\GP}}\right)
\left|\frac{\partial\vec{\boldsymbol{x}}}{\partial\GU}\times\frac{\partial\vec{\boldsymbol{x}}}{\partial\GP}\right|^{-1}\,,
\end{equation}
or equivalently
\begin{equation}
\hat{\boldsymbol{n}}_s = 
\frac{1}{\sqrt{h}}\left(\hat{\boldsymbol{r}}-\frac{1}{R}\frac{\partial R}{\partial \GU}\hat{\boldsymbol{\GU}}
             -\frac{1}{R\sin\GU}\frac{\partial R}{\partial\GP}\hat{\boldsymbol{\GP}} \right)\,,  
\label{eq:norvec}  \\
\end{equation}
where $\hat{\boldsymbol{r}}, \hat{\boldsymbol{\theta}},
\hat{\boldsymbol{\phi}}$ are the spherical unit vectors and
\begin{equation}
h(\GU,\GP) := 1+\left(\frac{1}{R}\frac{\pd R}{\pd \GU}\right)^2 +    
                \left(\frac{1}{R\sin\GU}\frac{\pd R}{\pd\GP}\right)^2
                \,. 
\end{equation}

Using Eqs. (\ref{eq:zeta}) and (\ref{eq:norvec}), the boundary
condition (\ref{eq:bcphi1}) is written as
\begin{equation}
\sum_{\ell=1}^{L}\sum_{m=1}^{\ell}a_{\ell m}F_{\ell m}(\GU,\GP) = H(\GU,\GP)\,,
\label{eq:bcphi2}
\end{equation}
where
\begin{equation}
H(\GU,\GP) := \GC^4 h u^t \GO^i m_i - m^i \pd_i\Phi_V - 
\GC^{4+A}\tilde{s}^i m_i\,,
\label{eq:Hbc}
\end{equation}
and
\begin{eqnarray}
&& \hskip -0.65cm F_{\ell m}(\GU,\GP) := [1+(-1)^{\ell+m}]\times   \nonumber\\
&& \left[ \ell R^{\ell-1}Y_\ell^m\sin(m\GP) - 
        \frac{\pd R}{\pd\GU} R^{\ell-2}\frac{\pd Y_\ell^m}{\pd\GU}
        \sin(m\GP) - \right. \nonumber\\
&& \left. \hskip 0.25cm
\frac{\pd R}{\pd\GP}\frac{R^{\ell-2}}{\sin^2\GU}Y_\ell^mm\cos(m\GP)\right]\,.
\label{eq:flm}
\end{eqnarray}
To solve for the coefficients $a_{\ell m}$, we consider the functional
\begin{equation}
\mathcal{E} := 
\sum_{\GU_i,\GP_j}\left[\sum_{\ell=1}^{L}\sum_{m=1}^{\ell}a_{lm}F_\ell^m(\GU_i,\GP_j) - 
                            H(\GU_i,\GP_j)\right]^2 = 0\,,
\label{eq:functE}
\end{equation}
of the discretized version of the boundary condition (\ref{eq:bcphi2}),
and demand that for fixed indices $p$ and $q$
\begin{equation}
\frac{\pd\mathcal{E}}{\pd a_{pq}}\ =\ 0\,.
\label{eq:Eeq}
\end{equation}
The minimizing condition Eq. (\ref{eq:Eeq}) then yields
\begin{eqnarray}
&& \sum_{\ell,m} a_{\ell m}\left[\sum_{i,j} F_{\ell m}(\GU_i,\GP_j) F_{pq}(\GU_i,\GP_j)\right] 
\qquad\qquad\\
&& \qquad\qquad\qquad\qquad = \sum_{i,j} H(\GU_i,\GP_j) F_{pq}(\GU_i,\GP_j)\,,
\end{eqnarray}
which is a linear system in terms of the $a_{\ell m}$ coefficients. For
$L$ even, the dimensions of the system is $M\times M$ with
$M=L(L+1)/2$. After determining the coefficients $a_{\ell m}$, the
solution for the velocity potential $\Phi$ is obtained from
Eqs. (\ref{eq:phisol}) and (\ref{eq:zeta}).

\begin{figure*}
\begin{tabular}{cc}
\begin{minipage}{.49\hsize}
\begin{center}
\includegraphics[height=60mm]{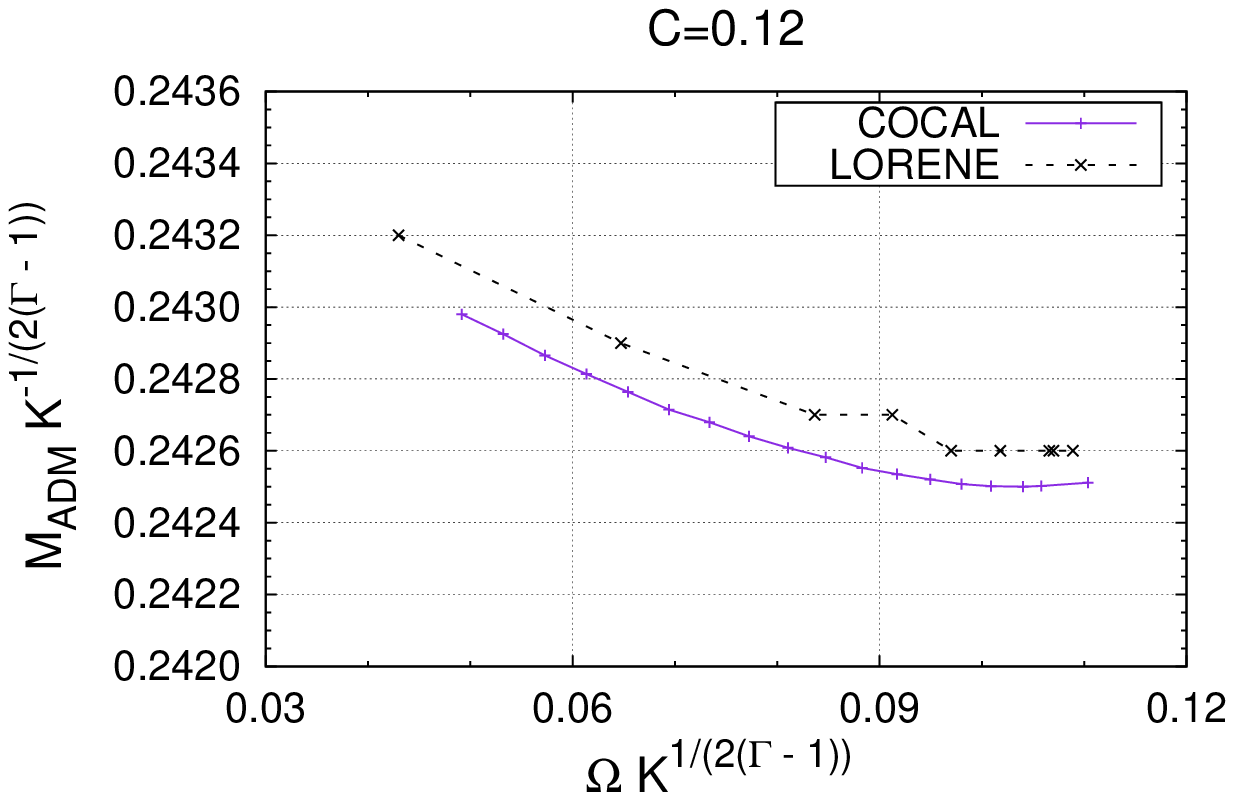}
\includegraphics[height=60mm]{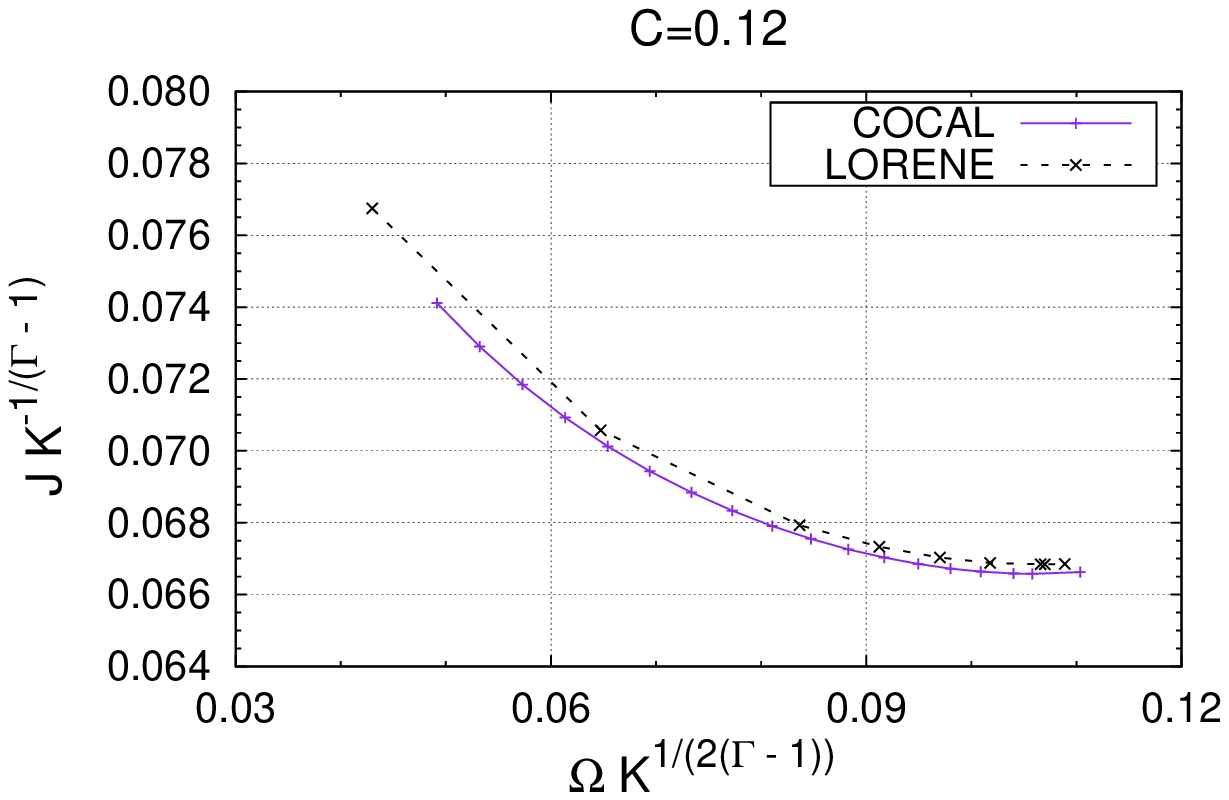}
\includegraphics[height=60mm]{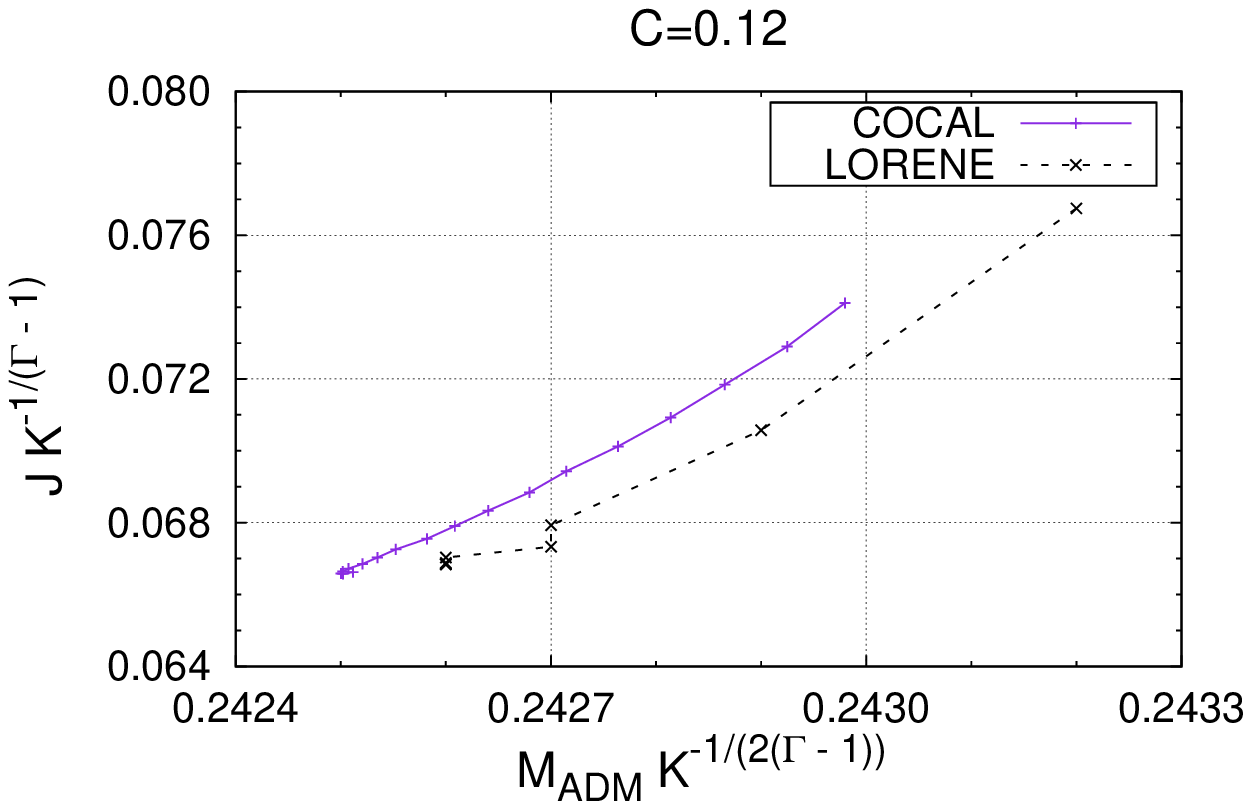}
\end{center}
\end{minipage} 
&
\begin{minipage}{.49\hsize}
\begin{center}
\includegraphics[height=60mm]{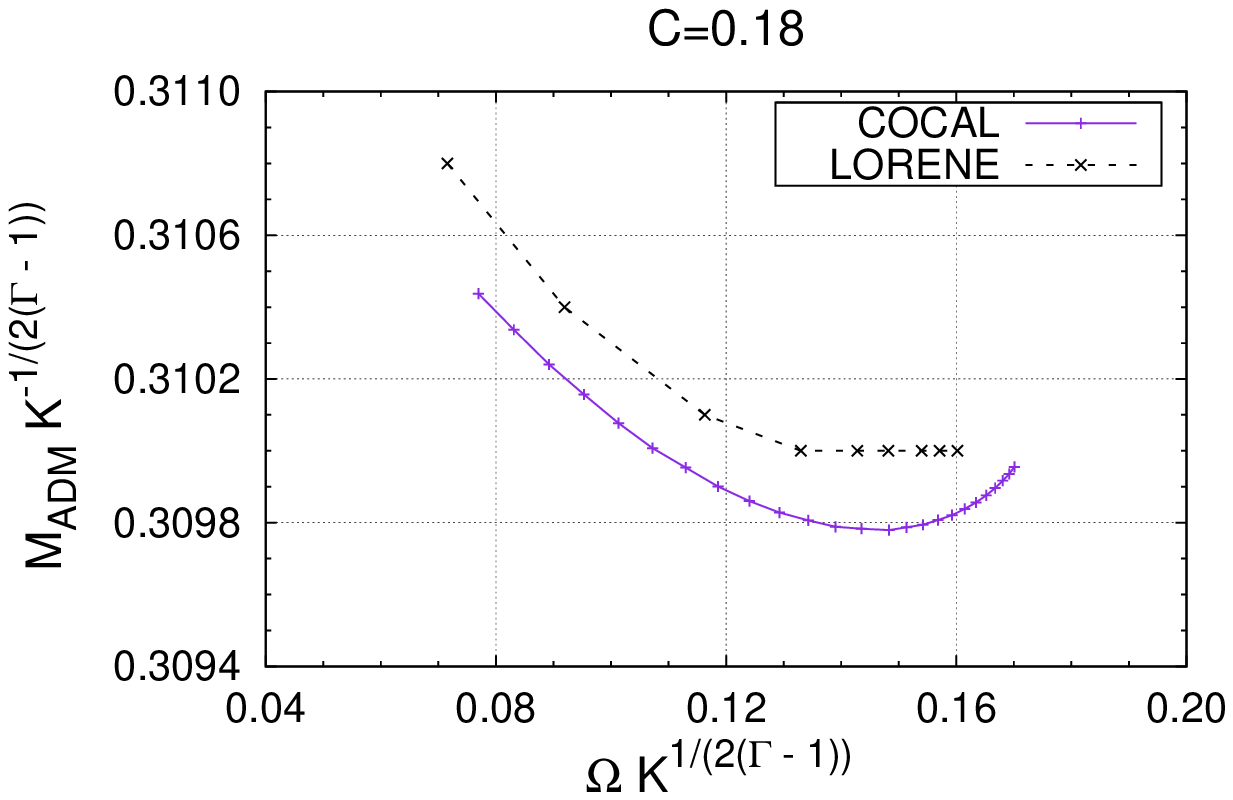}
\includegraphics[height=60mm]{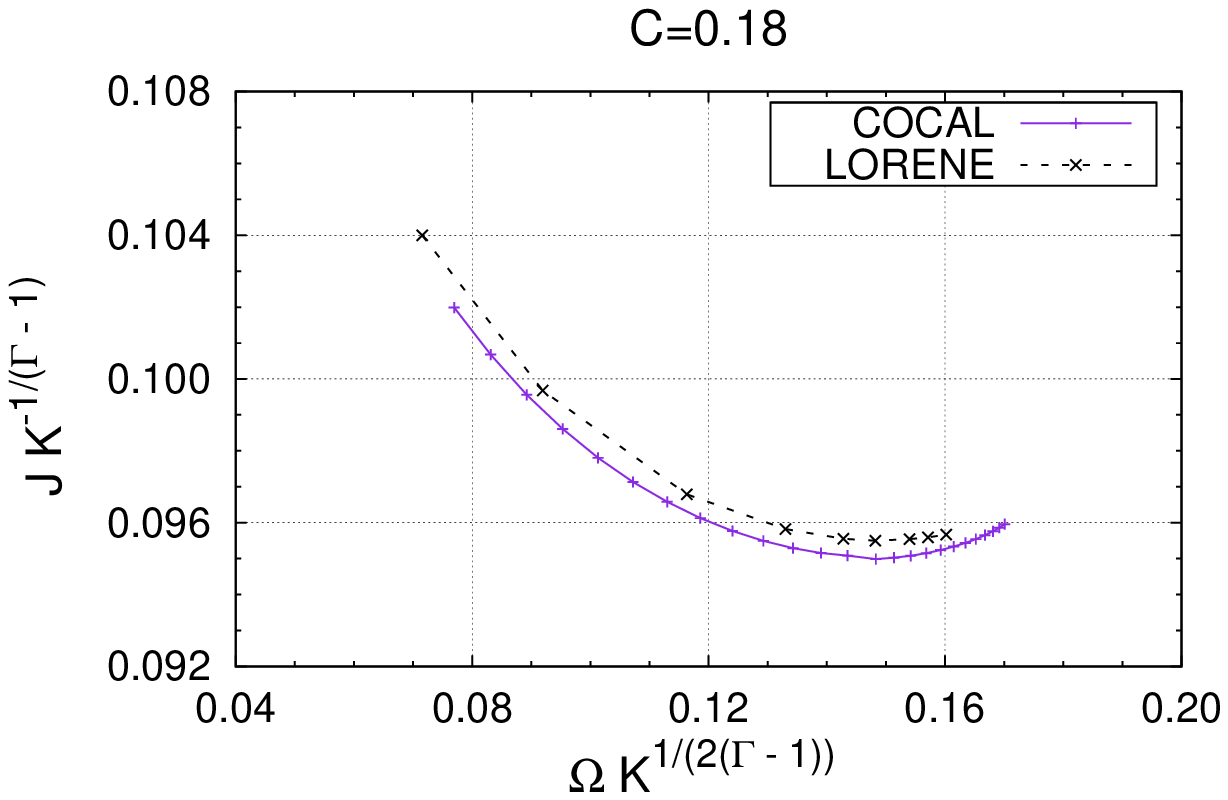}
\includegraphics[height=60mm]{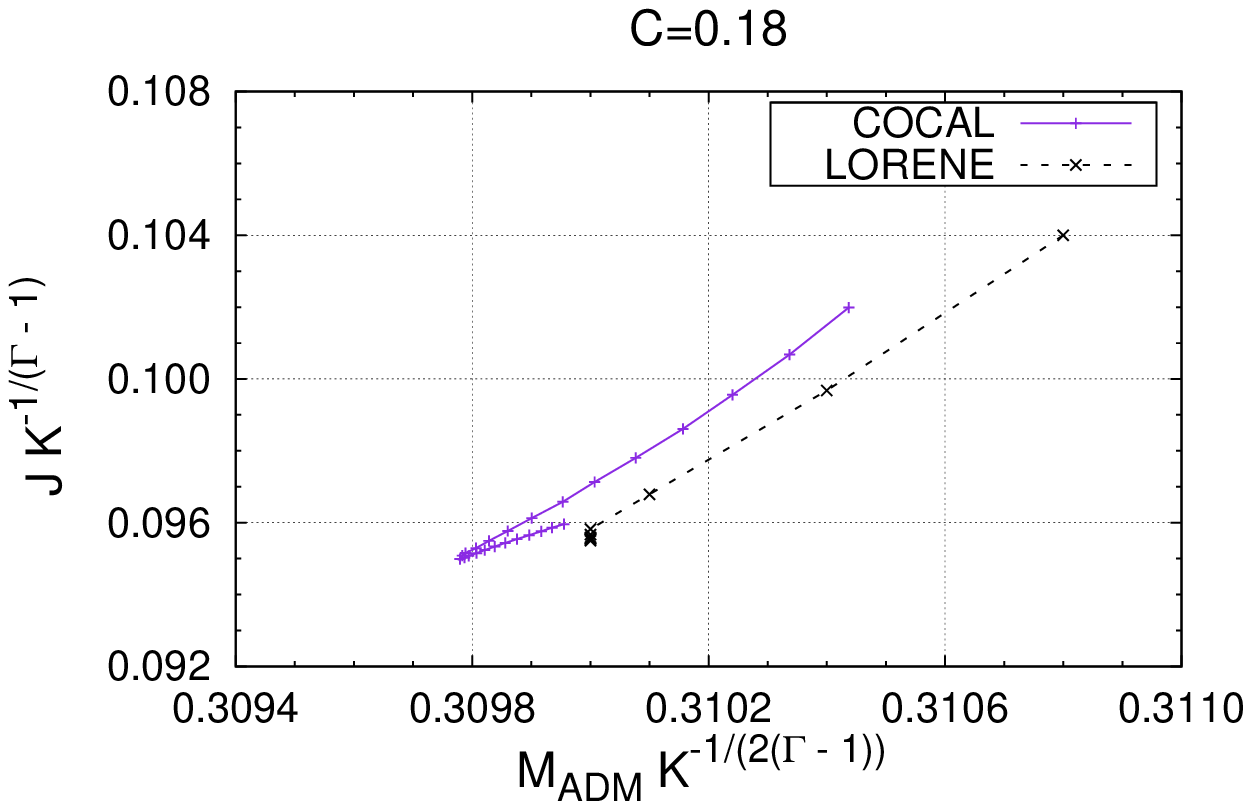}
\end{center}
\end{minipage} 
\end{tabular} 
\caption{Quasiequilibrium sequences for corotating binary neutron stars
  with EOS consisting of a single polytrope with $\Gamma=2$. The
    left column corresponds to models with compactness
    $\mathcal{C}=0.12$, while the right column corresponds to $\mathcal{C}=0.18$.
  $M$ and $J$ are the total ADM mass and angular momentum of the
  system. The resolutions used are those given in Table
  \ref{tab:corotsol}. A comparison is made with the results of
  Ref. \cite{TG2002b}, where similar solutions were obtained from the
  spectral code \lorene~\cite{LORENE}.}
\label{fig:corot_c0.12_c0.18}
\end{figure*}

\begin{table*}
\begin{tabular}{cl|ccccccccccccc}
\hline
Type & Patch  & $\ r_a\ $ & $\ r_s\ $ & $\ r_b\ $ & $\ r_c\ $ & $\ r_e\ $ & 
$\ \Nrf\ $ & $\ N_r^1\ $ & $\ \Nrm\ $ & $\ N_r\ $ & $\ N_\theta\ $ & $\ N_\phi\ $ & $\ L\ $ \\
\hline
${\rm Hs2d}$ & ${\rm COCP-1}$ & $0.0$ & ${\rm var}$ & $10^2$ & $1.25$ & $1.125$ & $50$  & $64$ & $80$ & $192$ & $48$  & $48$ & $12$ \\
             & ${\rm COCP-2}$ & $0.0$ & ${\rm var}$ & $10^2$ & $1.25$ & $1.125$ & $50$  & $64$ & $80$ & $192$ & $48$  & $48$ & $12$ \\
             & ${\rm ARCP}$   & $5.0$ & $-$         & $10^6$ & $6.25$ & $-$     & $16$  & $-$  & $20$ & $192$ & $48$  & $48$ & $12$ \\
\hline
${\rm Hs2b}$ & ${\rm COCP-1}$ & $0.0$ & ${\rm var}$ & $10^2$ & $1.0625$ & $1.03125$ & $60$  & $64$ & $68$ & $192$ & $48$  & $48$ & $12$ \\
             & ${\rm COCP-2}$ & $0.0$ & ${\rm var}$ & $10^2$ & $1.0625$ & $1.03125$ & $60$  & $64$ & $68$ & $192$ & $48$  & $48$ & $12$ \\
             & ${\rm ARCP}$   & $5.0$ & $-$         & $10^6$ & $6.25  $ & $-$       & $16$  & $-$  & $20$ & $192$ & $48$  & $48$ & $12$ \\
\hline
\hline
\end{tabular}
\caption{Grid parameters used for the corotating sequences of compactness
  $0.12,\ 0.18$, presented in Fig. \ref{fig:corot_c0.12_c0.18}. ${\rm
    Hs2d}$ refers to solutions at large separations, while ${\rm Hs2b}$
  refers to binaries with small separations. 	
	For the ${\rm Hs2d}$ case the
  separation between the two neutron stars is kept fixed at $d_s=2r_c=2.5$
  while the surface of the neutron star (maximum value $r_s$) varies from
  $r_s=0.5$ to $r_s=0.86$ creating effectively binaries with separations
  from $5.0=2.5/0.5$ to $2.91=2.5/0.86$. Similarly in the ${\rm Hs2b}$
  case $d_s=2r_c=2.125$, and the effective separations are from
  $2.125/0.74=2.87$ to $2.125/0.98=2.17$.}
\label{tab:corotsol}
\end{table*}

\section{Numerical Results}
\label{sec:test}

In what follows we report tests of our new code against previous results
obtained by other groups and then present some new ones. In particular,
we focus on the construction of quasiequilibrium sequences for
corotating, irrotational, and spinning binaries, and produce a binary white
dwarf solution in order to explore the weak-field limit of our code. We
compute two main global-error indicators to measure the accuracy of our
converged solutions; the first one is given by the relation $M_{_{\rm
    K}}=M_{_{\rm ADM}}$, where $M_{_{\rm K}}$ and $M_{_{\rm ADM}}$ are
the Komar and Arnowitt-Deser-Misner (ADM) mass, respectively \cite{K1959,
  K1962,B1978,AA1979,GB1994}. The second one is instead related to the
first law of binary thermodynamics $dM_{_{\rm ADM}}=\Omega dJ$, where
$\Omega$ is the orbital frequency and $J$ the orbital angular momentum
\cite{FUS2002}. Explicit definitions and computational algorithms for
these quantities within \cocal~are presented in Appendix \ref{sec:mam}.

Sequences of constant rest mass can be thought of as snapshots of an
evolutionary process that drives the two stars close to each other as a
result of gravitational radiation reaction. At every instant in time,
the rest mass of each star is conserved; furthermore, if the flow is
irrotational, the circulation of the fluid velocity on any loop is also
conserved (Kelvin-Helmholtz theorem \cite{RZ2013}). It is possible to
characterize these sequences via the properties of the stars, such as the
compactness or the ADM mass, when the binary has infinite separation and
each star is spherical.

By varying the separation between the two stars and solving each time all
the relevant equations, we obtain solutions of a given central rest-mass
density, and then another loop of solutions has to be invoked in order to
find the particular central rest-mass density that yields a star with the
desired rest mass. Typically, this is done by a Newton-Raphson method and
it takes a maximum of ten iterations depending on the starting solution.

In this way we can monitor important quantities like the binding energy
of the system, which is defined as
\begin{equation}
E_{\rm b} := M_{_{\rm ADM}} - M_{\infty} \,,
\label{eq:bindene}
\end{equation} 
and represents the total energy lost in gravitational waves by the
system, since $M_{\infty}$ is twice the ADM mass of a single isolated
spherical star.

\subsection{Corotating solutions}
\label{ssec:corotsol}

As mentioned in the Introduction, corotating states \cite{BCSST1998a,
  MMW1998a, UUE2000, UE2002}, \ie states with zero angular velocity of
the star with respect to a corotating observer, probably are not
physically realistic due to the low viscosity of the neutron-star
matter. Such solutions represent an important step in the numerical
solution of the binary problem, since they provide key insights for the
numerical implementation of a stable algorithm. In particular the
surface-fitted coordinates, as well as the solution of the Euler
integral, Eq. (\ref{eq:corotei}), can be thoroughly checked. This allows
us to perform a calibration without having to worry about the fluid flow
(\ie to solve the equation of conservation of rest mass). Corotating
evolutionary configurations are known to exhibit a minimum in the mass
and angular momentum versus the normalized angular velocity $\Omega
K^{1/(2(\Gamma-1))}$, which was taken to denote the putative innermost
stable circular orbit, beyond which the binary was thought to
proceed rapidly toward a merger. In practice, fully general-relativistic
simulations of inspiraling binary neutron stars do not show the
existence of such an instability, revealing instead that the inspiral and
merger is a smooth process \cite{SU2000, BGR2008}. Nevertheless, the
presence of such a minimum represents a useful test of numerical codes,
as does the appearance of a familiar spike, similar to the one
encountered in binary black-hole solutions, when plotting the binding
energy versus the angular momentum of the binary.

In Fig. \ref{fig:corot_c0.12_c0.18} we present sequences of binary
neutron stars that correspond to the compactness of $\mathcal{C}:=M_{_{\rm
    ADM}}/R=0.12$ and $0.18$, where $M_{_{\rm ADM}}$ and $R$ are the
(ADM) mass and radius of each star when taken at infinite separation. The
ADM and Komar mass, as well as other quantities used in \cocal, are
described in detail in Appendix \ref{sec:mam}. The adiabatic index is
$\Gamma=2$ and the polytropic constant is set to be $K=1$. In the various
plots a comparison is made between the results obtained with \cocal~and
those presented in Ref. \cite{TG2002b}, where the same initial data were
computed using \lorene, a pseudospectral code developed by the Meudon
group \cite{TG2002b}. As we can see, the relative difference in the
results between the two codes is of the order of $0.05\%$, even when a
medium resolution is used for \cocal. The grid structure used in these
calculations is the one described in Table \ref{tab:corotsol}.

\begin{figure}
  \includegraphics[height=60mm]{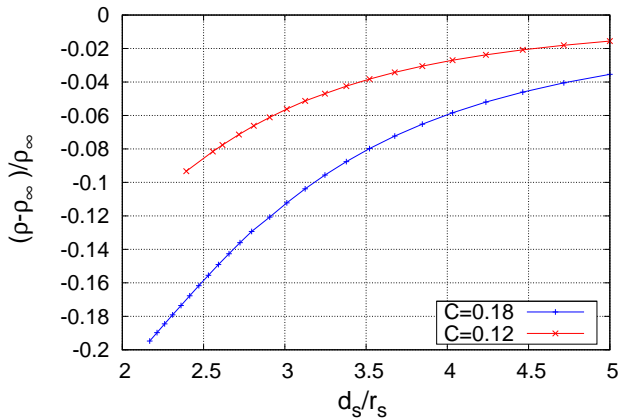}
	\caption{Relative change in the central rest-mass density for the
          corotating sequences in Fig. \ref{fig:corot_c0.12_c0.18}, shown
          as a function of separation.}
	\label{fig:corotrho}
\end{figure}

\begin{figure}
  \includegraphics[height=60mm]{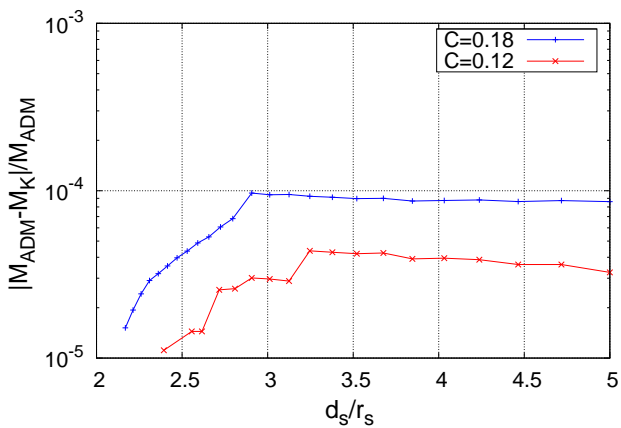}
	\caption{Measure of the virial error $M_{_{\rm K}}=M_{_{\rm ADM}}$ for
          the corotating sequences in Fig. \ref{fig:corot_c0.12_c0.18},
          as a function of separation.}
	\label{fig:corotve}
\end{figure}

Similarly, in Fig. \ref{fig:corotrho} we report the change in the central
rest-mass density with respect to the one at infinity, which is
$\GR_\infty=0.0922$ for compactness $\mathcal{C}=0.12$, while it is
$\GR_\infty=0.1956$ for $\mathcal{C}=0.18$. Clearly, the central
rest-mass density decreases as the binary comes closer, making the onset
of an instability to gravitational collapse very unlikely
\cite{WMM1996}. In addition, as a measure of accuracy of these corotating
sequences, we plot in Fig. \ref{fig:corotve} the relative difference
\begin{equation}
\Delta_M := \left|\frac{M_{_{\rm ADM}}-M_{_{\rm K}}}{M_{_{\rm ADM}}} \right|\,,
\end{equation}
as a function of the binary separation $d_s/r_s$. Note that all radii are here
normalized to the scaling factor $R_0$ and are therefore dimensionless,
so that, e.g., the physical distance between the two neutron stars is $d_s R_0$.
As we can see, even for
the medium resolution used in these calculations the error is below
$10^{-4}$. All of the quantities in the expression above have been
extracted from the ARCP patch, as integrals at infinity. We note that at
present \cocal~does not implement a unifying mesh, and this prevents us
from calculating the virial error as obtained by Friedman, Uryu, and
Shibata \cite{FUS2002}, since we are using overlapping coordinate
systems. We plan to revisit this issue in the future.

\begin{figure}
\includegraphics[height=60mm]{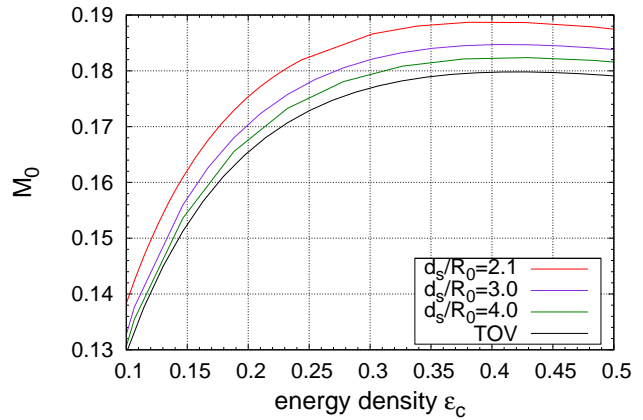}
	\caption{Corotating sequences versus central energy density for
          different separations. The TOV curve corresponds to infinite
          separation. Here we use $r_s=1$, and therefore $R_0$ is the radius
					of the star.}
	\label{fig:corotrhoseq}
\end{figure}

\begin{figure*}
\begin{tabular}{cc}
\begin{minipage}{.49\hsize}
\begin{center}
\includegraphics[height=60mm]{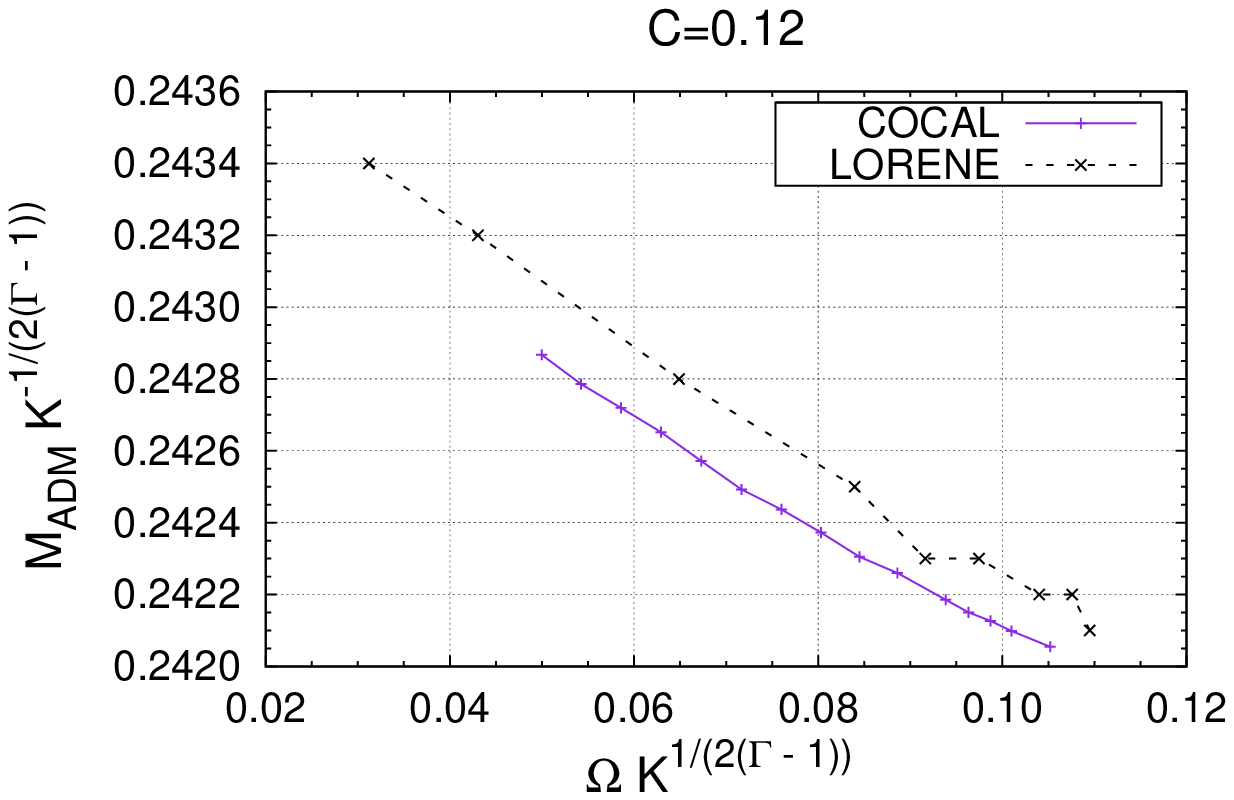}
\includegraphics[height=60mm]{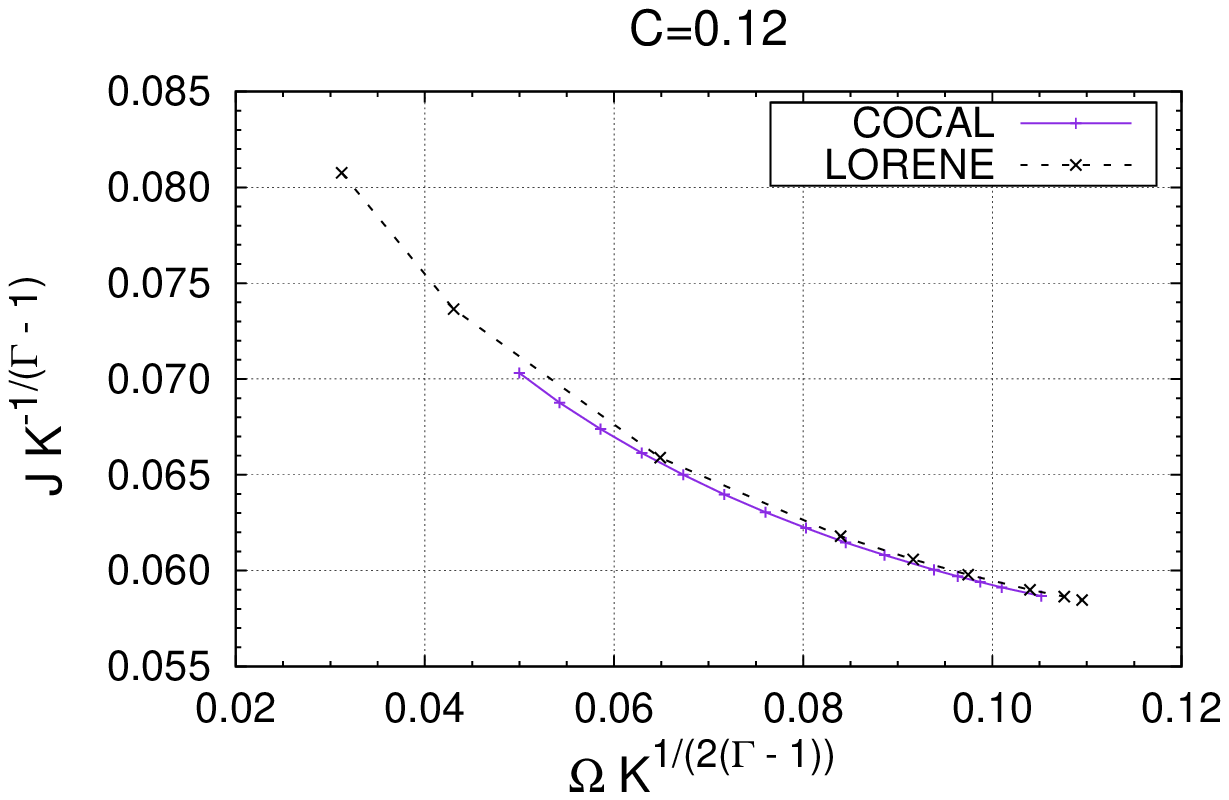}
\includegraphics[height=60mm]{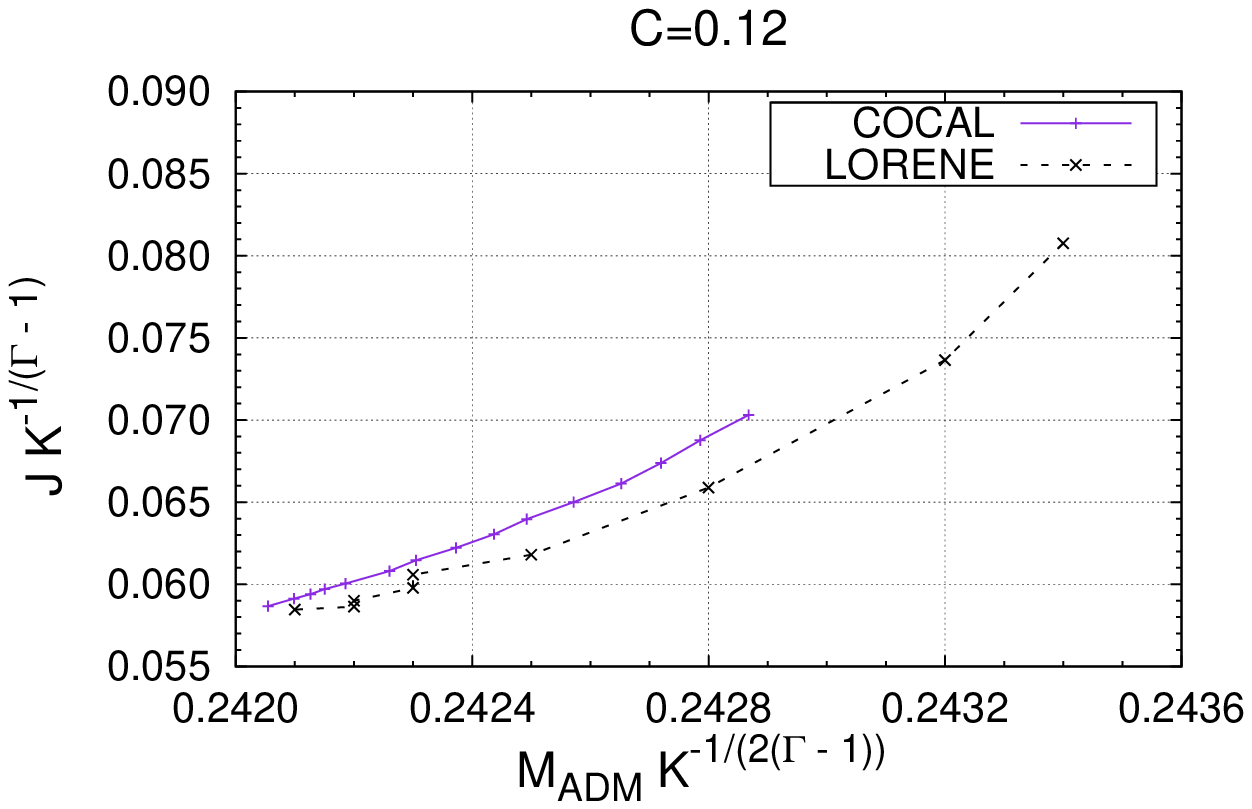}
\end{center}
\end{minipage} 
&
\begin{minipage}{.49\hsize}
\begin{center}
\includegraphics[height=60mm]{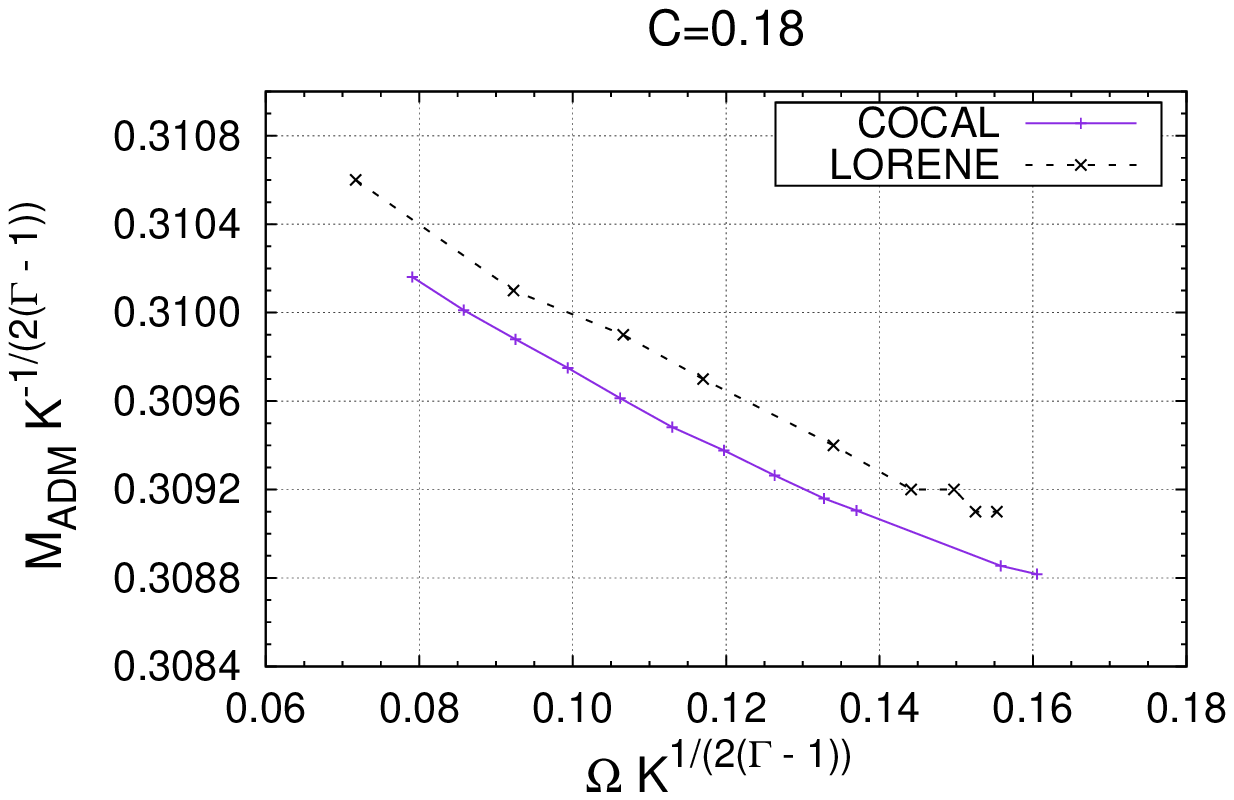}
\includegraphics[height=60mm]{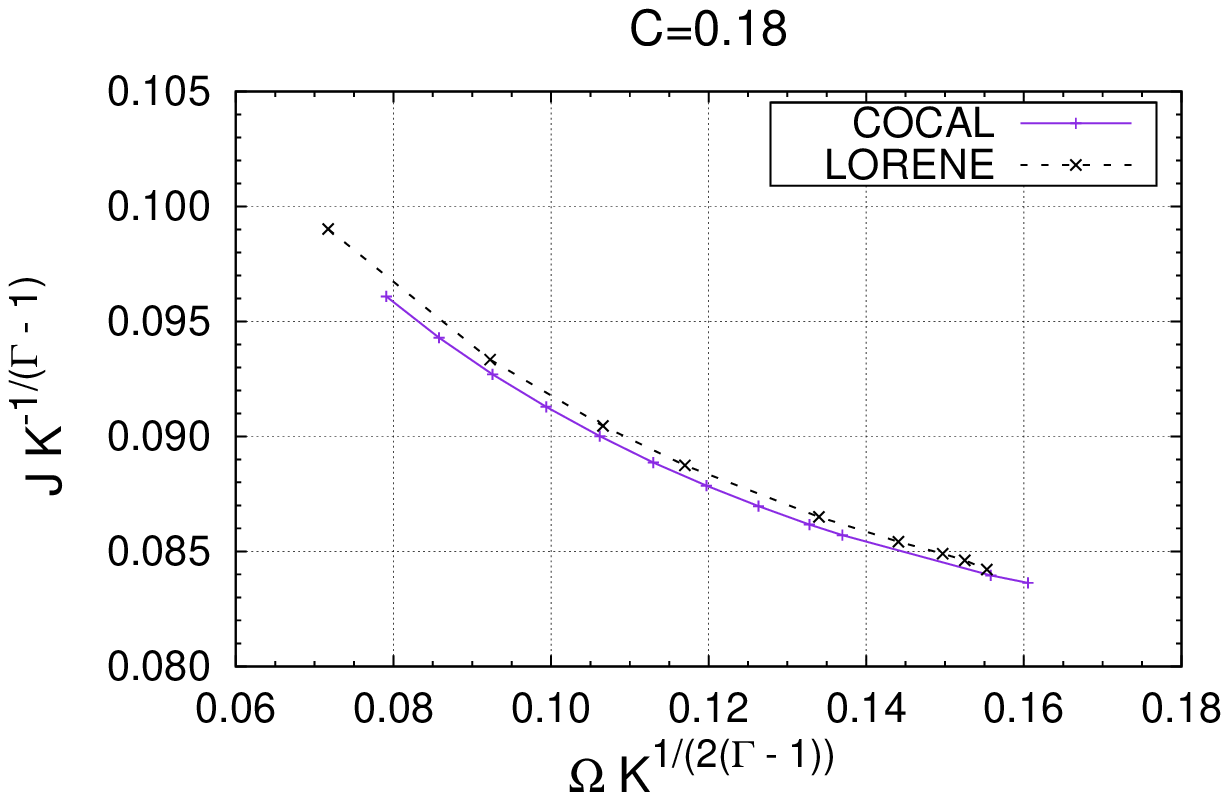}
\includegraphics[height=60mm]{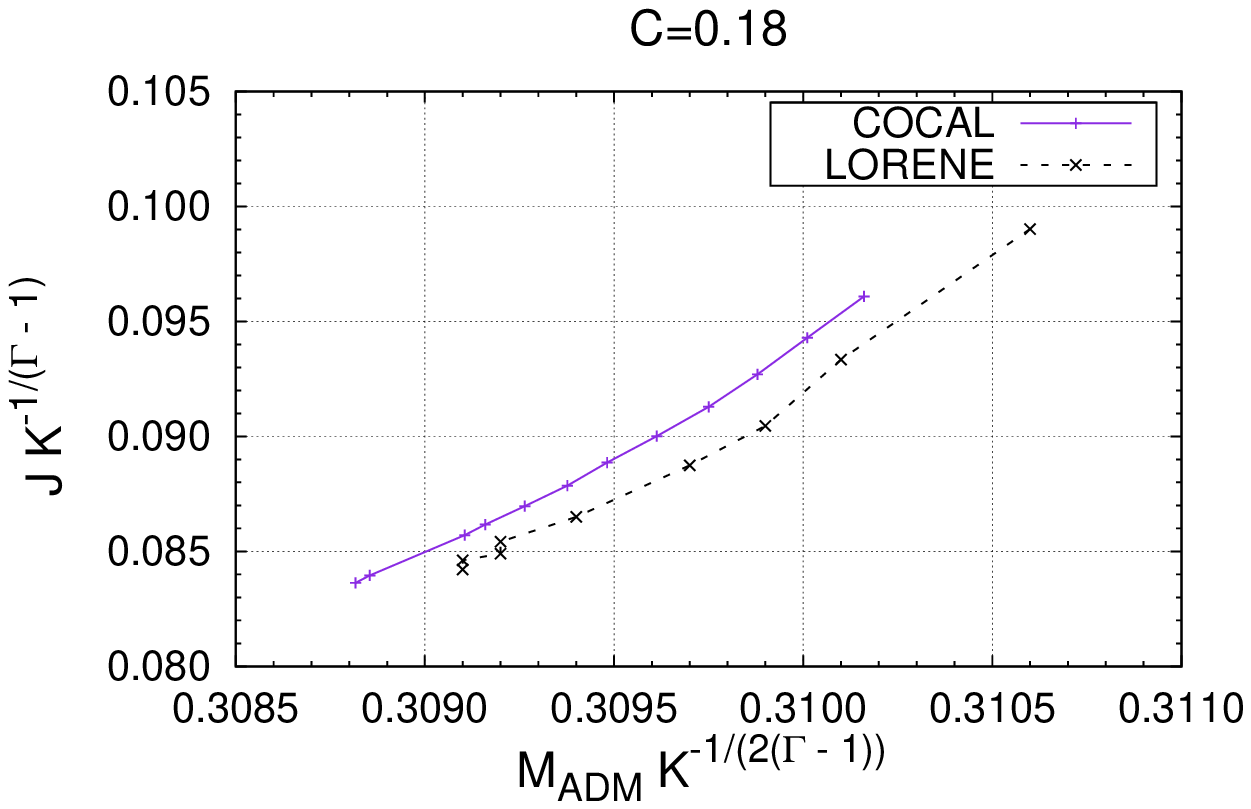}
\end{center}
\end{minipage} 
\end{tabular} 
\caption{Sequences of irrotational binary neutron stars, with an EOS
  consisting of a single polytrope with $\Gamma=2$.  The left column
    corresponds to models with compactness $\mathcal{C}=0.12$, while the
    right column corresponds to $\mathcal{C}=0.18$.  The resolution is ${\rm Hs2d}$
  from Table \ref{tab:corotsol}. A comparison is made with the results
  presented in \cite{TG2002b}, where similar solutions were obtained from
  the spectral code \lorene~\cite{LORENE}.}
\label{fig:irrot_c0.12_c0.18}
\end{figure*}

\begin{figure*}
\begin{center}
\includegraphics[width=0.75\columnwidth]{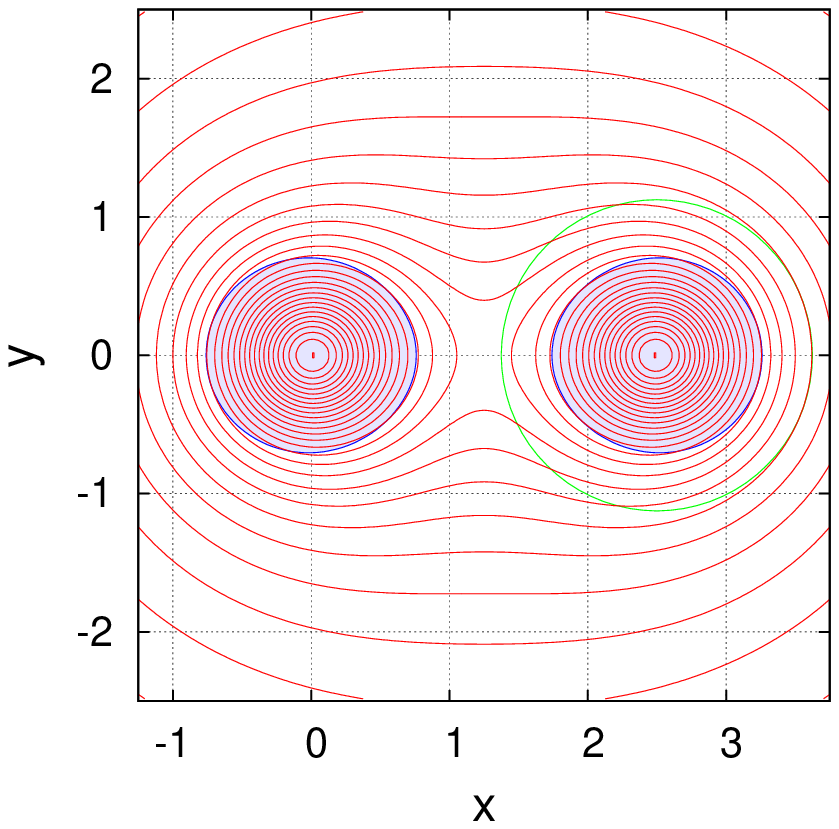}
\hskip 0.5cm
\includegraphics[width=0.75\columnwidth]{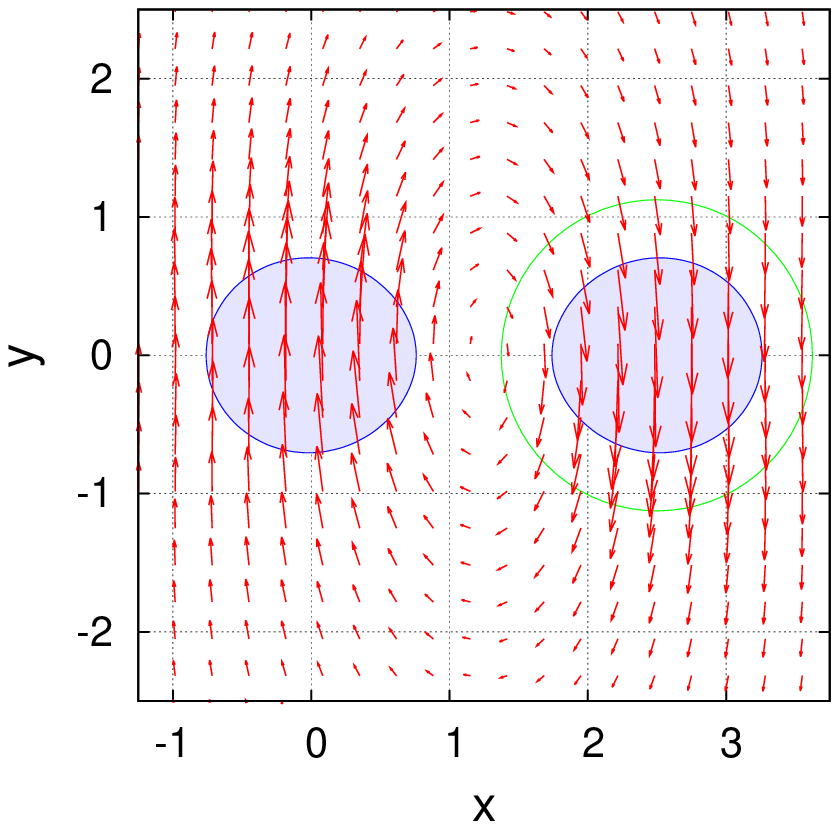}
\vskip 0.5cm
\includegraphics[width=0.75\columnwidth]{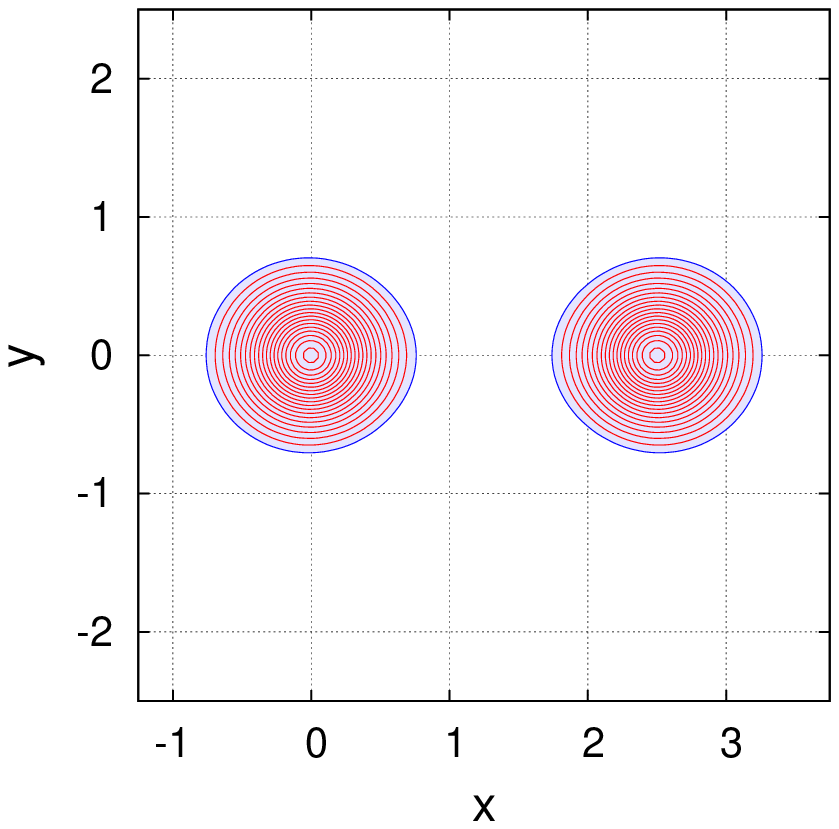}
\hskip 0.5cm
\includegraphics[width=0.75\columnwidth]{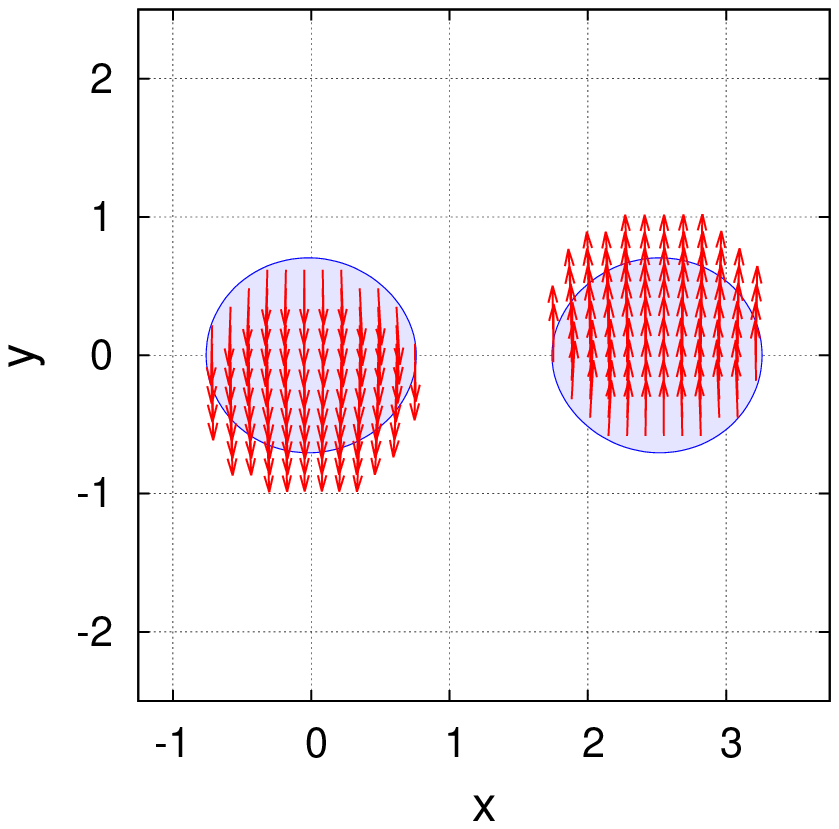}
\vskip 0.5cm
\includegraphics[width=0.75\columnwidth]{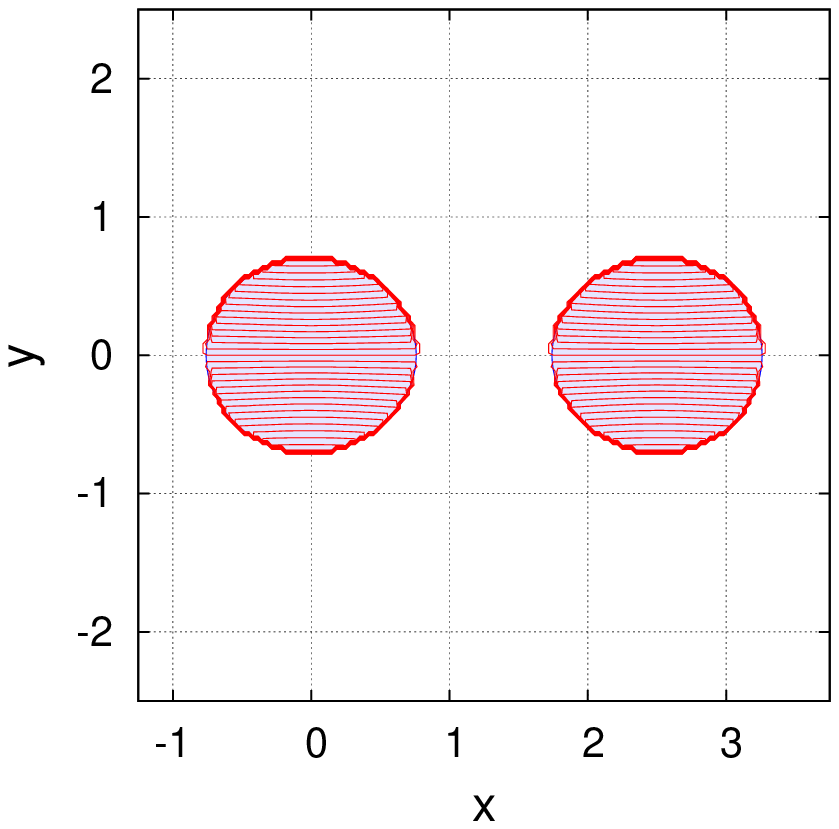}
\hskip 0.5cm
\includegraphics[width=0.75\columnwidth]{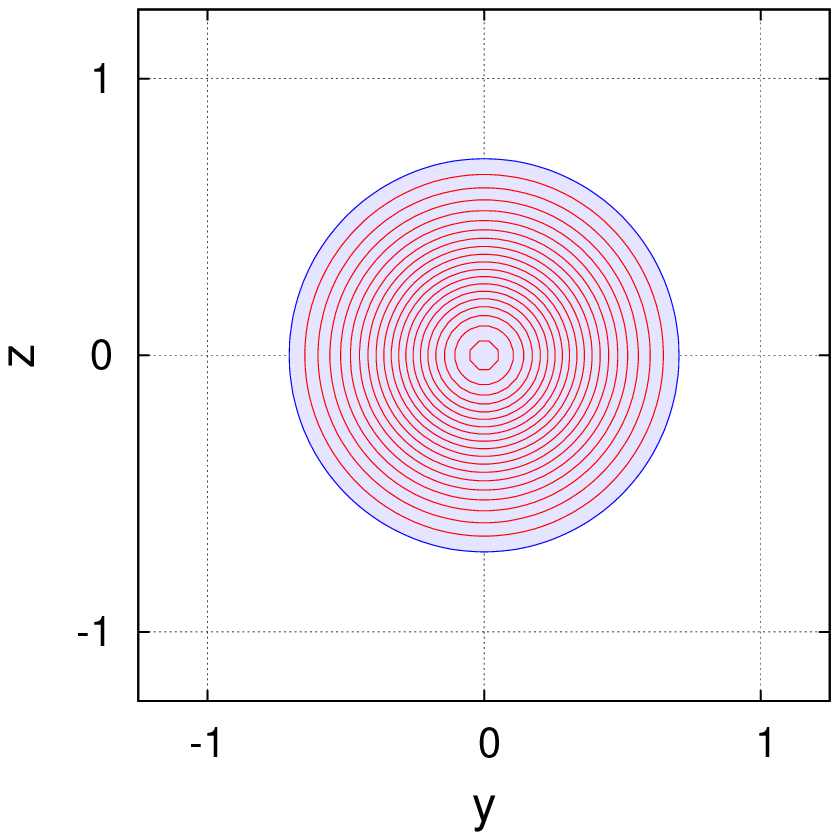}
\end{center}
\caption{Irrotational binary solution with $r_s=0.76$ and compactness
  $\mathcal{C}=0.18$. The separation between the two
  neutron stars is $d_s/r_s=2.5/0.76=3.29$.
	Left column: contour plots of the conformal factor
  $\psi$ from $1.0$ to $1.33$ with step $0.01$, of the rest-mass density
  from $0$ to $0.3$ with step $0.01$, and of the velocity potential
  $\Phi$ from $-0.2$ to $0.2$ with step $0.01$, all on the $(x,y)$
  plane. Right column: shift and fluid velocity vector fields on the
  $(x,y)$ plane, and contour plot of the rest-mass density on the $(y,z)$
  plane. Note that the green sphere corresponds to the excised sphere
  $S_e$ of COCP-1.}
\label{fig:irBNSa_rs0.76_c0.18}
\end{figure*}

\begin{table*}
\begin{tabular}{cl|ccccccccccccc}
\hline
\hline
Type & Patch  & $\ r_a\ $ & $\ r_s\ $ & $\ r_b\ $ & $\ r_c\ $ & $\ r_e\ $ & 
$\ \Nrf\ $ & $\ N_r^1\ $ & $\ \Nrm\ $ & $\ N_r\ $ & $\ N_\theta\ $ & $\ N_\phi\ $ & $\ L\ $ \\
\hline
${\rm Hs3d}$ & ${\rm COCP-1}$ & $0.0$ & $0.76$ & $10^2$ & $1.25$ & $1.125$ & $100$ & $128$ & $160$ & $384 $ & $96$ & $96$ & $12$ \\
             & ${\rm COCP-2}$ & $0.0$ & $0.76$ & $10^2$ & $1.25$ & $1.125$ & $100$ & $128$ & $160$ & $384 $ & $96$ & $96$ & $12$ \\
             & ${\rm ARCP}  $ & $5.0$ & $-$    & $10^6$ & $6.25$ & $-$     & $32 $ & $-$   & $40 $ & $384 $ & $96$ & $96$ & $12$ \\
\hline
\hline
\end{tabular}
\caption{Grid parameters used for the irrotational solution of
  compactness $\mathcal{C}=0.18$, presented in
  Fig. \ref{fig:irBNSa_rs0.76_c0.18}. The separation between the two
  neutron stars is $d_s/r_s=2.5/0.76=3.29$.}
\label{tab:irrotsol}
\end{table*}
 
Finally, in Fig. \ref{fig:corotrhoseq} we report sequences of corotating
binaries for increasing central rest-mass density at different
separations, from $d_s/R_0=4$, down to separation in which the two stars
are almost touching $d_s/R_0=2.125$. Here we use $r_s=1$, and therefore the 
radius of the neutron star is $R_0$.
Clearly, for any given central
energy density a larger mass is supported by the binary (supramassive solutions)
when we move to closer configurations, once again excluding the onset of an
instability to gravitational collapse to black hole
\cite{BCSST1998b,F1999,MW2000}.

\subsection{Irrotational sequences}
\label{ssec:irrotsol}

As anticipated in the Introduction, irrotational neutron stars have been
considered as a reasonable first approximation to describe the flow in
binary configurations. In such a case, the total angular momentum is less
than the corresponding of a corotating binary, since in each star there
is a flow in the counterdirection with respect to the orbital
motion\footnote{For this reason these binaries are also called
  \textit{counter-rotating configurations}.}. This has two consequences.
First the inspiral of irrotational binaries is faster than that of
corotating ones or, equivalently, the gravitational-wave frequency is
expected to increase with a faster rate for irrotational systems.  To
first order in the spins, the rate of inspiral is \cite{KWW1992,K1995}
\begin{eqnarray}
\frac{dr}{dt} = -\frac{64}{5}\nu\left(\frac{M}{r}\right)^3
\left\{ 1 - \frac{7}{12}\frac{1}{M^2}\left(\frac{M}{r}\right)^{3/2} \right. \nonumber\\
\left. \times \hat{\boldsymbol{L}}\cdot\left[\left(19+15\frac{M_2}{M_1}\right)\BS_1+
                                  \left(19+15\frac{M_1}{M_2}\right)\BS_2\right] \right\} \,, \quad
\label{eq:rdot}
\end{eqnarray}
where $\nu$ is the symmetric mass ratio and $\hat{\boldsymbol{L}}$ the unit angular momentum vector.
From \eqref{eq:rdot} we can see that
\be
 |\dot{r}^{\rm irr}|>|\dot{r}^{\rm cor}|    
\label{eq:coir_roi}
\ee
which is expected,
since spinning binaries have to radiate also the additional angular momentum before they merge
\cite{CLZ2006}. 
 
Second, in the light of the results obtained for binary black
holes, where binaries with larger spins lead to increasingly spinning
final black holes \cite{RDDPRSS2008,BR2009}, the irrotational binary
system will eventually lead to a Kerr black hole that is more slowly
rotating than the corresponding one produced by the corotating binary.

\begin{figure}
  \includegraphics[height=60mm]{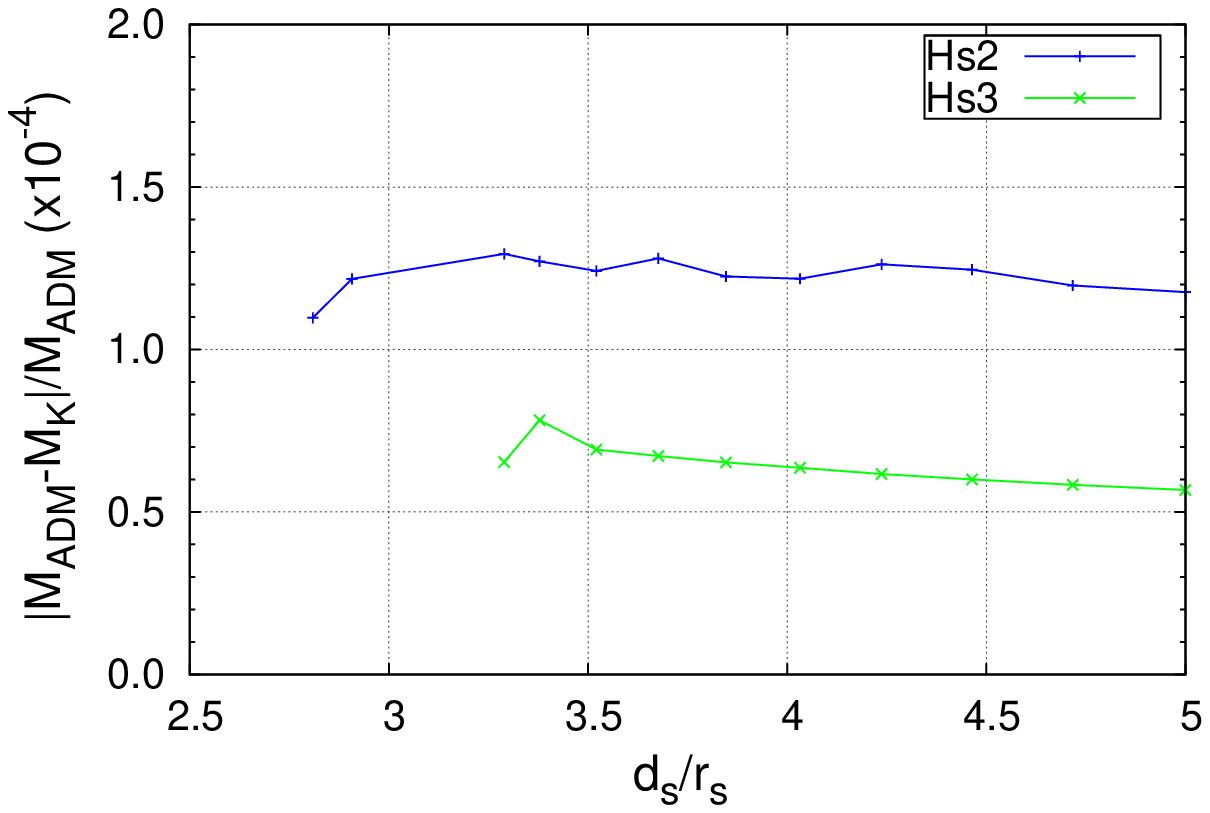}
  \includegraphics[height=60mm]{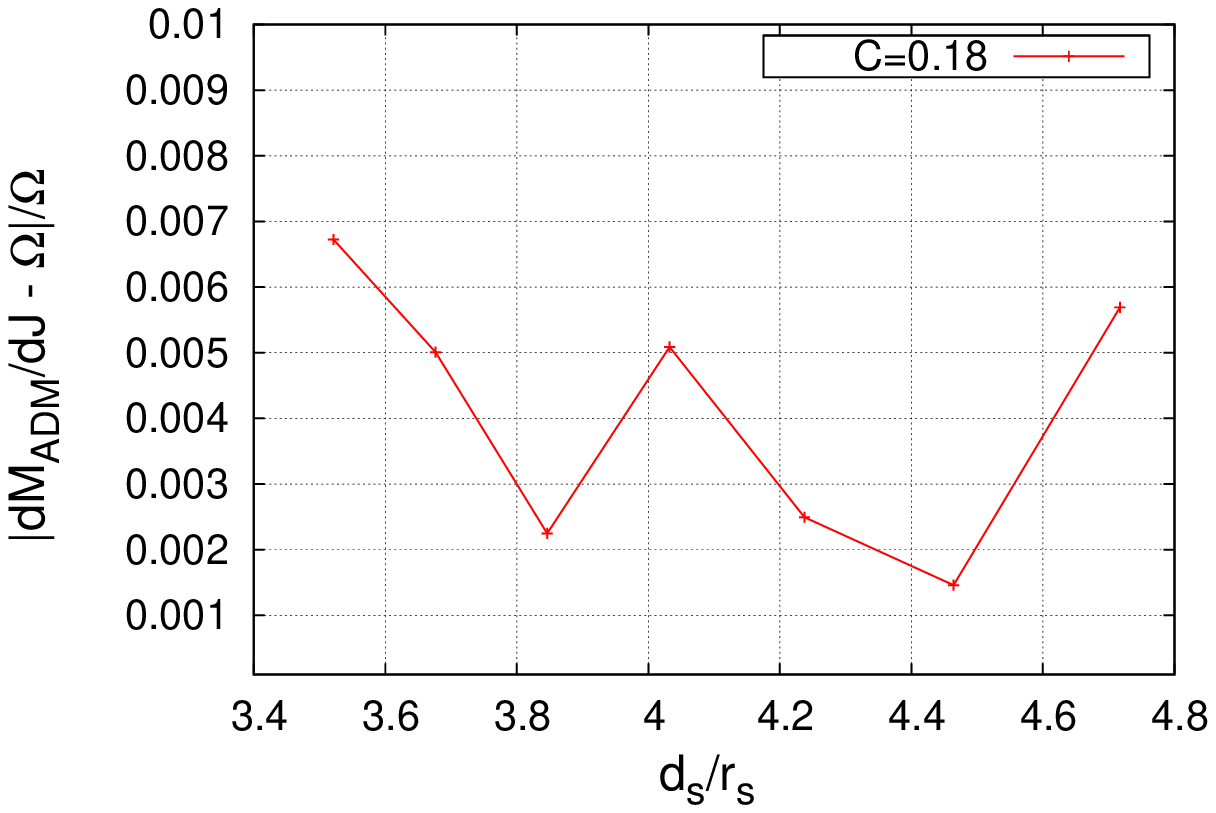}
	\caption{Top panel: relative error in the relation $M_{_{\rm
              K}}=M_{_{\rm ADM}}$ for irrotational sequences with
          $\mathcal{C}=0.18$ using the ${\rm Hs2d}$ grid of Table
          \ref{tab:corotsol} and the Hs3d grid of Table
          \ref{tab:irrotsol}. Bottom panel: relative error in the
          relation $dM_{_{\rm ADM}}=\Omega dJ$ for the irrotational
          sequences of compactness $\mathcal{C}=0.18$ as a function of
          the coordinate separation $d_s/r_s$.}
	\label{fig:irrotve}
\end{figure}

\begin{figure}
  \includegraphics[height=60mm]{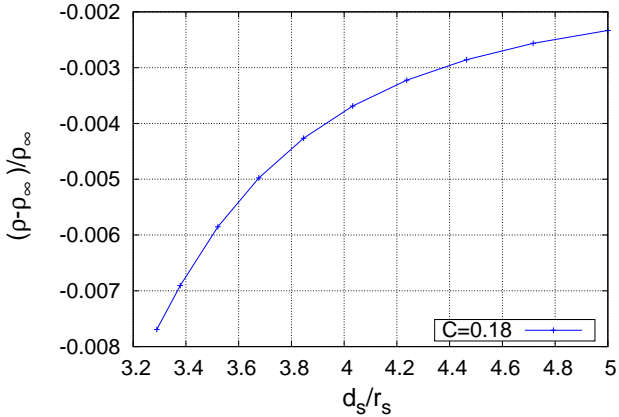}
	\caption{Relative change in the central rest-mass density for the
          irrotational sequences with $\mathcal{C}=0.18$ of
          Fig. \ref{fig:irrot_c0.12_c0.18} as a function of
          separation. Note that the decrease as the stars approach each
          other is 1 order of magnitude (or more) smaller than for
          corotating binaries (\cf Fig.~\ref{fig:corotrho}). }
	\label{fig:irrotrhorest}
\end{figure}

As done in Sec. \ref{ssec:corotsol} for corotating binaries, we compare
in Fig. \ref{fig:irrot_c0.12_c0.18} our irrotational solutions for
compactness $\mathcal{C}=0.12$ and $0.18$ against the corresponding
results presented in Ref. \cite{TG2002b}. Although we use here a
relatively small resolution, \ie ${\rm Hs2d}$ from Table
\ref{tab:corotsol}, the relative difference with Ref. \cite{TG2002b} is
again of the order of $0.05\%$. Note that for binaries with compactness
$\mathcal{C}=0.12$, the variable $r_s$ ranges from $r_s=0.5$ to
$r_s=0.87$, which corresponds to coordinate separations
$d_s/r_s=2.5/0.5=5$ and $d_s/r_s=2.5/0.87=2.87$, respectively. On the
other hand, for $\mathcal{C}=0.18$, $r_s$ varies from $r_s=0.5$ to
$r_s=0.79$, which corresponds to separations from $d_s/r_s=5$ to
$d_s/r_s=3.16$, respectively. Note also that the minimum in these plots
marks the mass-shedding limit and the creation on the equatorial plane of
a cusp in the rest-mass density.

In Fig. \ref{fig:irBNSa_rs0.76_c0.18} we present the results relative to
the irrotational binary solutions with $r_s=0.76$ and compactness
$\mathcal{C}=0.18$. More specifically, in the left column we report the
contour plots of the conformal factor $\psi$ from $1.0$ to $1.33$ with
step $0.01$, of the rest-mass density from $0.0$ to $0.3$ with step
$0.01$, and of the velocity potential $\Phi$ from $-0.2$ to $0.2$ with
step $0.01$, all on the $(x,y)$ plane. On the other hand, in the right
column we show the shift and the fluid velocity vector fields on the
$(x,y)$ plane, and a contour plot of the rest-mass density on the $(y,z)$
plane.

Similarly, in Fig. \ref{fig:irrotve} we report two global error
indicators computed for irrotational binaries with compactness
$\mathcal{C}=0.18$. More specifically, the top panel shows the fractional
difference in the Komar and ADM masses, while the bottom panel shows the
fractional error of the $dM_{_{\rm ADM}}=\Omega dJ$ relation;
note that the latter is a rather stringent test and that a fractional
error below $0.7\%$ gives us confidence on the accuracy of our solutions
already at an intermediate resolution. Finally, in
Fig. \ref{fig:irrotrhorest} we plot the relative change in the central
rest-mass density as the coordinate separation between the two stars is
reduced. This figure should be compared with the corresponding
Fig.~\ref{fig:corotrho} for corotating binaries and shows that again the
central rest-mass density decreases as the two stars approach, but also
that this decrease is smaller, of 1 order of magnitude or more, than in
the corotating case.

\begin{figure*}
\setlength{\abovecaptionskip}{30pt}
\includegraphics[width=0.99\columnwidth]{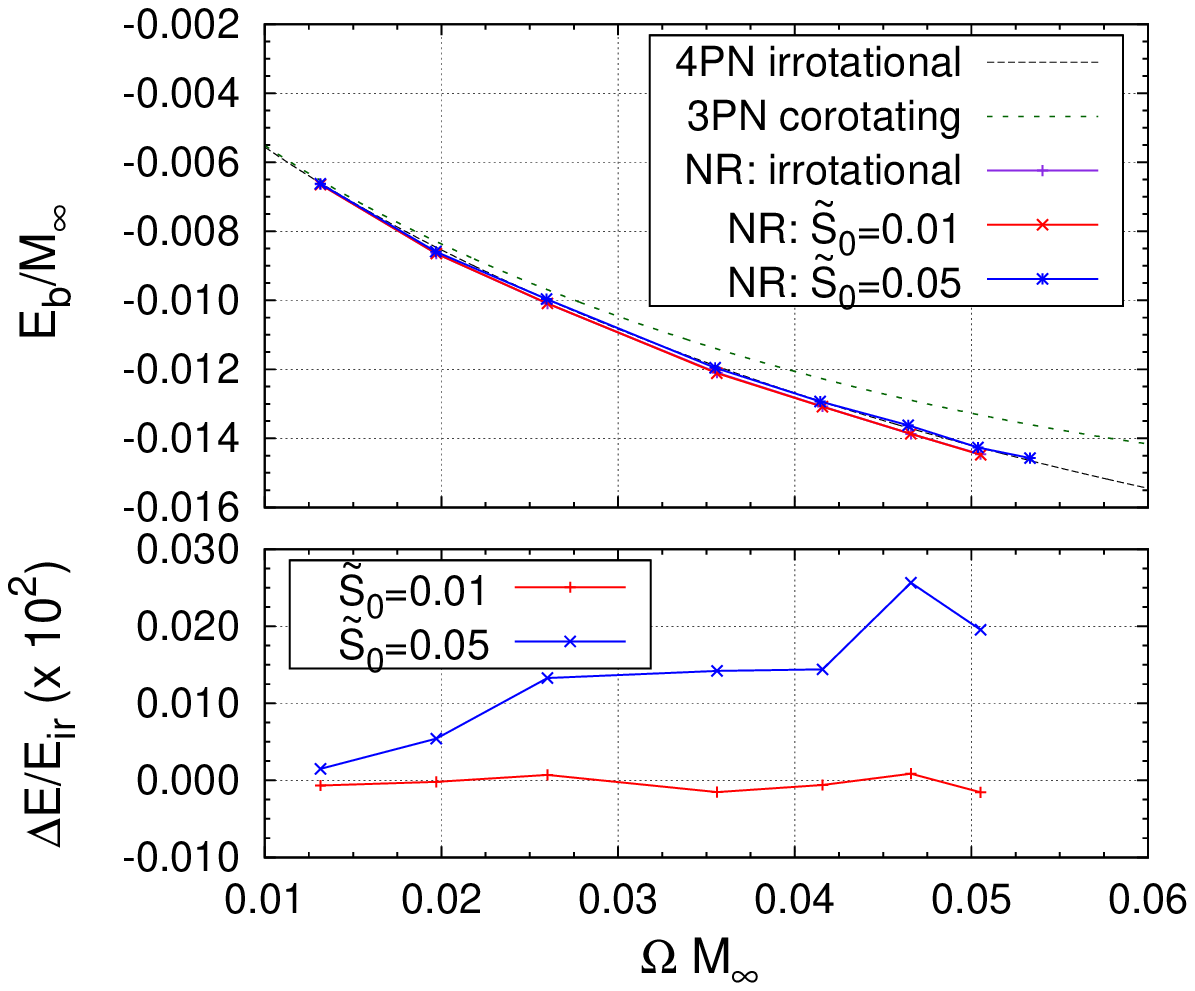}
\hskip 0.5cm
\includegraphics[width=0.99\columnwidth]{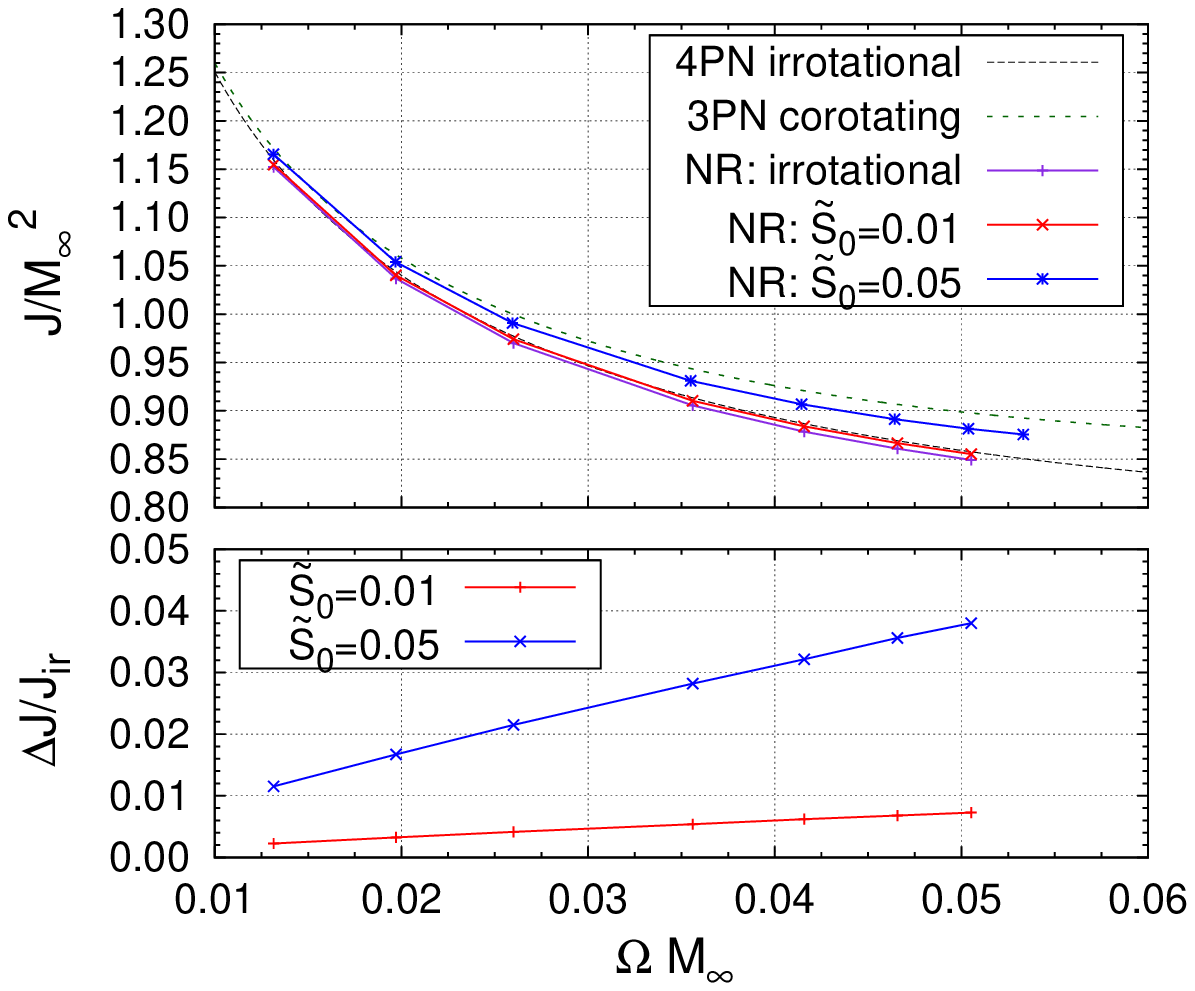}
	\caption{Top: dimensionless binding energy (left panel) and
          angular momentum (right panel) as a function of the
          dimensionless orbital frequency for sequences of constant rest
          mass neutron-star binaries with $M_0=1.5388\ M_\odot$. Shown
          with different lines are an: irrotational sequence (violet
          solid line) and two spinning ones with $A=0$ and either
          $\tilde{S}_0 = 0.01$ (red solid line) or $\tilde{S}_0 = 0.05$
          (blue solid line). All binaries are modeled with the APR1 EOS
          and are also shown for comparison is the 4PN irrotational (black
          dashed line), and the 3PN corotating (green dotted line)
          approximation.  Bottom: Fractional differences of the
          $\tilde{S}_0 = 0.01$ and $\tilde{S}_0 = 0.05$ spinning
          sequences with respect to the irrotational sequence, $\Delta
          E/E_{\rm ir} := E_{\rm sp}/E_{\rm ir}-1,\ \Delta J/J_{\rm ir}
          := J_{\rm sp}/J_{\rm ir}-1$.}
	\label{fig:spinAPR1_A0}
\end{figure*}

\begin{figure}
\includegraphics[height=60mm]{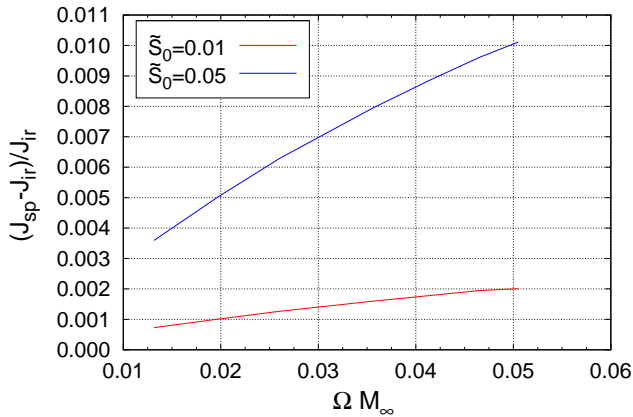}
	\caption{Spin contribution to the total angular momentum. Shown
          for the same binaries presented in Fig.~\ref{fig:spinAPR1_A0}
          but with $A=-6$, is the relative difference between the angular
          momentum of the spinning binaries and that of an irrotational
          binary.  Note that despite the short spin periods considered,
          the contribution of the spin angular momentum is at most $1\%$
          of the total, smaller than the $A=0$ case
          Fig.~\ref{fig:spinAPR1_A0}, where the same contribution was
          close to $4\%$.}
	\label{fig:spinAPR1_A0_A6}
\end{figure}

\subsection{Spinning sequences}
\label{sec:spinsol}

\begin{table*}
\begin{tabular}{ccccccccccc}
\hline
\hline
$\ \log p_1\ $ & $\ \Gamma_0\ $ & $\ \Gamma_1\ $ & $\ \Gamma_2\ $ & $\ \Gamma_3\ $ & 
$\ \log\GR_0\ $ & $\ \log\GR_1\ $ & $\ \log\GR_2\ $ & $\tilde {S}_0$ & $A$ & $(J_{\rm sp}-J_{\rm ir})/2$ \\
\hline
$33.943$ & $1.357$ & $2.442$ & $3.256$ & $2.908$ & $14.294$ & $14.700$ & $15.000$ & $\phantom{-}-$ & $-$ & $-$ \\
\hline
\hline
& & & & & & & & & & \\
\hline
\hline
$\ M_{_{\rm ADM}}\,[M_\odot]\ $ & $\ M_0\,[M_\odot]\ $ & $\ M_p\,[M_\odot]\ $ &
$\ R\,[{\rm km}]\ $ & 
$\quad\ M_{_{\rm ADM}}/R\quad $ & $\quad \log\rho_c\quad $ & $\quad \log p_c\quad $ & 
$\quad \log e_c\quad$ & $\tilde {S}_0$ & $A$ & $(J_{\rm sp}-J_{\rm ir})/2$ \\
\hline
%$1.990$ & $2.319$ & $2.553$ & $11.306$ & $0.260$ & $15.212$ & $35.740$ & $15.326$ & $0.00$ & $\phantom{-}-$  & $-$ \\
$1.666$ & $1.989$ & $2.294$ & $7.995$ & $0.307$ & $15.489$ & $36.340$ & $15.642$ & $\phantom{-}-$ & $\phantom{-}-$  & $-$ \\
$1.350$ & $1.539$ & $1.636$ & $9.138$  & $0.218$ & $15.221$ & $35.562$ & $15.276$ & $\phantom{-}-$ & $\phantom{-}-$  & $-$ \\
\hline
$2.661$ & $1.539$ & $1.636$ & $10.246$  & $-$ & $15.220$ & $35.559$ & $15.275$ & $0.01$ & $\phantom{-}0$ & $0.0225$ \\
$2.661$ & $1.539$ & $1.636$ & $10.249$  & $-$ & $15.220$ & $35.560$ & $15.275$ & $0.01$ & $-6$           & $0.0062$ \\
$2.661$ & $1.539$ & $1.636$ & $10.260$  & $-$ & $15.218$ & $35.556$ & $15.274$ & $0.05$ & $\phantom{-}0$ & $0.1176$ \\
$2.661$ & $1.539$ & $1.636$ & $10.248$  & $-$ & $15.219$ & $35.559$ & $15.275$ & $0.05$ & $-6$           & $0.0313$ \\
\hline                                                                                                     
\hline
\end{tabular}  
\caption{Top part: summary of the parameters in the piecewise polytropic
  EOS used to represent the APR1 EOS. Bottom part: the first lines report
  the main properties of the maximum-mass nonrotating
  configuration of the APR1 EOS; the second line reports the properties
  of the nonrotating configuration used to construct the constant
  rest-mass and spinning binary sequences computed in
  Sec.~\ref{sec:spinsol}. The last four lines report the stellar
  properties for the binary in the sequence having the smallest
  separation, with the last column providing a possible estimate of the
  maximum absolute contribution of the spin angular momentum, $(J_{\rm
    sp}-J_{\rm ir})/2$.}
\label{tab:sphdatamax}
\end{table*}

We conclude our discussion of the results with the new \cocal~by
presenting our first calculations of quasiequilibrium binary systems of
spinning neutron stars. The neutron-star matter is modeled using a
piecewise polytrope representation of the APR1 EOS \cite{APR1998}. As
mentioned in Sec.~\ref{ssec:eos}, an EOS with $N$ polytropic segments
requires $2N$ parameters to be specified, which can be thought of as one
adiabatic constant, $N-1$ dividing rest-mass densities, and $N$ adiabatic
indices. In Ref. \cite{RLOF2009} it was found that a number of tabulated
nuclear matter EOS can be modeled with three segments above nuclear
density and one in the crust, thus with a total of four polytropic
zones. The error in the approximation is $\sim 0.1\%$, or at worst $\sim
4\%$.  A fit with a minimum error was described in \cite{RLOF2009} that
had a fixed crust with $\Gamma_0=1.35692$, $K_0=3.59389\times 10^{13}$,
and three core zones with adiabatic exponents
$\{\Gamma_1,\Gamma_2,\Gamma_3\}$, joining the different pieces at
rest-mass densities $\GR_1=10^{14.7}\ \mathrm{gr/cm^3}$, and
$\GR_2=10^{15}\ \mathrm{gr/cm^3}$. Additional information on the
properties of the initial data are collected in Table
\ref{tab:sphdatamax}.

The spin contribution to the fluid velocity is expressed through the
spatial three-vector $\tilde{s}^i$ [\cf Eq. \eqref{eq:wi}], which we
express as
\begin{equation}
\tilde{s}^i = \tilde{S}_0(-y,x,0)\,,
\label{eq:spinconst0}
\end{equation}
where the Cartesian coordinates $x,y$ are centered in the COCP patch, and
the positive (negative) constant $\tilde{S}_0$ denotes the magnitude of
corotation (counter-rotation).

In Table \ref{tab:sphdatamax} we report the properties of a sequence of
binary neutron stars with constant rest mass $M_0=1.5388\,M_\odot$,
corresponding to an ADM mass $M_{_{\rm ADM}}=1.35\,M_\odot$ when the
stars are at infinite separation. The freedom in the choice of the spin
velocity vector defined in Eqs. (\ref{eq:wi}) and (\ref{eq:spinconst0})
has been fixed by taking $A=0$, while for $\tilde{S}_0$ we examine two
cases: $\tilde{S}_0=0.01$, which corresponds to a spinning period of
$\sim 3\, {\rm ms}$, and $\tilde{S}_0=0.05$, which corresponds to the
extreme case of a period $\sim 0.6\, {\rm ms}$. These choices correspond
to spinning periods that are more than twice smaller than those
considered in \cite{BDTB2014}, where the maximum value considered was
$6.7\,{\rm ms}$. Although it is rather unlikely that such small rotation
periods are encountered in reality in binaries about to merge, it is a
good consistency check for our new code and an exploration of its limits.

The top left panel of Fig. \ref{fig:spinAPR1_A0} reports the
dimensionless binding energy $E_{\rm b}/ M_{\infty}$ of the binary as a
function of the dimensionless orbital frequency $\Omega
M_{\infty}$. Considered and compared are an irrotational binary (violet
solid line) and two spinning binaries, one with $\tilde{S}_0 = 0.01$ (red
solid line) and another one with $\tilde{S}_0 = 0.05$ (blue solid
line). All binaries are 
modeled with the APR1 EOS using the grid parameters of Table \ref{tab:spinAPR1},
and both of the 
spinning binaries have velocity field with $A=0$. Also shown for
comparison is the irrotational fourth post-Newtonian (4PN) (black dashed
line) as well as the third post-Newtonian (3PN) corotating (green dotted
line) approximation \cite{BD2013,DJS2014,B2014,BMFB2013}. Explicit forms
for these curves are given in given in Appendix \ref{sec:pn}. In the top
right panel of Fig. \ref{fig:spinAPR1_A0} we report instead the analogue
curves for the dimensionless angular momentum $J/M^2_{\infty}$.

We also note that a closer comparison with a PN expression for spinning
neutron stars would be interesting. At the same time, it involves a
number of subtleties. In fact, in order to correctly plot such PN
spinning curves it is necessary to take into account two different
effects. First, at each separation (orbital frequency) the PN expressions
should use the correct values of the spins, which we recall increase (in
our case approximately linearly) along the constant rest-mass
sequence. This is made difficult by the fact that these spins cannot be
measured separately in our approach as they are part of the global
solution. These additional terms for aligned spins will typically move
the binding energy to more negative values relative to the irrotational
curve. Second, the masses appearing in the PN expression need to modified
to account for the spin kinetic energy. In contrast to binary black
holes, where it is possible to distinguish the irreducible mass from the
spin-induced mass \cite{BD2013,DJS2014,B2014,BMFB2013}, accounting for
this contribution is not easy for neutron stars. However, what is
important here is that these terms are positive and are will move the
quasiequilibrium curve upwards relative to the irrotational curve and
toward the corotating solution.

This is indeed the behavior that is shown by our solutions, which fall
between the irrotational sequence and the corotating sequence;
furthermore, the use of larger initial spins yields binding energies that
are systematically larger as part of the orbital kinetic energy is
channeled into ``spinning-up'' the stars. Clearly, the differences in the
binding energy with respect to the irrotational binaries are very small
even for these high spinning rates, with a maximum of $0.03\%$ as it can
be seen in the bottom left panel of Fig.~\ref{fig:spinAPR1_A0}. More
pronounced are the differences for the angular momentum which are
depicted on the bottom right panel of Fig.~\ref{fig:spinAPR1_A0} with a
maximum of $4\%$. Note also that the sign of $\tilde{S}_0$ determines the
relative position of the spinning sequence relative to the irrotational
one. In particular, with a negative value for $\tilde{S}_0$, the
angular-momentum curve for the spinning sequence would have appeared
below the irrotational one.

Finally, in Fig. \ref{fig:spinAPR1_A0_A6} we report an estimate of the
spin contribution to the total angular momentum for a different value of
$A$. For the same binaries presented in Fig.~\ref{fig:spinAPR1_A0} the
relative difference between the angular momentum of the spinning
binaries, $J_{\rm sp}$, and that of an irrotational binary, $J_{\rm ir}$
with $A=-6$ is at most $1\%$ of the total, smaller than the $4\%$ value
obtained with $A=0$ of Fig. \ref{fig:spinAPR1_A0}.  Note that this
quantity is not expected to be constant along this sequence, where only
the rest mass and the spinning coefficient $\tilde{S}_0$ are kept
constant.

If we estimate the dimensionless spin angular momentum to be
\begin{equation}
\chi:=\frac{S}{M_{_{\rm ADM}}^2} := 
\frac{1}{2}\left(\frac{J_{\rm sp}-J_{\rm ir}}{M_{_{\rm ADM}}^2}\right)\,,
\end{equation}
then because the ADM mass of the spinning binaries is very close to the
irrotational one and lies in the range $M_{_{\rm ADM}} \in [1.33,1.34]$,
the dimensionless spin takes values in the range $\chi \in [0.027,0.066]$
for the case with $A=0$ and $\tilde{S}_0=0.05$, and values in the
range $\chi \in [0.0017,0.0035]$ for the $A=-6$, $\tilde{S}_0=0.01$ case.
In all the cases considered we have found that the error estimates
discussed in the previous sections lead quite generically to relative
errors that are $\lesssim 0.7\%$, and that smaller errors can be obtained
with grids having a higher resolution.

\begin{table*}
\begin{tabular}{cl|ccccccccccccc}
\hline
\hline
Type & Patch  & $\ r_a\ $ & $\ r_s\ $ & $\ r_b\ $ & $\ r_c\ $ & $\ r_e\ $ & 
$\ \Nrf\ $ & $\ N_r^1\ $ & $\ \Nrm\ $ & $\ N_r\ $ & $\ N_\theta\ $ & $\ N_\phi\ $ & $\ L\ $ \\
\hline
${\rm Hs2.5d}$ & ${\rm COCP-1}$ & $0.0$ & ${\rm var}$ & $10^2$ & $1.25$ & $1.125$ & $76$ & $96$ & $120$ & $288 $ & $72$ & $72$ & $12$ \\
               & ${\rm COCP-2}$ & $0.0$ & ${\rm var}$ & $10^2$ & $1.25$ & $1.125$ & $76$ & $96$ & $120$ & $288 $ & $72$ & $72$ & $12$ \\
               & ${\rm ARCP}  $ & $5.0$ & $-$         & $10^6$ & $6.25$ & $-$     & $24$ & $-$  & $30 $ & $192 $ & $72$ & $72$ & $12$ \\
\hline
\hline
\end{tabular}
\caption{Grid parameters used for the irrotational and spinning sequences
  with the APR1 EOS, presented in Fig. \ref{fig:spinAPR1_A0}. }
\label{tab:spinAPR1}
\end{table*}

\section{Conclusion}

We have presented the extension of the \cocal~code to treat binary
configurations of compact stars within the IWM formalism of general
relativity. As with the work done for binary black holes, we have used
multiple coordinate patches so as to be able to treat asymmetric
binaries. Also in the spirit of previous work, we have introduced a
particular normalization scheme that allows us to accurately compute
binary systems that have small or large separations, recovering the
spherical limit at large distances. This is done by keeping the stars at
fixed coordinate positions, but artificially reducing their
radius. Furthermore, we have made use of surface-fitted coordinates to
describe accurately the stellar shape as it varies along sequences of
constant rest mass.

Also for the nonvacuum spacetimes considered here, we have employed the
KEH method \cite{OM1968,H1986a,H1986b,HEN1986a,HEN1986b,KEH1989}, in
which the gravitational equations are solved using Poisson solvers with
appropriate Green's functions, while for the conservation of rest mass we
employ a least-squared algorithm. The code makes use of a piecewise
polytropic description to represent the EOS of stellar matter and, for
the specific cases considered here, we have adopted the representation of
the APR1 EOS \cite{APR1998}.

Making use of a suitably adapted formulation described in
Ref. \cite{T2011}, the code is able to describe fluid flows within the
stars that corresponds to corotating, irrotational, but also spinning
binaries, As a validation of the numerical solutions, we have constructed
a number of sequences of corotating and irrotational binary neutron stars
having the same mass. The results for corotating and irrotational
binaries have been compared with those published from the pseudospectral
code \lorene~\cite{TG2002b}, and they revealed that the relative difference in
the results between the two codes is of the order of $0.05\%$, even when
a medium resolution is used for \cocal.

When considering spinning binaries, and although the code can handle
arbitrary rotation prescriptions for the individual stars, we have
concentrated here on the case of fluid flows in which the spins are
parallel to the orbital angular momentum. For this class of solutions,
and to explore the possible range of behaviors, we have considered
sequences with stars that either are slowly spinning or are spinning
at rates that are 10 times larger than those observed in binary pulsars
systems.  In all the cases considered, we have found that error estimates
of different types lead to relative errors that are $\lesssim 0.7\%$.

A number of applications of these results and of additional developments
of the code are expected to take place in the coming months. First, we
will explore the impacts of stellar spins in numerical simulations of
binary neutron stars; more specifically, by exploiting the high
convergence order of our new numerical general-relativistic code
\cite{RRG2014a}, we plan to extend the work carried out in
\cite{RRG2014b} for the inspiral part and the one recently published in
\cite{TRB2014a,TRB2014b} for the postmerger signal. Second, by
combining the approaches followed in the solution of binary black holes
and binary neutron stars, we will extend the code to handle also binaries
comprising a black hole and a neutron star of different masses and spin
orientation. Third, we will explore the space of solutions in which the
spins of the neutron stars are oriented arbitrarily as these are likely
to correspond to the most realistic configurations. Finally, working on a
parallelization of the code will allow us to obtain results with much
smaller computational costs, enabling us to provide public initial data
for spinning binary neutron stars under a variety of conditions.

\acknowledgments 

We thank John Friedman for carefully reading the manuscript and providing
useful input. We also thank Luc Blanchet, Alejandro Boh\'e, and Gerhard Sch\"afer for useful
discussions regarding the post-Newtonian approximation.
Partial support comes from the DFG Grant SFB/Transregio~7
and by ``NewCompStar'', COST Action MP1304. A.\,T. is supported by the
LOEWE-Program in HIC for FAIR.
This work was supported by JSPS Grant-in-Aid for Scientific Research(C) 23540314, 
15K05085, and 25400262.

\appendix

\section{Mass and angular momentum}
\label{sec:mam}

In this appendix we review the mathematical definitions of several of the
quantities that have been used to characterize the properties of the
binaries. We start with the rest mass of each star, $M_0$, defined as
an integral over the spacelike hypersurface $\Sigma_t$ of the rest-mass
density as measured by the comoving observers 
\begin{eqnarray}
M_0 & := & \int_{\Sigma_t} \GR u^\GA dS_\GA  =
          \int_{\Sigma_t} \GR u^\GA \nabla_\GA t \sqrt{-g} d^3x  \nonumber\\
& = & \int_{\Sigma_t} \GR u^t \GA \GC^6 r^2 \sin\GU dr d\GU d\GP \,.
\label{eq:m0}
\end{eqnarray}
In \cocal, integrals like this are computed in dimensionless form using
normalized coordinates. With the help of Eq. (\ref{eq:rho0}),
Eq. \eqref{eq:m0} is rewritten as 
\begin{equation}
\hat{M}_0  :=  R_0^3 \int_{\Sigma_t}  K_i^{{1}/(1-\Gamma_i)} q^{{1}/(\Gamma_i-1)} 
u^t \GA \GC^6 \hat{r}^2 \sin\GU d\hat{r} d\GU d\GP\,, 
\label{eq:barm0}
\end{equation}
where $K_i$ depends on $\hat{r}$. The integrand in Eq. (\ref{eq:barm0})
is evaluated on the gravitational coordinates therefore an interpolation
to the surface-fitted coordinates is needed before the integral
evaluation.

Next, a measure of the total energy of the system is given by the ADM mass,
$M_{_{\rm ADM}}$, which is defined as a surface integral at spatial
infinity as
\begin{eqnarray}
M_{_{\rm ADM}} & := & \frac{1}{16\pi}\int_\infty (f^{im}f^{jn}-f^{ij}f^{mn})\pd_j\GG_{mn} dS_i \nonumber\\
& = & -\frac{1}{2\pi}\int_\infty \pd^i\GC dS_i 
= -\frac{1}{2\pi}\int_\infty \frac{\pd\GC}{\pd r} r^2\sin\GU d\GU d\GP\,,
\nonumber \\
\label{eq:madm}
\end{eqnarray}
and which in normalized coordinates becomes
\begin{eqnarray}
\hat{M}_{\rm ADM} & := & -\frac{R_0}{2\pi} \int_{r=r_b}  \frac{\pd\GC}{\pd \hat{r}} \hat{r}^2\sin\GU d\GU d\GP\,.
\label{eq:barmadm}
\end{eqnarray}
Note that spatial infinity in \cocal~is represented by a spherical
surface with radius $r \approx 0.8 r_b$ of the ARCP coordinate
patch. Closely related to the ADM is the Komar mass of the binary, which
is related to the timelike Killing field $t^\GA$ and is defined as
\begin{eqnarray}
M_{_{\rm K}} & := & -\frac{1}{4\pi}\int_\infty \nabla^\GA t^\GB dS_{\GA\GB} 
= \frac{1}{4\pi}\int_\infty \pd^\GA dS_\GA \,,
\label{eq:mk}
\end{eqnarray}
or in normalized form as
\begin{eqnarray}
\hat{M}_{\rm K} & := & \frac{R_0}{4\pi} \int_{r=r_b}  \frac{\pd\GA}{\pd \hat{r}} \hat{r}^2\sin\GU d\GU d\GP \,.
\label{eq:barmk}
\end{eqnarray}

The angular momentum of the system is also calculated from a surface integral
at spatial infinity
\begin{eqnarray}
J & := & \frac{1}{8\pi} \int_\infty K^i_{\ j}\GP^j dS_i 
= \frac{1}{8\pi} \int_\infty A_{ij}\GP^j x^i r_\infty \sin\GU d\GU d\GP \,,
\nonumber \\
\label{eq:J}
\end{eqnarray}
where $\GP^i$ is the generator of the orbital trajectories and we have
used the maximal slicing gauge. The corresponding normalized quantity is
\begin{eqnarray}
\hat{J} & := & R_0^2\ \frac{1}{8\pi} \int_{r=r_b} 
							\hat{A}_{ij}\hat{\GP}^j \hat{x}^i \hat{r}_b \sin\GU d\GU d\GP \,.
\label{eq:barJ}
\end{eqnarray}
Finally, we also compute the ``proper mass'' of each star as the
integral of the total energy density measured by the comoving observer
\begin{equation}
M_p :=  \int_{\rm star} \epsilon u^\GA dS_\GA \,.
\label{eq:mp}
\end{equation}

\section{Isotropic coordinates TOV solver}
\label{sec:iTOV}

In this appendix we describe our implementation for obtaining spherical
solutions and the related rescaling that is used in \cocal. We can obtain
the same solutions using a one-dimensional KEH solver that mimics the
full three-dimensional code in a 3+1 setting. However, because most of
the time the TOV equations are presented in terms of Schwarzschild
coordinates while the actual calculations are performed in isotropic
coordinates, in what follows we show how to transform the system of
equations from Schwarzschild to isotropic coordinates without having to
go through a new derivation of equations and automatically obtaining a
smooth solution at the stellar surface.  The results are of course
identical to machine precision, at least for simple polytropes we have
checked. To the best of our knowledge this approach has not been
presented before in the literature.

We recall that the line element in Schwarzschild and in isotropic
coordinates is given, respectively, by
\begin{eqnarray}
ds^2 & = &  -A(r) dt^2 + B(r)dr^2 + r^2 d\Omega^2 \,, \label{eq:sds2}   \\ 
     & = &  -\GA^2(\bar{r})dt^2 +
\GC^4(\bar{r})(d\bar{r}^2+\bar{r}^2 d\Omega^2) \,, \label{eq:ids2}
\end{eqnarray}
with well known expressions for the functions
$A(r),B(r),\GA(\bar{r}),\GC(\bar{r})$ for the exterior of the star. For
the interior, instead, we need to solve the TOV equations
\begin{eqnarray}
\frac{dA}{dr} & = &  -\frac{2A}{\epsilon+p}\frac{dp}{dr} \,, 
\label{eq:dAdr}   \\ 
\frac{dp}{dr} & = &  -\frac{(\epsilon+p)(m+4\pi r^3p)}{r^2-2mr} \,, 
\label{eq:dPdr}
\end{eqnarray}
where
\begin{equation}
\frac{dm}{dr}  =  4\pi r^2 \epsilon \,, \qquad {\rm and} \qquad
B(r)  =  \frac{1}{1-2m(r)/r} \,.
\end{equation}
Of course it not difficult to derive the TOV equations in the isotropic
coordinates (\ref{eq:ids2}) and then perform a direct numerical
integration in these coordinates. However, this is not necessary and it
is possible to always work in Schwarzschild coordinates rescaling the
radial profile of the solution so as to make the surface of the star
appear at the correct position and automatically obtain a smooth solution
in $[0,\infty)$ without resorting to a postprocessing rescaling.

\begin{figure}
  \includegraphics[height=60mm]{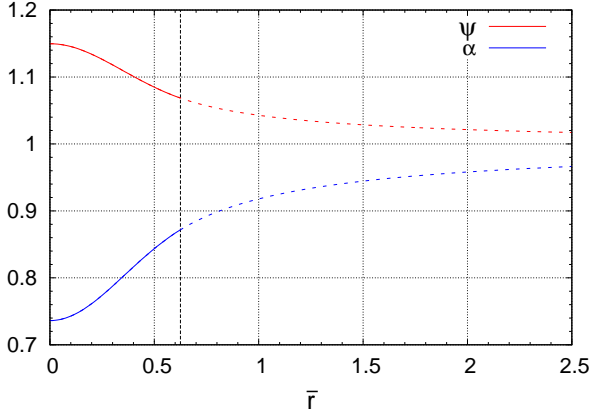}
	\caption{Solution of isotropic TOV using correct boundary
          conditions. A rescaling was done at the surface to be at
          $\bar{r}=0.625$.}
	\label{fig:tov}
\end{figure}

Comparing Eqs. (\ref{eq:sds2}) and (\ref{eq:ids2}) it is easy to deduce
that
\begin{equation}
\GC^2(\bar{r}) d\bar{r} =\sqrt{B} dr,\qquad\mbox{and}\qquad \GC^2(\bar{r}) \bar{r}=r\,,
\label{eq:alpsi_barr}
\end{equation}
which yield
\begin{equation}
\frac{dr}{d\bar{r}}=\frac{r}{\bar{r}}\sqrt{1-\frac{2m(r)}{r}} \,.
\label{eq:drdbarr}
\end{equation}
Using Eq. (\ref{eq:drdbarr}), we can rewrite the TOV system in terms of
the isotropic radial coordinate $\bar{r}$ as 
\begin{eqnarray}
\frac{dm}{d\bar{r}} &\ =\ & (4\pi r^2\epsilon)\frac{r}{\bar{r}}\sqrt{1-\frac{2m}{r}}\,,  \label{eq:f2}\\
\frac{dp}{d\bar{r}} &\ =\ &  -\frac{(\epsilon+p)(m+4\pi r^3p)}{\sqrt{1-{2m}/{r}}} \frac{1}{r\bar{r}}\,, \label{eq:f3} \\
\frac{d\GC}{d\bar{r}} &\ =\ &  \frac{\GC}{2\bar{r}}\left(\sqrt{1-\frac{2m}{r}}-1\right)\,,  \label{eq:f4}\\
\frac{d\GA}{d\bar{r}} &\ =\ &  \GA \frac{m+4\pi
  r^3p}{\sqrt{1-{2m}/{r}}}\frac{1}{r\bar{r}} \,, \label{eq:f5}
\end{eqnarray}
where we used Eq. (\ref{eq:alpsi_barr}) to derive Eqs. (\ref{eq:f4}) and
(\ref{eq:f5}). It is possible to simply integrate the system above to
obtain the star profile in isotropic coordinates. Initial values at
$\bar{r}=0$ are needed and although this is not a problem for $r$, $m$,
and $p$ (the latter being in general a free parameter), but the values of
$\GA,\ \GC$ are not available to have a smooth matching at the surface of
the star $\bar{r}=\bar{R}$, whose position is still unknown. On the other
hand, one way to obtain a smooth solution across the star's surface is to
exploit the coordinate transformations
\begin{eqnarray}
r & = & \bar{r}\left(1+\frac{M}{2\bar{r}}\right)^2 \,, \\
\bar{r} & = &  \frac{r}{2}\left[1+\sqrt{1-\frac{2M}{r}}-\frac{M}{r}\right]\,,
\end{eqnarray}
to derive the expressions of $\GC,\ \GA$ in the Schwarzschild coordinates
\begin{eqnarray}
\GC(r) & = &\frac{r}{M}\left(1-\sqrt{1-\frac{2M}{r}}\right)\,,  \\
\GA(r) & = & \sqrt{1-\frac{2M}{r}}\,.
\end{eqnarray}
Again making use of Eq. \eqref{eq:drdbarr} and of the analytic
integration of its left-hand side, Eq. \eqref{eq:f4} can be written as
\begin{equation}
d \ln \psi = F(r,m(r)) dr\,,
\end{equation}
where
\begin{equation}
F(r,m(r)):=\frac{1}{2r}\left(1-\frac{1}{\sqrt{1-{2m(r)}/{r}}}\right)\,,
\end{equation}
As a result, an integration between $r=0$ and $r=R$ of $F(r,m)$ leads to
the following condition for the conformal factor $\psi$
\begin{equation}
\GC(0)=\GC(R)\exp\left[-\int_0^R F(r,m(r)) dr \right]\,,
\end{equation}
so that $\GC$ (and similarly $\GA$) at the star's surface are guaranteed to
match smoothly the exterior solution, as it can be seen in
Fig. \ref{fig:tov}.

\begin{figure*}
  \includegraphics[width=0.8\textwidth]{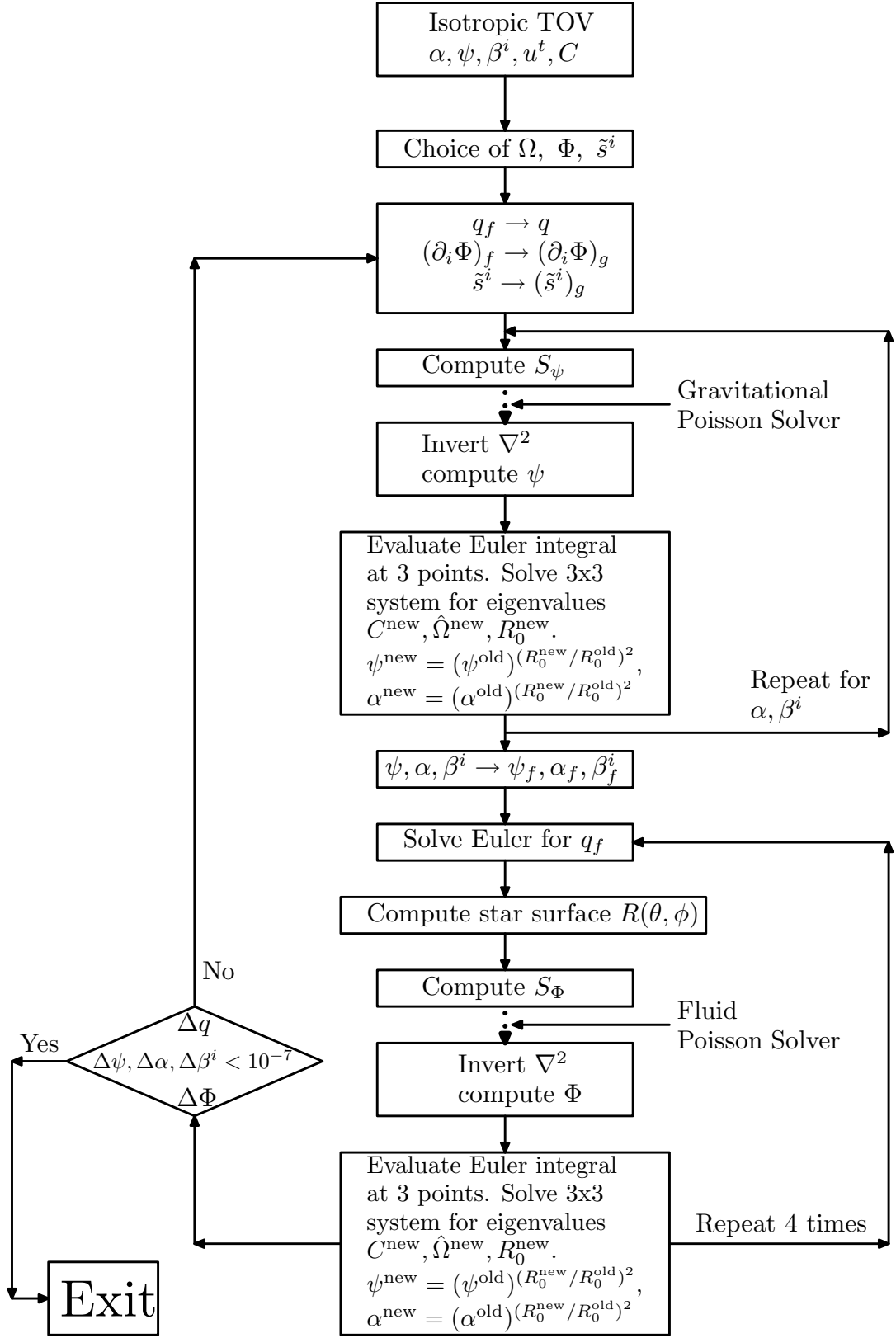}
	\caption{The \cocal~iteration for spinning binary stars.}
	\label{fig:iter_fc}
\end{figure*}

\section{Iteration scheme}
\label{ssec:is}

The iteration procedure for binary stars is similar to the one for binary
black holes described in Ref. \cite{UT2012}, but it also contains the
fluid coordinates, and a solution for the extra fluid variables in a
multipatch setting. An overall picture of this procedure is shown in
Fig. \ref{fig:iter_fc}, where the steps of the gravitational Poisson
solver (essentially everything from "Compute $S_\GC$" up to "Invert
$\nabla^2$, compute $\GC$") have been described in detail in
\cite{UT2012}.
 
The code first initializes the lapse function and the conformal factor
from some initial spherical solution. For that purpose we have developed
two methods. One is an isotropic TOV solver (see
Appendix~\ref{sec:iTOV}), and another is a one-dimensional KEH
method. This last choice reproduces the KEH approach used in
three-dimensional computations, but in a one-dimensional mesh. Included
in this method are all the important ingredients of the three-dimensional
code, such as the renormalization of variables. Comparing the results
from these two independent schemes gives us confidence about the
robustness of the \cocal~iterative solutions.
 
After a choice of the velocity fluid potential, of the orbital angular
velocity, and in the case of spinning binaries, also of the rotational
states of each compact object, the code proceeds to the main part of the
iteration, which always starts by interpolating $q=p/\GR,\ \pd_i \Phi$,
and $\tilde{s}^i$ from the surface-fitted coordinates to the
gravitational coordinates. The interpolated quantities are then used in
the gravitational Poisson solver, which is executed in addition to the
root-finding routine explained in Sec. \ref{ssec:dimnorm}. As discussed
there, the constants related to the Euler integral, the orbital angular
velocity, and the scaling of our grids $C,\ \hat{\Omega},\ R_0$, are
calculated at this point, and the lapse function, as well as the
conformal factor, are updated according to
\begin{equation}     
\psi^{\rm new}=(\psi^{\rm old})^{(R_0^{\rm new}/R_0^{\rm old})^2}, \quad 
		\alpha^{\rm new}=(\alpha^{\rm old})^{(R_0^{\rm
                    new}/R_0^{\rm old})^2}\,. 
\end{equation}

When the gravitational solver ends, $\GC,\GB^i,\GA$ are interpolated to
the surface-fitted coordinates in preparation for the fluid Poisson
solver. The main steps now are the computation of the new value of $q$,
by the use of Eq. (\ref{eq:irsph}) or Eq. (\ref{eq:corotei}), and then
the solution of the conservation of rest mass, Eq. (\ref{eq:cbm4}). At
this point, also
the surface of the star is computed. 

At each iteration step, the fluid computation is performed a few times
(4 times for the results presented here) since this results in a more
stable final computation. A relaxation parameter $\xi$ is used when
updating a newly computed variable. If $\Phi^{(n)}(x)$ is the $n$-th step
value, and $\widehat{\Phi}(x)$ the result of the Poisson solver, then the
$(n+1)$-th step value will be
\begin{equation} 
\Phi^{(n+1)}(x)\,:=\,\xi\widehat{\Phi}(x)+(1-\xi)\Phi^{(n)}(x)\,,
\end{equation}
where $0.1\leq \xi\leq 0.4$. Usually $\xi=0.4$ for $\GA,\GC,\GB^i,q$,
while $\xi=0.1$ for $\GP$.

\begin{table*}
\begin{tabular}{cl|ccccccccccccc}
\hline
\hline
Type & Patch  & $\ r_a\ $ & $\ r_s\ $ & $\ r_b\ $ & $\ r_c\ $ & $\ r_e\ $ & 
$\ \Nrf\ $ & $\ N_r^1\ $ & $\ \Nrm\ $ & $\ N_r\ $ & $\ N_\theta\ $ & $\ N_\phi\ $ & $\ L\ $ \\
\hline
${\rm WD}$ & ${\rm COCP-1}$ & $0.0$ & $1.0$ & $10^2$ & $1.50$ & $1.25$ & $64$  & $64$ & $96$   & $192$  & $144$ & $144$ & $12$ \\
           & ${\rm COCP-2}$ & $0.0$ & $1.0$ & $10^2$ & $1.50$ & $1.25$ & $64$  & $64$ & $96$   & $192$  & $144$ & $144$ & $12$ \\
           & ${\rm ARCP}$   & $5.0$ & $-$   & $10^6$ & $6.25$ & $-$    & $16$  & $-$  & $20$   & $192$  & $144$ & $144$ & $12$ \\
\hline
\hline
\end{tabular}
\caption{Grid parameters used for the white-dwarf solutions with
  $\Gamma=5/3$.}
\label{tab:WDcorotsol}
\end{table*}
\begin{figure*}
\begin{center}
\includegraphics[width=0.66\columnwidth]{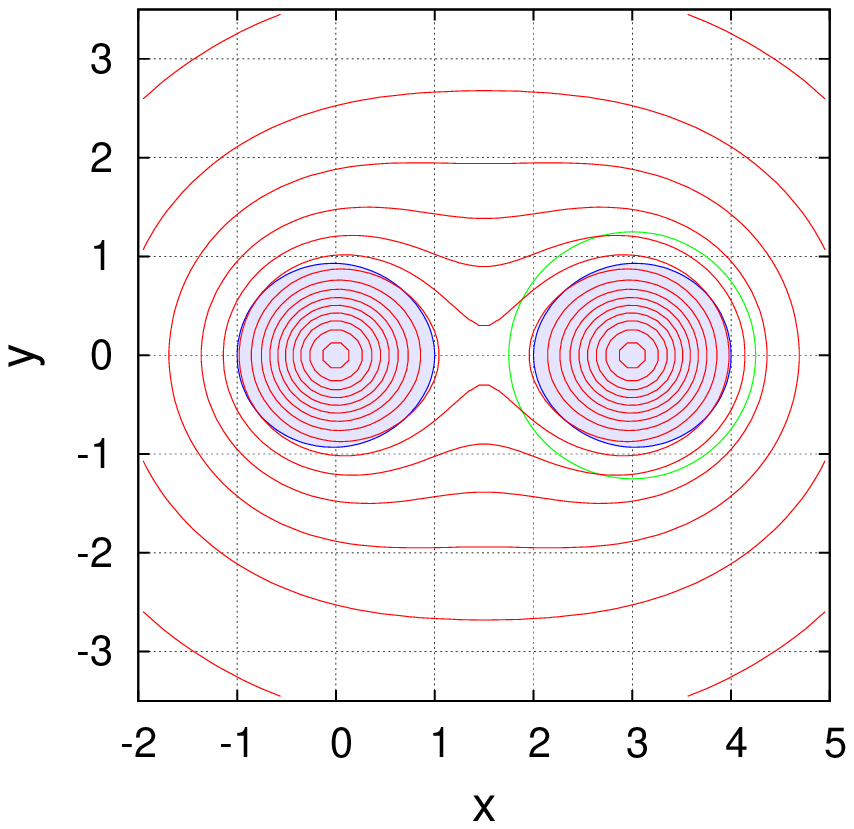}
\includegraphics[width=0.66\columnwidth]{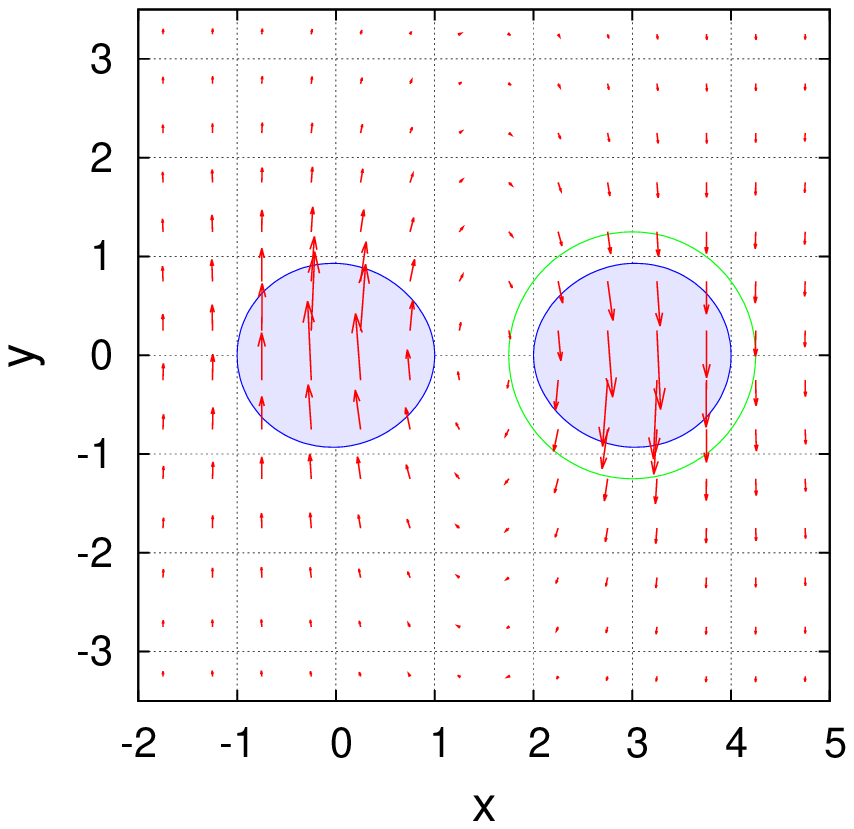}
\includegraphics[width=0.66\columnwidth]{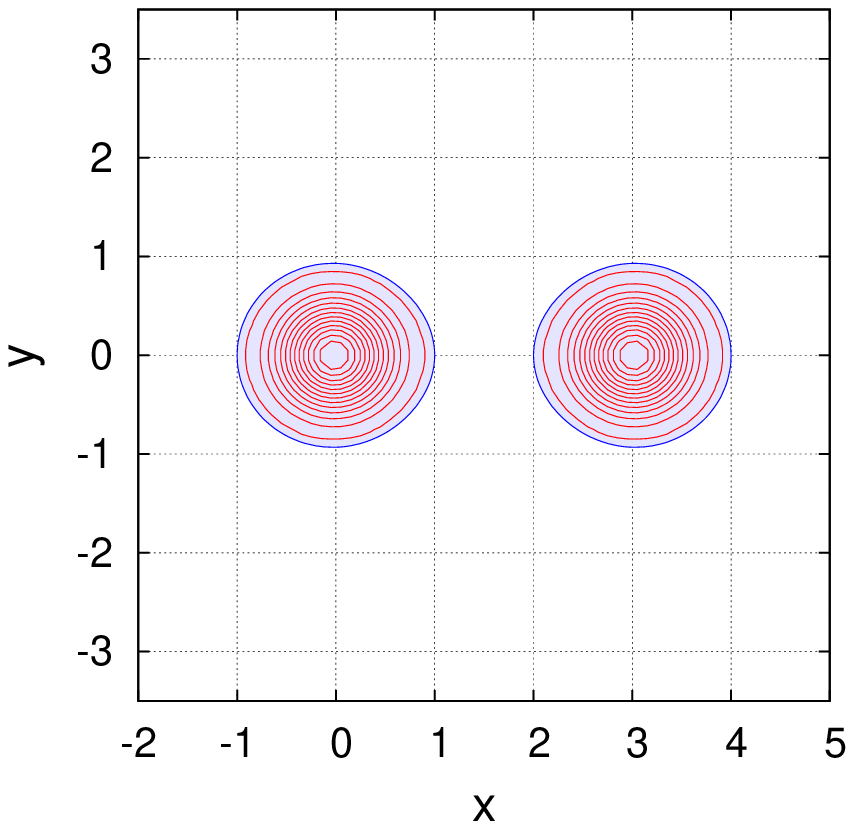}
\end{center}
\caption{Binary white dwarfs solution with compactness
  $\mathcal{C}=2\times\ 10^{-4}$, stellar centers placed at $x=\pm 1.25$
  and unit radii. Shown from left to right are: a contour plot of the
  lapse function from $0.9994$ to $1.0$ with step of $2\times\ 10^{-5}$,
  the shift vector field with maximum value $8.4\times 10^{-6}$, and a
  contour plot of the rest-mass density from $2\times 10^{-8}$ to $
  10^{-4}$ with step $8\times 10^{-8}$. Note that the green sphere
  corresponds to the excised sphere $S_e$ of COCP-1.}
\label{fig:WDsol}
\end{figure*}

The criterion used by \cocal~to stop the iteration is given by
\begin{equation}     
2\frac{|\Phi^{(n)}-\Phi^{(n-1)}|}{|\Phi^{(n)}|+|\Phi^{(n-1)}|}<
\epsilon_c \,, 
\end{equation}     
for all points of the grids, and all variables
$\GA,\ \GC,\ \GB^i,\ q,\ \Phi$, where we used $\epsilon_c = 10^{-6}$ in
this paper. In almost all of our calculations, $\GC$ and $\GA$ converge
to machine precision, while the error in the fluid variables $q$ and
$\Phi$ decreases to $\approx 10^{-12}$ before the error in the shift
reaches $10^{-7}$. This is due to the existence of points in the
gravitational mesh where the shift has almost zero values, and
convergence is much slower there. Neglecting such points can speed up a
solution by a factor of at least 2. Currently, \cocal~is running on a
serial processor and it needs around 3-4GB of RAM to produce the
solutions presented in this work. With an Intel Xeon $3.60\ \mathrm{GHz}$
processor, about two days are needed for these computations, with the
irrotational configurations taking longer than the spinning ones; this is
not a surprise since convergence is faster for corotating binaries.

\section{Corotating binary white dwarfs}
\label{sec:WDs}

To test the sensitivity of our code and to prepare for future work
concerning neutron star-white dwarf or black hole-white dwarf binaries,
we also compute a corotating binary white dwarf solution. Here the fields
are orders of magnitude less than the ones encountered in typical binary
neutron star binaries and greater resolution is required in order to
acquire smooth solutions. The resolution used is reported in Table
\ref{tab:WDcorotsol} where we can see that an increase in $N_\GU,\ N_\GP$
by a factor of 3 relative to the solutions obtained in
Fig. \ref{fig:corot_c0.12_c0.18}, Table \ref{tab:corotsol}, was used. In
Fig. \ref{fig:WDsol} we show a representative binary white dwarfs
solution with compactness $\mathcal{C}=2\times\ 10^{-4}$, with centers
placed at $x=\pm 1.25$ and unit radii. From left to right, the different
panels report the contour plot of the lapse function from $0.9994$ to
$1.0$ with step of $2\times\ 10^{-5}$, the shift vector field, and the
contour plot of the rest mass density. Note that the plots are centered
at the origin of the COCP-1 patch and the green circle refers to the excised sphere 
$S_e$. Values inside $S_e$ are taken from the COCP-2 patch.
The shift vector in
binary white dwarfs is approximately 4 orders of magnitude smaller
than the one typically encountered in neutron stars, while the quantity
$|\GA-1|$ is about 3 orders of magnitude smaller. Overall we see a
good convergence between the different coordinate systems even for these
small values of the metric quantities.

\section{Post-Newtonian approximation}
\label{sec:pn}

The 4PN approximation for the binding energy 
of a system of two nonspinning bodies with masses
$M_1,\ M_2$ and in quasicircular orbit has been used in
Fig.~\ref{fig:spinAPR1_A0} to compare with the numerical results of
irrotational and spinning binaries. The explicit expression
 is given by \cite{BD2013,DJS2014,B2014}
\begin{widetext}
\begin{eqnarray}
\frac{E_{\rm irr}}{Mc^2} & = & -\frac{\nu x}{2}\left\{1 + 
\left(-\frac{3}{4}-\frac{1}{12}\nu\right)x  +
\left(-\frac{27}{8}+\frac{19}{8}\nu-\frac{1}{24}\nu^2\right)x^2 
+ \left[-\frac{675}{64}+\left(\frac{34445}{576}-\frac{205}{96}\pi^2\right)\nu
          -\frac{155}{96}\nu^2 - \frac{35}{5184}\nu^3 \right]x^3 \right.  \nonumber\\
& & \qquad\qquad  + \left[-\frac{3969}{128}+\left(-\frac{123671}{5760} 
+ \frac{9037}{1536}\pi^2 + \frac{1792}{15}\ln2 +\frac{896}{15}\gamma_E\right)\nu \right. \nonumber\\
& &  \qquad\qquad  + \left. \left. \left(-\frac{498449}{3456}+\frac{3157}{576}\pi^2 \right)\nu^2 
+ \frac{301}{1728}\nu^3 + \frac{77}{31104}\nu^4 +\frac{448}{15}\nu\ln x\right]x^4\right\} \,, 
\label{eq:eb} 
\end{eqnarray}
\end{widetext}
where $\gamma_{_E}$ is the Euler constant, $M:=M_1+M_2$ is the total mass
of the system, $\nu:=M_1M_2/M^2$ is the symmetric mass ratio and $x$ is the
dimensionless orbital frequency
\begin{equation}
x:=\left(\frac{\Omega
  GM}{c^3}\right)^{2/3}\, .  
\label{eq:xpar} 
\end{equation}
For a corotating binary the binding energy also includes terms from the kinetic
energy of the spins as well as from the spin orbit interaction. To 3PN this
extra contribution is   
\be
\frac{\Delta E_{\rm cor}}{Mc^2} = (2-6\nu)x^3+(-10\nu+25\nu^2)x^4 \, ,
\label{eq:ebcor}
\ee
and therefore the total binding energy is given by the sum of \eqref{eq:eb} and \eqref{eq:ebcor}.

For a system that satisfies the first
law of binary mechanics and has binding energy of the form \be
E_b(x)=\sum_{i=1}^{N}(A_i+B_i\ln x) x^{i/2}
\label{eq:ebgen}
\ee
the angular momentum will be
\be
J(x)=\sum_{i=1}^{N}\left[\frac{i}{i-3}(A_i+B_i\ln x)-\frac{6B_i}{(i-3)^2}\right] x^{\frac{i-3}{2}}  \,.
\label{eq:jgen}
\ee
Using \eqref{eq:jgen} the irrotational part of the angular momentum is
\begin{widetext}
\begin{eqnarray}
\frac{J_{\rm irr}}{GM^2/c} & = & \frac{\nu}{\sqrt{x}}\left\{1 + 
\left(\frac{3}{2}+\frac{1}{6}\nu\right)x  +
\left(\frac{27}{8}-\frac{19}{8}\nu+\frac{1}{24}\nu^2\right)x^2 
+ \left[\frac{135}{16}+\left(-\frac{6889}{144}+\frac{41}{24}\pi^2\right)\nu
+\frac{31}{24}\nu^2 + \frac{7}{1296}\nu^3 \right]x^3  \right. \nonumber\\
& & \qquad\quad\  + \left[\frac{2835}{128}+\left(\frac{98869}{5760} 
- \frac{6455}{1536}\pi^2 - \frac{256}{3}\ln2 -\frac{128}{3}\gamma_E\right)\nu \right. \nonumber\\
& &  \qquad\quad\  + \left. \left. \left(\frac{356035}{3456}-\frac{2255}{576}\pi^2 \right)\nu^2 
- \frac{215}{1728}\nu^3 - \frac{55}{31104}\nu^4 -\frac{64}{3}\nu\ln x\right]x^4\right\}\,,  
\label{eq:jpn} 
\end{eqnarray}
\end{widetext}
while the corotating additional part is
\be
\frac{\Delta J_{\rm cor}}{GM^2/c} = (4-12\nu)x^{3/2}+(-16\nu+40\nu^2)x^{5/2} \, .
\label{eq:jcor}
\ee

%--------------------------------------------------------------
\bibliographystyle{apsrev4-1}

%--------------------------------------------------------------

\end{document}